\documentclass[
reprint,
groupedaddress,
amsmath,amssymb,
aps,
prb,
floatfix,
]{revtex4-2}

\usepackage{CJK}
\usepackage{amsmath}
\usepackage{physics}
\usepackage{amsfonts}
\usepackage{graphicx}
\usepackage[caption=false]{subfig}
\usepackage{float}
\usepackage{color}
\usepackage{empheq}
\usepackage{xcolor} % Colors
\usepackage{soul} % Highlighting and strikeout
\usepackage{tikz}
\usetikzlibrary{arrows.meta,positioning,calc}

\newcommand{\qbinom}[2]{\genfrac{[}{]}{0pt}{}{#1}{#2}_{2}}
\newtheorem{proposition}{Proposition}

\begin{document}

\title{Coherent error induced phase transition}

\author{Hanchen Liu}
\email[]{hanchen.liu@bc.edu}
\author{Xiao Chen}
\email{chenaad@bc.edu}

\affiliation{Department of Physics, Boston College, Chestnut Hill, Massachusetts 02467, USA}

\date{\today}

\begin{abstract}
We investigate the stability of logical information in quantum stabilizer codes subject to coherent unitary errors. Beginning with a logical state, we apply a random unitary error channel and subsequently measure stabilizer checks, resulting in a syndrome-dependent post-measurement state. By examining both this \emph{syndrome state} and the associated syndrome distribution, we identify a phase transition in the behavior of the logical information. In the Clifford/stabilizer setting, the change of logical stabilizer structure in a syndrome branch determines the state-level MAP Pauli-frame return probability for a chosen logical-basis input. We separately use quantum coherent information to diagnose recoverability of arbitrary encoded inputs at the channel level. Below a critical error threshold \(p_c\), the syndrome branches remain compatible with the original logical sector, enabling a high state-return probability. Above \(p_c\), the data are consistent with effective scrambling of the encoded subspace relative to the stabilizer checks, where the syndrome-resolved logical maps approach a global-Clifford-induced stabilizer instrument. This common post-threshold picture organizes both the toric-code and finite-rate random-stabilizer-code results within the same scrambling mechanism. The syndrome distribution supplies complementary local and global structural diagnostics whose scaling depends on the code family. We refer to this phenomenon as a \emph{coherent error induced phase transition}. To illustrate this transition, we study two classes of quantum error-correcting codes, the toric code and finite-rate random stabilizer codes, thereby shedding light on the design and performance limits of quantum error correction under coherent errors.
\end{abstract}

\maketitle

\section{Introduction}
\label{sec:intro}

Quantum error-correcting codes (QECCs) are central to fault-tolerant quantum computation, particularly as quantum platforms progress from the noisy intermediate-scale quantum (NISQ) era toward fully fault-tolerant quantum computing (FTQC) \cite{preskill2018quantum, nielsen2010quantum, aharonov1997fault}. Preserving encoded quantum information against noise and decoherence is essential for scalable and reliable quantum technologies. Over the past few decades, a broad range of quantum error-correcting codes (QECCs) have been developed, including topological codes such as the surface code \cite{kitaev2003fault,fowler2012surface, dennis2002topological}, hypergraph-product (HGP) quantum low-density parity-check (qLDPC) codes \cite{tillich2013quantum, leverrier2015quantum}, and random Clifford codes \cite{brown2013short, nelson2023fault, PhysRevResearch.6.023055, PhysRevX.11.031066, turkeshi2024error}. These code families differ in their resource requirements, threshold error rates, and practical implementation prospects.

A central theme in the field is the characterization of \emph{error thresholds}---the critical error probabilities or strengths at which a QECC transitions from effective correction to failure \cite{aharonov1997fault, knill1998resilient, kitaev2003fault}. Such thresholds often admit an interpretation analogous to phase transitions in statistical mechanics \cite{dennis2002topological, chubb2021statistical}, thereby providing a useful framework for benchmarking error-correction protocols and guiding the design of new codes and decoding algorithms.

Early work focused primarily on incoherent noise models, such as random Pauli channels \cite{RevModPhys.87.307, nielsen2010quantum, dennis2002topological}. Despite their simplified form, these models have yielded deep insight into fault tolerance and threshold behavior, including the existence of finite thresholds for many codes in the thermodynamic limit \cite{knill1998resilient, aharonov1997fault, dennis2002topological, PhysRevX.2.021004, PhysRevResearch.6.023055, PhysRevX.11.031066, turkeshi2024error, putz2025learning, wang2025decoherence}. More recently, growing attention has been devoted to \emph{coherent} unitary errors \cite{bravyi2018correcting, chen2024nishimori, PhysRevLett.131.200201, gskb-t5ql, behrends2024surface, marton2023coherent, PhysRevLett.131.060603, PhysRevResearch.6.013137, bao2024phases, cheng2024emergent, eckstein2024robust}.
Compared with incoherent noise, coherent errors can produce qualitatively different effects. Under suitable time-reversal-symmetry and code-parity conditions, syndrome measurement can induce unitary scrambling, driving the ensemble of syndrome-conditioned logical unitaries toward a unitary design~\cite{cheng2024emergent}. More general coherent-noise models present additional challenges: continuous rotations can lead to complex-weight statistical-mechanics descriptions, while generic dynamics can produce severe computational bottlenecks associated with volume-law entanglement \cite{skinner2019measurement, gullans2020dynamical, li2019measurement, li2018quantum, Fisher2022qey, bravyi2018correcting, zhao2021analytic, PhysRevA.77.042322, PhysRevA.76.022304, liu2023quantum, gskb-t5ql, behrends2024surface, bao2024phases}. Much of the earlier toric- and surface-code literature on coherent noise focused on continuous single-qubit over-rotation channels and on the resulting logical noise after decoding \cite{iverson2020coherence, gskb-t5ql, behrends2024surface, bao2024phases, cheng2024emergent}. Important progress has been made through tensor-network and transfer-matrix methods, statistical-mechanics mappings, and mixed-state topological diagnostics to characterize coherent-noise thresholds and associated phase structure in
topological codes
\cite{darmawan2017tensor,gskb-t5ql,behrends2024surface,bao2024phases,
cheng2024emergent,hchr-rqq9}.

In this work, we study this problem from a complementary perspective by considering coherent-error-induced transitions within a random Clifford \(q\)-unitary error ensemble. This framework naturally includes multi-qubit coherent errors and extends straightforwardly to finite-rate random stabilizer codes. By remaining within the Clifford/stabilizer regime, we retain computational tractability: the encoding, coherent error channel, and syndrome-resolved state analysis all stay inside a stabilizer-compatible setting, thereby avoiding volume-law simulation bottlenecks and enabling simulations at relatively large system sizes.

The Clifford restriction also delimits the scope of our conclusions.  Our ensemble does not include generic continuous non-Clifford over-rotations~\cite{bravyi2018correcting,iverson2020coherence,gskb-t5ql,behrends2024surface,bao2024phases}, coherent errors accumulated over repeated QEC cycles~\cite{marton2023coherent}, or temporally correlated noise~\cite{schultz2021schwarma}; it also excludes nonunitary processes such as amplitude damping~\cite{darmawan2017tensor} and leakage~\cite{marshall2025leakage}.  Consequently, the exact stabilizer identities, discrete logical-group statistics, and syndrome-independence properties derived below rely on the Clifford structure, and the numerical thresholds and critical exponents reported here are not claimed to be universal for arbitrary coherent noise.  Nevertheless, the post-threshold Clifford data are consistent with a simple coarse-grained picture: for \(p>p_c\), the coherent error effectively scrambles the encoded state relative to the stabilizer checks, and the resulting syndrome-resolved logical maps approach the statistics of a global-random-Clifford-induced stabilizer-instrument ensemble.  For a fixed stabilizer input, the normalized branches correspondingly exhibit random-stabilizer-state statistics~\cite{DahlstenPlenio2006,SmithLeung2006}. We expect a global-Haar/Wishart instrument benchmark, with Clifford stabilizer statistics replaced by Haar/Page-type statistics~\cite{ZyczkowskiSommers2001,Page1993}, to describe sufficiently scrambling non-Clifford coherent errors; establishing this beyond the stabilizer-simulable setting remains an open question.

Our approach formulates the threshold problem directly in terms of two objects: the syndrome-resolved post-measurement state and the corresponding syndrome distribution. The latter possesses a statistical-mechanics structure governed by emergent hard \(\delta\)-function constraints, yielding a zero-temperature hard-constraint \(\mathbb{Z}_2\)-type model with nonnegative weights. This differs conceptually from the complex-weight partition functions that often arise in continuous coherent-noise settings. Moreover, the transition studied here is distinct from standard measurement-induced entanglement transitions in monitored random circuits~\cite{skinner2019measurement, gullans2020dynamical, li2019measurement, li2018quantum, Fisher2022qey, PhysRevLett.127.235701}: rather than arising from competition between unitaries and measurements in a dynamically monitored system, it concerns the recoverability and scrambling of \emph{pre-encoded} logical information under coherent errors.

A key theoretical ingredient developed in this work is a simple stabilizer diagnostic for the post-measurement logical structure. In the Clifford/stabilizer setting, we derive an exact relation between the sign-free change of logical stabilizer structure and the state-level MAP Pauli-frame return probability for a fixed logical-basis input. This gives a direct operational interpretation to the logical-group diagnostic without identifying it with a state-independent decoder. We address the stronger question of channel-level reversibility separately through coherent information.

Our results show that beyond a critical error threshold \(p_c\), syndrome measurements can map the original logical stabilizer structure to a different one, signaling logical-state conversion or scrambling of the encoded information. By combining several diagnostics across quantum and classical descriptions---including logical-group changes, the state-level MAP return probability, quantum coherent information, and conditional mutual information---we show that coherent errors can induce effective unitary scrambling within the logical subspace, in addition to conventional information loss. The reduced free entropy of the syndrome distribution provides a complementary measure of the global constraint density, without being assumed to locate the same transition. Below \(p_c\), the encoded logical information remains stable within the original logical sector. We refer to this global restructuring as a \emph{coherent error induced phase transition}.

To illustrate these ideas, we study two representative classes of QECCs: the toric code, as a canonical topological code \cite{dennis2002topological, kitaev2003fault}, and finite-rate random stabilizer-code ensembles, including HGP codes and random Clifford codes \cite{tillich2013quantum,leverrier2015quantum, brown2013short,nelson2023fault, PhysRevResearch.6.023055, PhysRevX.11.031066}. Through numerical simulations, we find evidence for a critical error density \(p_c\) separating a regime in which encoded logical information remains accessible from one in which it is lost or scrambled beyond straightforward syndrome-based recovery.

The remainder of this paper is organized as follows. In Sec.~\ref{sec:prelim}, we review the fundamentals of quantum error-correcting codes and formulate the coherent-error setting studied in this work. In Sec.~\ref{sec:TC}, we apply our methods to the toric code and present numerical evidence for the coherent error induced phase transition. In Sec.~\ref{sec:rsce}, we investigate the same phenomenon in two representative examples of finite-rate random stabilizer-code ensembles, namely HGP codes and random Clifford codes. Finally, Sec.~\ref{sec:conclusion} summarizes our conclusions and outlines directions for future research.

\section{Preliminaries}\label{sec:prelim}

We begin by reviewing the basic concepts of quantum stabilizer codes~\cite{nielsen2010quantum,RevModPhys.87.307}.  Let
\begin{equation}
    \mathcal P_N
    =
    \left\{
        i^m P_1\otimes\cdots\otimes P_N
        \,\middle|\,
        m\in\mathbb Z_4,\;
        P_j\in\{I,X,Y,Z\}
    \right\}
\end{equation}
denote the full, phase-inclusive \(N\)-qubit Pauli group.  An
\([[N,k,d]]\) stabilizer code encodes \(k\) logical qubits into \(N\)
physical qubits and is specified by \(N-k\) independent commuting Hermitian
Pauli check operators \(\{h_i\}_{i=1}^{N-k}\).  These operators generate the
Abelian stabilizer group
\begin{equation}
    \mathcal S
    =
    \langle h_1,\ldots,h_{N-k}\rangle
    < \mathcal P_N,
    \quad
    -I\notin\mathcal S.
\end{equation}
Throughout this section, \(H<G\) denotes that \(H\) is a subgroup of \(G\);
the inclusion need not be proper.
The corresponding \(2^k\)-dimensional code space is their simultaneous \(+1\)-eigenspace,
\begin{equation}
    \mathcal H_{\mathrm{code}}
    =
    \left\{
        \ket{\psi}
        \,\middle|\,
        h_i\ket{\psi}=\ket{\psi},
        \quad i=1,\ldots,N-k
    \right\}.
\end{equation}

The logical Pauli group is defined as the centralizer of
\(\mathcal S\) in \(\mathcal P_N\), modulo \(\mathcal S\):
\begin{equation}
    G_{\mathcal L}
    =
    \operatorname{Cent}_{\mathcal P_N}(\mathcal S)/\mathcal S,
\end{equation}
where
\begin{equation}
    \operatorname{Cent}_{\mathcal P_N}(\mathcal S)
    =
    \left\{
        P\in\mathcal P_N
        \,\middle|\,
        Ph=hP,\quad \forall h\in\mathcal S
    \right\}.
\end{equation}
An element 
\begin{equation}
\bar P=P\mathcal S\in G_{\mathcal L}
\end{equation} 
is an equivalence class of physical Pauli operators that differ by a
stabilizer and therefore act identically on the code space.
Including its central phases, \(G_{\mathcal L}\cong\mathcal P_k\)
and is non-Abelian for \(k>0\).

For a Pauli symbol, a bar denotes its logical equivalence class:
\(\bar P=P\mathcal S\in G_{\mathcal L}\). The physical identity is
\(I_N\), while the identity element of \(G_{\mathcal L}\) is the class
\begin{equation}
    \bar I=I_N\mathcal S=\mathcal S.
\end{equation}
We denote the central logical phase class by
\begin{equation}
    \zeta_{\mathcal L}\equiv(iI_N)\mathcal S.
\end{equation}
Let \(\mathcal H_{\mathcal L}\cong(\mathbb C^2)^{\otimes k}\) be an abstract
logical Hilbert space and let
\(V:\mathcal H_{\mathcal L}\to\mathcal H_{\mathrm{code}}\) be a fixed
encoding isometry.  For a class \(\bar P=P\mathcal S\), we distinguish the
class itself from its induced operator on the abstract logical space by writing
\begin{equation}
    \widehat P\equiv V^\dagger P V.
    \label{eq:induced_logical_operator}
\end{equation}
This operator is independent of the chosen representative \(P\in\bar P\),
because every stabilizer acts as the identity on the code space.

To define a logical basis, choose \(k\) independent, commuting, order-two
logical Pauli classes \(\{\bar g_i\}_{i=1}^k\subset G_{\mathcal L}\), together
with commuting Hermitian representatives
\(g_i\in\operatorname{Cent}_{\mathcal P_N}(\mathcal S)\) satisfying
\(g_i^2=I_N\). A logical basis state is, by definition, a state
\(\ket{c}\in\mathcal H_{\mathrm{code}}\), with
\(c=(c_1,\ldots,c_k)\in\{0,1\}^k\), satisfying
\begin{equation}\label{eq:logical_state}
    g_i\ket{c}
    =
    (-1)^{c_i}\ket{c},
    \quad i=1,\ldots,k.
\end{equation}
This action is independent of the chosen representative because any two
representatives of \(\bar g_i\) differ by an element of \(\mathcal S\),
which acts as the identity on \(\mathcal H_{\mathrm{code}}\).

The logical stabilizer group of \(\ket{c}\) is the Abelian subgroup
\begin{equation}
    \mathcal G_{\mathcal L}(c)
    =
    \left\langle
        (-1)^{c_1}\bar g_1,\ldots,
        (-1)^{c_k}\bar g_k
    \right\rangle
    < G_{\mathcal L}.
\end{equation}
Thus \(G_{\mathcal L}\) is the logical Pauli group of the code, whereas
\(\mathcal G_{\mathcal L}(c)\) stabilizes the particular logical state
\(\ket{c}\).

For a physical Pauli operator \(P\), let \(\operatorname{wt}(P)\) denote
the number of qubits on which \(P\) acts nontrivially.  The code distance
is the minimum weight over all physical representatives of nontrivial
logical Pauli classes:
\begin{equation}
\begin{aligned}
    d
    &=
    \min_{
        P\in
        \operatorname{Cent}_{\mathcal P_N}(\mathcal S)
        \setminus
        \langle iI_N\rangle\mathcal S
    }
    \operatorname{wt}(P).
\end{aligned}
\end{equation}
Here
\(\langle iI_N\rangle\mathcal S
=\{i^m s:m\in\mathbb Z_4,\ s\in\mathcal S\}\)
is the physically trivial sector consisting of stabilizers multiplied by
global phases.

\subsection{Coherent errors, syndrome measurement, and syndrome states}\label{sec:qce}

In this work, we consider quantum codes with code space \(\mathcal H_{\mathrm{code}}\) subjected to coherent unitary errors described by the channel
\begin{equation}
    \mathcal{E}[\cdot] = \mathcal{U}(\cdot)\mathcal{U}^\dagger,
\end{equation}
where \(\mathcal{U}\) is a unitary operation. The specific form of \(\mathcal{U}\) depends on the underlying error model and will be specified later for each code family. Acting on the logical code state 
\begin{equation}
    \rho_{\mathrm{enc}} = \ket{c}\bra{c},
    \quad \ket{c}\in\mathcal H_{\mathrm{code}},
\end{equation}
it produces the errored state
\begin{equation}\label{eq:error_state}
    \rho_E = \mathcal{U}\rho_{\mathrm{enc}}\mathcal{U}^\dagger.
\end{equation}

A conceptual quantum error-correction (QEC) protocol proceeds by measuring the
independent stabilizer checks \(\{h_i\}_{i=1}^{N-k}\) after the error occurs,
as illustrated in Fig.~\ref{fig:qec_procedure}.

\begin{figure}[t]
    \centering
    \begin{tikzpicture}[
        >=Latex,
        font=\footnotesize,
        node distance=4.5mm,
        stage/.style={
            text width=3.25cm,
            minimum height=9mm,
            inner sep=2pt,
            align=center,
            rounded corners=1.5pt
        },
        state/.style={stage,fill=cyan!18},
        error/.style={stage,fill=green!24},
        measurement/.style={stage,fill=teal!48},
        correction/.style={stage,fill=black!8},
        flow/.style={->,line width=0.55pt}
    ]
        \node[state] (input) {
            encoded state\\[-2pt]
            \(\rho_{\mathrm{enc}}\)
        };

        \node[error,above=of input] (error) {
            coherent error\\[-2pt]
            \(\mathcal U\)
        };

        \node[measurement,above=of error] (measure) {
            syndrome projection\\[-2pt]
            \(\Pi_s\)
        };

        \node[correction,above=of measure] (correct) {
            syndrome-sector identification\\[-2pt]
            \(R_s^\dagger\)
        };

        \draw[flow] (input) -- (error);

        \draw[flow]
            (error) --
            node[midway,right=2pt] {\(\rho_E\)}
            (measure);

        \draw[flow]
            (measure) --
            node[midway,right=2pt] {\(\rho_s\)}
            (correct);

        \draw[flow]
            (correct.north) -- ++(0,4.5mm)
            node[above] {\(\widetilde\rho_s\)};

        \coordinate (syndrome-lower) at ($(measure.west)+(-5mm,0)$);
        \coordinate (syndrome-upper) at ($(correct.west)+(-5mm,0)$);
        \draw[flow,draw=blue!55!black]
            (measure.west) -- (syndrome-lower) -- (syndrome-upper) --
            (correct.west);
        \node[left=2pt] at ($(syndrome-lower)!0.5!(syndrome-upper)$)
            {\(s\sim\mathbb P[s]\)};
    \end{tikzpicture}

    \caption{\label{fig:qec_procedure}
    Conceptual quantum error-correction procedure for the fixed logical-basis
    input \(\rho_{\mathrm{enc}}=\ket c\!\bra c\). The coherent error produces
    \(\rho_E=\mathcal U\rho_{\mathrm{enc}}\mathcal U^\dagger\). Stabilizer
    measurement yields outcome \(s\) with probability
    \(\mathbb P[s]=\Tr(\Pi_s\rho_E)\) and conditional branch
    \(\rho_s=\Pi_s\rho_E\Pi_s/\mathbb P[s]\). The chosen syndrome
    representative \(R_s^\dagger\) identifies this branch with the reference
    code sector, yielding
    \(\widetilde\rho_s=R_s^\dagger\rho_sR_s\). This final identification is
    used only for fixed-input branch-state diagnostics; it is not an optimized
    recovery, and no recovery is applied in the channel-level
    coherent-information calculation.}
\end{figure}

The measurement outcomes form the tuple
\(s=(s_1,\ldots,s_{N-k})\in\{\pm1\}^{N-k}\).  For a fixed input label \(c\)
and error realization \(\mathcal U\), define
\begin{equation}\label{eq:syndrome_state}
    \rho_s=\mathbb{P}[s\mid c,\mathcal U]^{-1}\Pi_s\rho_E\Pi_s,
    \quad
    \mathbb{P}[s\mid c,\mathcal U]=\Tr(\Pi_s\rho_E),
\end{equation}
for every outcome with \(\mathbb{P}[s\mid c,\mathcal U]>0\), where
\begin{equation}
    \Pi_s
    =
    \frac{1}{2^{N-k}}\prod_{i=1}^{N-k}(I_N+s_i h_i)
\end{equation}
is the orthogonal projector onto the syndrome sector labeled by \(s\).  The
family \(\{\Pi_s\}_{s\in\{\pm1\}^{N-k}}\) satisfies
\begin{equation}
    \Pi_s\Pi_t=\delta_{s,t}\Pi_s,
    \quad
    \sum_s\Pi_s=I_N.
\end{equation}
We refer to \(\rho_s\) as the \emph{syndrome state}.  When \((c,\mathcal U)\)
is fixed, we suppress this conditioning and write simply \(\mathbb P[s]\) and
\(\rho_s\).

The two primary objects in our analysis are the conditional state \(\rho_s\)
and the syndrome distribution \(\mathbb P[s]\), obtained after a single
coherent-error step followed by stabilizer measurement.  For a fixed
logical-basis input \(\ket c\) and error realization \(\mathcal U\), we also
use \(\rho_s\) to define a state-specific MAP Pauli-frame correction, as
discussed in Sec.~\ref{sec:recoverability}.  This construction quantifies the
return fidelity of that chosen state; it is not a state-independent decoder
for an arbitrary encoded input.  Channel-level reversibility is considered
separately through coherent information.  Thus the two post-measurement
objects support three complementary uses: syndrome-distribution diagnostics
built from \(\mathbb P[s]\), branch-state and fixed-input return diagnostics
built from \(\rho_s\), and the channel-level coherent information of the
unrecovered effective channel.

\paragraph*{Syndrome state.}
The first question concerns the logical structure of the post-measurement state:
\begin{quote}
\emph{Is the post-measurement syndrome state \(\rho_s\) stabilized by a logical stabilizer group \(\mathcal{G}_\mathcal{L}'\) that differs from the initial logical stabilizer group \(\mathcal{G}_\mathcal{L}\)?}
\end{quote}
Here, \(\mathcal{G}_\mathcal{L}'\) denotes the logical stabilizer group associated with the post-measurement syndrome state after quotienting out the local stabilizer redundancy appropriate to the code. When \(\mathcal{G}_\mathcal{L}'\neq \mathcal{G}_\mathcal{L}\), the syndrome branch no longer corresponds to the original encoded logical state. This signals either logical-state conversion or logical scrambling and, in general, obstructs straightforward syndrome-based recovery of the original logical information.

\paragraph*{Syndrome distribution.}
The second question concerns the probability distribution of syndrome outcomes:
\begin{quote}
\emph{How does the coherent error alter the structure of the syndrome distribution \(\mathbb{P}[s]\)?}
\end{quote}
The syndrome distribution captures how information about the error and the encoded state is imprinted into the measurement outcomes. As we will show, qualitative changes in \(\mathbb{P}[s]\) provide a complementary diagnostic of coherent-error-induced phase transitions.

\subsection{Logical-group diagnostic and state-level MAP return probability}
\label{sec:recoverability}

We first construct the post-measurement logical stabilizer structure and then
relate it to a state-level Pauli-frame recoverability benchmark.  Fix a
logical-basis input \(\ket c\), a Clifford error realization \(\mathcal U\),
and an allowed syndrome \(s=(s_1,\ldots,s_{N-k})\).  The syndrome-\(s\)
sector is stabilized by the signed physical group
\begin{equation}
    \mathcal S_s
    =
    \left\langle
        s_1h_1,\ldots,s_{N-k}h_{N-k}
    \right\rangle .
\end{equation}
For the pure stabilizer branches considered in this work, let
\begin{equation}
    \mathcal T_s
    =
    \left\{
        P\in\mathcal P_N:
        P\rho_s=\rho_sP=\rho_s
    \right\}
    \label{eq:full_branch_stabilizer}
\end{equation}
be the full signed physical stabilizer group of the branch.  In particular,
\(\mathcal S_s<\mathcal T_s\).  Appendix~\ref{app:stab_measurement}
constructs a generating set for \(\mathcal T_s\) explicitly by solving the
binary commutator equations; see Eq.~\eqref{eq:post_mea_stab}.

The stabilizer simulation first extracts the logical structure without
performing any recovery operation.  Let
\begin{equation}
    \eta:
    \mathcal P_N
    \rightarrow
    \underline{\mathcal P}_N
    =
    \mathcal P_N/\langle iI_N\rangle
\end{equation}
discard the global Pauli phase, and write
\(\underline{\mathcal T}_s=\eta(\mathcal T_s)\) and
\(\underline{\mathcal S}=\eta(\mathcal S)\).  Because
\(\eta(\mathcal S_s)=\underline{\mathcal S}\), one has
\(\underline{\mathcal S}<\underline{\mathcal T}_s\).  Under the
standard identification \(\underline{\mathcal P}_N\cong\mathbb F_2^{2N}\),
let
\begin{equation}
    \omega\bigl((\mathbf x\mid\mathbf z),(\mathbf x'\mid\mathbf z')\bigr)
    =\mathbf x\!\cdot\!\mathbf z'+\mathbf z\!\cdot\!\mathbf x'
    \pmod 2
\end{equation}
be the binary symplectic form.  The projective logical Pauli space is
\begin{equation}
    \underline G_{\mathcal L}
    \equiv
    \underline{\mathcal S}^{\perp_\omega}/\underline{\mathcal S}
    \cong
    G_{\mathcal L}/\langle\zeta_{\mathcal L}\rangle
    \cong\mathbb F_2^{2k}.
    \label{eq:projective_logical_pauli_space}
\end{equation}
For any logical subgroup \(\mathcal G_{\mathcal L}\), an underline denotes
its image in this phase-free quotient.  The phase-free
post-measurement logical stabilizer subspace is therefore
\begin{equation}
    \underline{\mathcal G}'_{\mathcal L}(s;c,\mathcal U)
    =
    \underline{\mathcal T}_s/
    \underline{\mathcal S}
    < \underline G_{\mathcal L}.
    \label{eq:postmeasurement_logical_group_signfree}
\end{equation}
This is the object used in the logical-group diagnostic.  Numerically, it is
constructed directly in the binary symplectic representation: we evolve the
projective input stabilizer subspace under the Clifford symplectic map, update
it by the projective check-measurement subspace, row-reduce the resulting
phase-free branch subspace \(\underline{\mathcal T}_s\) against
\(\underline{\mathcal S}\), and retain the remaining \(k\) logical rows.

For the branch-specific Pauli-frame correction, the stabilizer signs must
also be retained and the syndrome sector must be identified with the original
code space.  Choose, by a fixed deterministic convention, a Pauli syndrome
representative \(R_s\) from
\begin{equation}
    \mathcal R_s
    =
    \left\{
        R\in\mathcal P_N:
        R h_i R^\dagger=s_i h_i,\ 
        i=1,\ldots,N-k
    \right\}.
\end{equation}
Then
\begin{equation}
    \widetilde\rho_s
    =
    R_s^\dagger\rho_sR_s
\end{equation}
is supported on the original code space, and its signed physical stabilizer
group is
\begin{equation}
    \widetilde{\mathcal T}_s
    =
    R_s^\dagger\mathcal T_sR_s .
\end{equation}
Because
\(R_s^\dagger\mathcal S_sR_s=\mathcal S\), one has
\[
    \mathcal S
    <
    \widetilde{\mathcal T}_s
    <
    \operatorname{Cent}_{\mathcal P_N}(\mathcal S).
\]
Let
\begin{equation}
    \mathfrak q_{\mathcal L}:
    \operatorname{Cent}_{\mathcal P_N}(\mathcal S)
    \rightarrow
    G_{\mathcal L},
    \quad
    \mathfrak q_{\mathcal L}(P)=P\mathcal S ,
\end{equation}
be the logical quotient map.  The post-measurement logical stabilizer group is
then
\begin{equation}
    \mathcal G'_{\mathcal L}(s;c,\mathcal U)
    =
    \mathfrak q_{\mathcal L}\!\left(\widetilde{\mathcal T}_s\right)
    =
    \widetilde{\mathcal T}_s/\mathcal S
    < G_{\mathcal L}.
    \label{eq:postmeasurement_logical_group}
\end{equation}
Thus \(\mathcal G'_{\mathcal L}(s;c,\mathcal U)\) is a signed Abelian
subgroup of the logical Pauli group, rather than a set of physical
representatives.  Its phase-free image is precisely
\(\underline{\mathcal G}'_{\mathcal L}(s;c,\mathcal U)\) in
Eq.~\eqref{eq:postmeasurement_logical_group_signfree}, because conjugation
by the Pauli operator \(R_s\) changes stabilizer signs but not their binary
symplectic vectors.

We then optimize only over logical Pauli frames.  Since logical Paulis act
transitively on the logical-basis labels, the optimal return probability is
\begin{equation}
    P_{\mathrm{rec}}^{\mathrm{opt}}(s;c,\mathcal U)
    =
    \max_{\bar P=P\mathcal S\in G_{\mathcal L}}
    \bra c P\,\widetilde\rho_sP^\dagger\ket c
    =
    \max_{c'}\bra{c'}\widetilde\rho_s\ket{c'} ,
\end{equation}
where one fixed physical representative \(P\) is used for each class; the
matrix element is representative-independent.  If the class represented by
the physical Pauli \(P_{s;c,\mathcal U}^{\mathrm{MAP}}\) attains the maximum,
the corresponding physical correction is
\begin{equation}
    U_{s;c,\mathcal U}^{\mathrm{MAP}}
    =
    P_{s;c,\mathcal U}^{\mathrm{MAP}}R_s^\dagger.
    \label{eq:map_recovery_unitary}
\end{equation}
Its return fidelity is exactly
\(P_{\mathrm{rec}}^{\mathrm{opt}}(s;c,\mathcal U)\).  This is a
per-state benchmark because the maximizing logical Pauli may depend on the
known input \(c\); it is not a decoder for an arbitrary unknown logical
state.  In particular, it differs from degenerate maximum-likelihood decoding
of Pauli errors~\cite{iyer2015hardness} and from optimization of a
state-independent recovery channel~\cite{chamberland2017hard,beale2018quantum}.
We address channel-level reversibility separately in
Sec.~\ref{sec:effective_channel}.

We next express this benchmark through the logical-group difference
signature using the phase-free images introduced above.  Let
\(\underline{\mathcal G}_{\mathcal L}(c)\) denote the binary symplectic span
of the input logical stabilizers with their signs discarded.  Together with
\(\underline{\mathcal G}'_{\mathcal L}(s;c,\mathcal U)\) from
Eq.~\eqref{eq:postmeasurement_logical_group_signfree}, these are subspaces of
the projective logical Pauli space
\(\underline G_{\mathcal L}\cong\mathbb F_2^{2k}\).  Their combined subspace
is
\begin{equation}
    G_{\mathrm{comb.}}^{\mathrm{sf}}(s;c,\mathcal U)
    =
    \operatorname{span}_{\mathbb F_2}
    \left(
        \underline{\mathcal G}_{\mathcal L}(c),
        \underline{\mathcal G}'_{\mathcal L}(s;c,\mathcal U)
    \right).
\end{equation}
The corresponding sign-free logical-group difference is
\begin{equation}\label{eq:delta_logical_general}
    \Delta_{\mathrm{Logi.}}(s;c,\mathcal U)
    =
    \dim_{\mathbb F_2}
    G_{\mathrm{comb.}}^{\mathrm{sf}}(s;c,\mathcal U)
    -
    \dim_{\mathbb F_2}
    \underline{\mathcal G}_{\mathcal L}(c).
\end{equation}
Algebraically, \(\Delta_{\mathrm{Logi.}}\) is the number of independent
logical directions in the branch stabilizer group that are absent from the
input stabilizer group.  The stabilizer signs can change the affine offset of
the supported logical-basis labels, and hence the maximizing label \(c'\) and
the corresponding MAP Pauli frame, but they do not change the number of
supported labels or their Born weights.  Because
\(P_{\mathrm{rec}}^{\mathrm{opt}}\) maximizes over \(c'\), its optimal value
is therefore determined completely by the phase-free subspaces.

After the syndrome is removed, the branch is a logical stabilizer state.  Its
logical-basis expansion has support on
\(2^{\Delta_{\mathrm{Logi.}}(s;c,\mathcal U)}\) labels, all with equal Born
weight \(2^{-\Delta_{\mathrm{Logi.}}(s;c,\mathcal U)}\).
Consequently,
\begin{equation}\label{eq:P_from_delta_branch}
    P_{\mathrm{rec}}^{\mathrm{opt}}(s;c,\mathcal U)
    =
    2^{-\Delta_{\mathrm{Logi.}}(s;c,\mathcal U)}.
\end{equation}
Thus \(\Delta_{\mathrm{Logi.}}=0\) gives perfect Pauli-frame return,
\(\Delta_{\mathrm{Logi.}}=1\) gives \(P_{\mathrm{rec}}^{\mathrm{opt}}=1/2\),
and so forth.

Finally, the explicit \(s\) and \(c\) dependence can be suppressed in the
Clifford setting.  For fixed \(\mathcal U\), changing \(c\), changing among
allowed syndromes \(s\), or applying the Pauli representative \(R_s\) changes
only stabilizer signs.  The underlying binary symplectic subspaces, and hence
\(\Delta_{\mathrm{Logi.}}\), are unchanged.  Equivalently, different
\(c\) select different affine offsets of the same constraint space, and for
each \(c\) the allowed syndromes are uniformly weighted.  Therefore, for every
input label \(c\) and every syndrome \(s\) satisfying
\(\mathbb P[s\mid c,\mathcal U]>0\),
\begin{equation}
    \Delta_{\mathrm{Logi.}}(s;c,\mathcal U)
    \equiv \Delta_{\mathrm{Logi.}}(\mathcal U),
    ~
    P_{\mathrm{rec}}^{\mathrm{opt}}(s;c,\mathcal U)
    \equiv P_{\mathrm{rec}}^{\mathrm{opt}}(\mathcal U).
\end{equation}
Only the optimal value is independent of \(s\) and \(c\); the maximizing
Pauli frame can still depend on both.  The algebraic derivation is given in
App.~\ref{app:map_delta}.

\subsection{Effective channel and recoverability}\label{sec:effective_channel}

The effects of coherent errors and syndrome measurements can be formalized via a quantum channel. The overall map corresponding to a single error-and-recovery cycle is
\begin{equation}
\begin{aligned}\label{eq:error_channel_decomposition}
    \mathcal{E}_{re}(\cdot) &= \sum_s U_s \Pi_s\,\mathcal{U}\,(\cdot)\,\mathcal{U}^\dagger \Pi_s U_s^\dagger \\
    &= \sum_s U_s \Pi_s \left( \sum_{s'} \Pi_{s'}\,\mathcal{U}\,(\cdot)\,\mathcal{U}^\dagger \Pi_{s'} \right)\Pi_s U_s^\dagger \\
    &= \mathcal E_r\circ\mathcal E_e(\cdot),
\end{aligned}
\end{equation}
where \(U_s\) denotes a general recovery unitary conditioned on the measured
syndrome \(s\), chosen independently of the unknown encoded input. The channel
decomposes into the recovery channel
\begin{equation}
\mathcal E_r(\cdot)=\sum_s U_s \Pi_s(\cdot)\Pi_s U_s^\dagger
\end{equation}
and the \emph{uncorrected error-and-syndrome channel}
\begin{equation}\label{eq:eff_noise_channel}
    \mathcal E_e(\cdot)=\sum_s \Pi_s\,\mathcal U\,(\cdot)\,\mathcal U^\dagger \Pi_s.
\end{equation}
Although \(\mathcal U\) is unitary, \(\mathcal E_e\) is generally nonunitary
because it includes the nonselective syndrome measurement.  For a fixed input,
its output is the nonselective syndrome state
\begin{equation}
    \rho_{\mathrm{ns}}
    \equiv\sum_s \mathbb{P}[s]\,\rho_s
    =\sum_s \Pi_s \rho_E \Pi_s.
    \label{eq:nonselective_syndrome_state}
\end{equation}

The state-specific MAP Pauli correction in
Eq.~\eqref{eq:map_recovery_unitary} is used only to evaluate the return
fidelity of the specified logical-basis input.  By contrast, the channel-level
calculations below leave \(\mathcal E_e\) unrecovered: we neither insert that
state-specific operation into \(\mathcal E_e\) nor construct an optimized
state-independent recovery channel for arbitrary encoded inputs.  We instead
assess whether \(\mathcal E_e\) is reversible in principle through its
\emph{quantum coherent information}
(qCI)~\cite{PhysRevA.54.2629,PhysRevA.55.1613,colmenarez2024accurate}.
Throughout the paper we denote this quantity by
\begin{equation}\label{eq:coherent_information}
    I_c
    \equiv
    S(\rho_{Q'})-S(\rho_{\mathsf RQ'}),
\end{equation}
where
\begin{equation}
    \rho_{Q'}=\mathcal E_e(\rho_Q),
\end{equation}
and
\begin{equation}
    \rho_{\mathsf RQ'}=(\mathcal I_{\mathsf R}\otimes \mathcal E_e)
    (\rho_{\mathsf RQ}),
\end{equation}
with \(\rho_{\mathsf RQ}\) a purification of the code state \(\rho_Q\).  The
upright label \(\mathsf R\) denotes the reference system and is distinct from
the code rate introduced below.  For an
\([[N,k,d]]\) code, let \(\Pi_0\) denote the code-space projector and
choose the maximally mixed code-space input
\begin{equation}
    \rho_Q=\frac{\Pi_0}{2^k},
    \quad
    S(\rho_Q)=k.
    \label{eq:maximally_mixed_code_input}
\end{equation}

The qCI provides a channel-level diagnostic of whether the encoded logical information remains recoverable after coherent errors and syndrome measurement. In particular, perfect quantum error correction is possible if and only if
\begin{equation}
    I_c=S(\rho_Q)=k.
\end{equation}
For code families with a growing number of logical qubits, we report the
coherent-information density
\begin{equation}
    i_c\equiv\frac{I_c}{k}.
    \label{eq:normalized_coherent_information}
\end{equation}
Thus the two diagnostics answer complementary questions.  The state-level MAP
rule gives the best logical Pauli-frame return probability for the specified
logical-basis state, whereas the coherent information tests whether the
effective channel preserves arbitrary encoded quantum information in
principle.  The latter is a reversibility diagnostic and does not itself
construct the corresponding recovery algorithm.

\subsection{Random-instrument benchmark for the post-threshold phase}
\label{sec:random_instrument}

The high-error phase \(p>p_c\) admits a useful random-state interpretation. For the syndrome-resolved observables studied here, the accumulated physical error can be modeled effectively as a globally scrambling random unitary on the \(N\) physical qubits.  It rotates the encoded subspace to a generic orientation relative to the fixed decomposition into syndrome sectors, so that stabilizer readout probes an effectively random encoded state.  This is an emergent high-error ansatz, not an assertion that the microscopic local circuit is exactly Haar-random or uniformly distributed over the global Clifford group at every \(p>p_c\).

Crucially, a globally random physical unitary is not the same as a random logical rotation within the original code space.  Projection onto a syndrome sector can reveal logical information and reduce the rank of the conditional map.  The appropriate logical object is therefore the full \emph{syndrome-resolved instrument}.  In the Clifford setting, each physical-error realization produces a stabilizer quantum instrument in the standard sense~\cite{bombin2023logicalblocks}.  We use \emph{global-Clifford-induced stabilizer-instrument ensemble} for the distribution over these complete instruments obtained by sampling the physical Clifford; the syndrome branches belonging to one realization remain correlated and are not sampled independently.

Recall the abstract logical space \(\mathcal H_{\mathcal L}\), code subspace
\(\mathcal H_{\mathrm{code}}\), and encoding isometry
\(V:\mathcal H_{\mathcal L}\to\mathcal H_{\mathrm{code}}\) introduced in
Sec.~\ref{sec:prelim}.  After identifying syndrome sector \(s\) with the code
space, define
\begin{equation}
    A_s
    =
    V^\dagger R_s^\dagger\Pi_s\mathcal U V .
    \label{eq:logical_branch_map}
\end{equation}
Physically, \(A_s\) is the logical Kraus operator for the syndrome-\(s\)
branch: it encodes the input, applies the physical error and syndrome
projection, removes the measured syndrome, and maps the state back to the
logical Hilbert space, without applying any additional optimized logical
recovery.  The corresponding trace-nonincreasing completely positive branch
map acting on a logical density operator is
\begin{equation}
    \mathcal A_s(\rho_{\mathcal L})
    =
    A_s\rho_{\mathcal L}A_s^\dagger.
    \label{eq:logical_branch_superoperator}
\end{equation}
For an input logical state \(\rho_{\mathcal L}\), the probability of obtaining
syndrome \(s\) is
\begin{equation}
    p_s(\rho_{\mathcal L})
    =
    \Tr\!\left[\mathcal A_s(\rho_{\mathcal L})\right].
\end{equation}
For any branch with \(p_s(\rho_{\mathcal L})>0\), the normalized conditional
logical state is
\begin{equation}
    \rho_{\mathcal L\mid s}
    =
    \frac{\mathcal A_s(\rho_{\mathcal L})}
    {p_s(\rho_{\mathcal L})}.
\end{equation}
The positive operator that determines the probability of this branch is its
logical POVM element,
\begin{equation}
    E_s
    =
    A_s^\dagger A_s
    =
    V^\dagger\mathcal U^\dagger\Pi_s\mathcal U V ,
    \label{eq:logical_povm_element}
\end{equation}
so that
\begin{equation}
    p_s(\rho_{\mathcal L})
    =
    \Tr(E_s\rho_{\mathcal L}).
    \label{eq:logical_branch_probability}
\end{equation}
Because the syndrome projectors resolve the identity,
\begin{equation}
    \sum_s E_s
    =
    V^\dagger\mathcal U^\dagger
    \left(\sum_s\Pi_s\right)
    \mathcal U V
    =
    I_{\mathcal L},
    \label{eq:logical_instrument_completeness}
\end{equation}
and hence \(\{\mathcal A_s\}_s\) is a quantum instrument: each branch is
trace nonincreasing, while the sum of the branches is trace preserving.
Retaining the classical syndrome outcome gives the associated flagged channel.
To obtain Eq.~\eqref{eq:logical_povm_element}, use
\(\Pi_0=VV^\dagger\) and
\(R_s\Pi_0R_s^\dagger=\Pi_s\).  Hence the correction cancels from
\(A_s^\dagger A_s\), and \(E_s\) is independent of the chosen syndrome
representative \(R_s\).

For the coherent-information benchmark, let \(D=2^k\) and denote the
maximally mixed state on the abstract logical space by
\begin{equation}
    \pi_{\mathcal L}
    =
    \frac{I_{\mathcal L}}{D}.
    \label{eq:logical_maximally_mixed_state}
\end{equation}
Its encoded physical state is exactly the code-space input used in
Sec.~\ref{sec:effective_channel},
\begin{equation}
    \rho_Q
    =
    V\pi_{\mathcal L}V^\dagger.
    \label{eq:encoded_maximally_mixed_state}
\end{equation}
For this input, Eq.~\eqref{eq:logical_branch_probability} reduces to
\begin{equation}
    p_s^{(\pi)}
    =
    \Tr(E_s\pi_{\mathcal L})
    =
    \frac{\Tr E_s}{D}.
    \label{eq:maximally_mixed_branch_probability}
\end{equation}
The superscript \((\pi)\) distinguishes this maximally mixed-input branch law
from the fixed-input distribution \(\mathbb P[s\mid c,\mathcal U]\) in
Eq.~\eqref{eq:syndrome_state}.  Normalized conditional objects and all
branchwise sums below are understood only for outcomes with
\(p_s^{(\pi)}>0\), unless zero-probability terms are explicitly harmless.

For the Clifford ensemble, \(E_s\) can be related directly to the stabilizer
Paulis.  Writing \(m=N-k\), define
\begin{equation}
    h(\mathbf a)
    =
    \prod_{i=1}^{m}h_i^{a_i},
    \quad
    \chi_s(\mathbf a)
    =
    \prod_{i=1}^{m}s_i^{a_i}.
\end{equation}
The syndrome projector and its logical POVM element then have the Fourier
expansions
\begin{equation}
    \begin{aligned}
        \Pi_s
        &=
        \prod_{i=1}^{m}\frac{I_N+s_i h_i}{2}
        =
        2^{-m}
        \sum_{\mathbf a\in\mathbb F_2^m}
        \chi_s(\mathbf a)h(\mathbf a),\\
        E_s
        &=
        2^{-m}
        \sum_{\mathbf a\in\mathbb F_2^m}
        \chi_s(\mathbf a)
        V^\dagger P_{\mathbf a}V,
        \quad
        P_{\mathbf a}
        =
        \mathcal U^\dagger h(\mathbf a)\mathcal U .
    \end{aligned}
    \label{eq:logical_povm_pauli_expansion}
\end{equation}
Because \(\mathcal U\) is Clifford, every \(P_{\mathbf a}\) is a signed
physical Pauli.  Its restriction to the code space is
\begin{equation}
    V^\dagger P_{\mathbf a}V
    =
    \begin{cases}
        0,
        & P_{\mathbf a}\notin
        \operatorname{Cent}_{\mathcal P_N}(\mathcal S),\\[3pt]
        \widehat P_{\mathbf a},
        & P_{\mathbf a}\in
        \operatorname{Cent}_{\mathcal P_N}(\mathcal S),
    \end{cases}
    \label{eq:physical_to_logical_pauli_restriction}
\end{equation}
where \(\widehat P_{\mathbf a}=V^\dagger P_{\mathbf a}V\) is the induced
logical Pauli operator, including
the logical identity when \(P_{\mathbf a}\in\mathcal S\).  Therefore
\begin{equation}
    E_s
    =
    2^{-m}
    \sum_{\substack{\mathbf a\in\mathbb F_2^m\\
        P_{\mathbf a}\in
        \operatorname{Cent}_{\mathcal P_N}(\mathcal S)}}
    \chi_s(\mathbf a)\widehat P_{\mathbf a}.
    \label{eq:logical_povm_surviving_paulis}
\end{equation}
Thus \(E_s\) is a signed sum of precisely those pulled-back stabilizer-check
products that induce well-defined logical Paulis on the code space.  It is
not itself a Pauli rotation: \(E_s\) is a positive POVM element.  Because
the surviving logical Paulis commute, let \(r\) be the binary rank of the
Abelian subgroup that they generate after logical identities are removed.
Equivalently, \(r\) is the number of independent commuting logical Pauli
observables learned by the syndrome, so
\(0\le r\le\min\{k,m\}\).  This channel-level measurement rank is distinct
from the state-level basis-mismatch quantity
\(\Delta_{\mathrm{Logi.}}\), even though both are ranks of related binary
subspaces.  The sum can then be reorganized as
\begin{equation}
    \begin{aligned}
        E_s
        &=
        2^r p_s^{(\pi)}\,
        C_{\mathrm{meas}}^\dagger P_{\lambda(s)}C_{\mathrm{meas}},\\
        P_{\lambda(s)}
        &=
        \prod_{j=1}^{r}
        \frac{I_{\mathcal L}+(-1)^{\lambda_j(s)}\widehat Z_j}{2},
    \end{aligned}
    \label{eq:logical_povm_projector_form}
\end{equation}
The overall coefficient is fixed by
\(\Tr E_s=D p_s^{(\pi)}\), with \(D=2^k\), together with
\(\Tr P_{\lambda(s)}=2^{k-r}\).  The logical Clifford
\(C_{\mathrm{meas}}\) diagonalizes \(E_s\) in the logical space by mapping
its \(r\) independent commuting logical Pauli constraints to the canonical
\(\widehat Z_1,\ldots,\widehat Z_r\) basis.  Thus, up to normalization and
a logical Clifford change of basis, \(E_s\) is a joint-eigenspace projector
that fixes \(r\) logical qubits according to \(\lambda(s)\), leaves the
remaining \(k-r\) logical qubits unconstrained, and has rank \(2^{k-r}\).

\begin{proposition}[Branchwise logical unitarity]
\label{prop:branch_unitarity}
For a nonzero syndrome branch, the following statements are equivalent:
\begin{enumerate}
    \item \(A_s\) is proportional to a unitary on all of
    \(\mathcal H_{\mathcal L}\);
    \item \(E_s=p_s^{(\pi)} I_{\mathcal L}\) for some
    \(p_s^{(\pi)}>0\);
    \item all \(2^k\) singular values of \(A_s\) are equal and nonzero;
    \item the syndrome probability
    \(p_s(\rho_{\mathcal L})
    =\Tr(E_s\rho_{\mathcal L})\) is independent of the logical input state.
\end{enumerate}
\end{proposition}

\emph{Proof.}
Suppose first that statement 1 holds, so that
\(A_s=\alpha_s U_{\mathcal L,s}\), where \(U_{\mathcal L,s}\) is unitary and
\(\alpha_s\neq0\).  Then
\begin{equation}
    E_s=A_s^\dagger A_s
    =
    |\alpha_s|^2 I_{\mathcal L},
\end{equation}
which is statement 2 with
\(p_s^{(\pi)}=|\alpha_s|^2>0\).
Conversely, if \(E_s=p_s^{(\pi)}I_{\mathcal L}\), the polar decomposition
gives
\begin{equation}
    A_s=W_s\sqrt{E_s}
    =
    \sqrt{p_s^{(\pi)}}\,W_s .
\end{equation}
Because \(p_s^{(\pi)}>0\), \(A_s\) has full rank.  The polar factor \(W_s\) is
therefore a unitary, rather than only a partial isometry, on the
\(2^k\)-dimensional logical Hilbert space.  Hence statement 2 implies
statement 1.

The singular values of \(A_s\) are the eigenvalues of
\(\sqrt{A_s^\dagger A_s}=\sqrt{E_s}\).  Thus
\(E_s=p_s^{(\pi)}I_{\mathcal L}\) if and only if all \(2^k\) singular
values equal \(\sqrt{p_s^{(\pi)}}\).  The condition
\(p_s^{(\pi)}>0\) is precisely the requirement that none of them vanish.
This proves the equivalence of statements 2 and 3.

Finally, statement 2 gives, for every normalized logical state,
\begin{equation}
    p_s(\rho_{\mathcal L})
    =
    \Tr(p_s^{(\pi)}I_{\mathcal L}\rho_{\mathcal L})
    =
    p_s^{(\pi)},
\end{equation}
and hence implies statement 4.  Conversely, assume statement 4 and denote
the common probability by \(p_s^{(\pi)}\).  In particular,
\(\bra{\psi}E_s\ket{\psi}=p_s^{(\pi)}\) for every normalized pure logical state
\(\ket{\psi}\).  Choosing \(\ket{\psi}\) successively as the eigenvectors of
the positive operator \(E_s\) shows that every eigenvalue of \(E_s\) equals
\(p_s^{(\pi)}\), and therefore
\(E_s=p_s^{(\pi)}I_{\mathcal L}\).  Since the branch is nonzero,
\(E_s\neq0\) and \(p_s^{(\pi)}>0\).  This proves statement 2 and completes
the equivalence.
\hfill\(\square\)

Proposition~\ref{prop:branch_unitarity} is thus both necessary and
sufficient.  Time-reversal symmetry, when present, is one possible
sufficient mechanism for enforcing it~\cite{cheng2024emergent}.

\paragraph*{Clifford specialization.}
The projector form in Eq.~\eqref{eq:logical_povm_projector_form}, together
with the polar decomposition, gives every nonzero branch the normal form
\begin{equation}
    A_s
    =
    \sqrt{2^r p_s^{(\pi)}}\,
    W_s P_{\lambda(s)}C_{\mathrm{meas}},
    \label{eq:logical_instrument_normal_form}
\end{equation}
Here \(W_s\) is a syndrome-dependent logical Clifford operator, while
\(P_{\lambda(s)}\) is the rank-\(2^{k-r}\) projector defined above.
Because \(C_{\mathrm{meas}}\) and \(W_s\) are unitary and
\(P_{\lambda(s)}\) is a
projector, the singular-value spectrum is
\begin{equation}
    \operatorname{Spec}\!\sqrt{A_s^\dagger A_s}
    =
    \left\{
        \underbrace{\sqrt{2^r p_s^{(\pi)}},\ldots,
        \sqrt{2^r p_s^{(\pi)}}}_{2^{k-r}},
        \underbrace{0,\ldots,0}_{2^k-2^{k-r}}
    \right\}.
    \label{eq:clifford_branch_singular_values}
\end{equation}
Thus every Clifford branch automatically has equal nonzero singular values.
The only possible obstruction to Proposition~\ref{prop:branch_unitarity} is
the zero part of the spectrum: the branch is proportional to a unitary
exactly when \(r=0\), and it is a rank-deficient logical projection whenever
\(r>0\).

To connect this rank defect to the coherent information, first normalize the
effect associated with each nonzero branch.  Here
\begin{equation}
    S(X)
    \equiv
    -\Tr(X\log_2X)
    \label{eq:von_neumann_entropy_definition}
\end{equation}
denotes the von Neumann entropy of a normalized positive operator \(X\).
From Eq.~\eqref{eq:logical_instrument_normal_form},
\begin{equation}
    \widehat E_s
    \equiv
    \frac{E_s}{\Tr E_s}
    =
    \frac{
        C_{\mathrm{meas}}^\dagger
        P_{\lambda(s)}C_{\mathrm{meas}}
    }{
        2^{k-r}
    },
    \quad
    S(\widehat E_s)=k-r.
    \label{eq:clifford_normalized_povm_entropy}
\end{equation}
Although \(\widehat E_s\) is a normalized POVM element rather than the
logical input state, its positivity and unit trace allow it to be interpreted
as a density operator.  Its operational meaning follows by purifying the
same logical input \(\pi_{\mathcal L}\) used above.  Introduce a reference
system \(\mathsf R\cong\mathcal L\) and the canonical purification of the
maximally mixed logical state,
\begin{equation}
    \ket{\Phi_D}_{\mathsf R\mathcal L}
    =
    \frac{1}{\sqrt D}
    \sum_{i=1}^{D}
    \ket{i}_{\mathsf R}\ket{i}_{\mathcal L}.
    \label{eq:maximally_mixed_purification}
\end{equation}
To distinguish this channel-level reference--output state from the
fixed-input logical branch \(\ket{\psi_s}_{\mathcal L}\) used in
App.~\ref{app:map_delta}, denote the conditional Choi state by
\(\ket{\Omega_s}_{\mathsf R\mathcal L'}\), where \(\mathcal L'\) is the
decoded logical output and is distinct from the physical output \(Q'\).
Applying the branch operator \(A_s\) to the logical half gives the
unnormalized conditional state
\begin{equation}
    \ket{\widetilde\Omega_s}_{\mathsf R\mathcal L'}
    =
    (I_{\mathsf R}\otimes A_s)\ket{\Phi_D}_{\mathsf R\mathcal L}.
\end{equation}
Its squared norm is the syndrome probability for the maximally mixed logical
input:
\begin{equation}
    \begin{aligned}
        p_s^{(\pi)}
        &=
        \langle\widetilde\Omega_s\mid\widetilde\Omega_s\rangle\\
        &=
        \bra{\Phi_D}(I_{\mathsf R}\otimes A_s^\dagger A_s)\ket{\Phi_D} \\
        &=
        \frac{1}{D}\sum_{i=1}^{D}
        \bra{i}A_s^\dagger A_s\ket{i}
        =
        \frac{\Tr E_s}{D}.
    \end{aligned}
    \label{eq:conditional_choi_probability}
\end{equation}
After normalizing the branch, define the normalized conditional Choi state by
\begin{equation}
    \ket{\Omega_s}
    =
    \bigl(p_s^{(\pi)}\bigr)^{-1/2}\ket{\widetilde\Omega_s}.
\end{equation}
The branchwise partial traces are then
\begin{equation}
        \rho_{\mathsf R\mid s}
        =
        \Tr_{\mathcal L'}\!\left(\ket{\Omega_s}\!\bra{\Omega_s}\right)
        =
        \frac{E_s^{T}}{\Tr E_s}
        =
        \widehat E_s^{\,T}
    \label{eq:conditional_reference_povm}
\end{equation}
The transpose follows from the maximally entangled-state identity
\(\Tr_{\mathcal L'}[(I_{\mathsf R}\otimes A_s)
\ket{\Phi_D}\!\bra{\Phi_D}
\,(I_{\mathsf R}\otimes A_s^\dagger)]=(A_s^\dagger A_s)^T/D\).
The transpose is taken in the basis used to define \(\ket{\Phi_D}\) and does
not change the spectrum.  Consequently,
\begin{equation}
    S(\widehat E_s)
    =
    S(\rho_{\mathsf R\mid s})
    =
    S(\rho_{\mathcal L'\mid s}),
    \label{eq:branch_entanglement_entropy}
\end{equation}
where the second equality holds because
\(\ket{\Omega_s}_{\mathsf R\mathcal L'}\) is pure.  Thus
\(S(\widehat E_s)\) is the
reference--output entanglement that survives in branch \(s\).

To combine the syndrome branches, define the decoding isometry that maps each
physical syndrome sector to an orthogonal syndrome flag and a decoded logical
output:
\begin{equation}
    \mathcal J
    \equiv
    \sum_s
    \ket{s}_{\mathsf S}\otimes V^\dagger R_s^\dagger\Pi_s .
    \label{eq:syndrome_decoding_isometry}
\end{equation}
Using \(\sum_s\Pi_s=I_N\) and
\(R_sVV^\dagger R_s^\dagger=\Pi_s\), this map satisfies
\begin{equation}
    \mathcal J^\dagger\mathcal J
    =
    I_N,
    \qquad
    \mathcal J\mathcal U V
    =
    \sum_s\ket{s}_{\mathsf S}\otimes A_s .
\end{equation}

As an isometry, \(\mathcal J\) does not change either entropy entering the
coherent information.  Hence
the physical channel acting on
\(\rho_Q=V\pi_{\mathcal L}V^\dagger\) has the same coherent information as
the flagged logical instrument with Kraus operators \(A_s\).  Let
\(S_{\mathrm{cl}}(p)\equiv-\sum_s p_s\log_2p_s\) denote the classical
Shannon entropy of a probability distribution \(p\).  We use the standard
entropy identity for a block-diagonal classical--quantum state
\begin{equation}
    \rho_{\mathrm{cq}}
    =
    \sum_s p_s\ket{s}\!\bra{s}\otimes\rho_s .
\end{equation}
Its entropy is
\begin{equation}
    S(\rho_{\mathrm{cq}})
    =
    S_{\mathrm{cl}}(p)+\sum_s p_s S(\rho_s).
\end{equation}
The flagged output has entropy
\(S_{\mathrm{cl}}(p^{(\pi)})
+\sum_s p_s^{(\pi)}S(\rho_{\mathcal L'\mid s})\), whereas
the joint
reference--output state is a mixture of orthogonal pure branch states and
therefore has entropy \(S_{\mathrm{cl}}(p^{(\pi)})\).  Their classical syndrome
contributions cancel:
\begin{equation}
    \begin{aligned}
        I_c
        &=
        S(\rho_{Q'})-S(\rho_{\mathsf RQ'})\\
        &=
        \left[
            S_{\mathrm{cl}}(p^{(\pi)})
            +\sum_s p_s^{(\pi)}S(\rho_{\mathcal L'\mid s})
        \right]
        -S_{\mathrm{cl}}(p^{(\pi)})\\
        &=
        \sum_s p_s^{(\pi)}S(\widehat E_s).
    \end{aligned}
    \label{eq:coherent_information_from_povm_entropy}
\end{equation}
For a fixed Clifford realization, \(r\) is independent of the allowed
syndrome outcome, although \(\lambda(s)\) and \(W_s\) can depend on
\(s\).  Therefore
\begin{equation}
    I_c
    =
    \sum_s p_s^{(\pi)} S(\widehat E_s)
    =
    k-r.
    \label{eq:logical_measurement_rank}
\end{equation}

For the random Clifford ensemble, taking the expectation value gives
\begin{equation}
    \mathbb E_{\mathcal U}[I_c]
    =
    k-\mathbb E_{\mathcal U}[r].
    \label{eq:rmt_rank_to_coherent_information}
\end{equation}
Thus the random-matrix problem becomes a finite-field rank-counting problem: determining the distribution of \(r\) determines both the full distribution and the average of the coherent information.

The global-Clifford calculation below supplies this theoretical benchmark.
Agreement with a local \(q\)-unitary ensemble requires that its
logical-measurement-rank distribution flow to the corresponding
global-random distribution.

\paragraph*{Random-stabilizer counting at fixed \(k\).}
The random-state ansatz predicts that a normalized fixed-input Clifford
branch approaches a uniformly random logical stabilizer state.  After Pauli phases are discarded, the number of distinct \(k\)-qubit stabilizer groups, equivalently maximal isotropic subspaces of
\(\mathbb F_2^{2k}\), is
\begin{equation}
    N_{\mathrm{stab}}(k)
    =
    \prod_{j=1}^{k}(2^j+1).
    \label{eq:number_stabilizer_groups}
\end{equation}
For the toric code \(k=2\), this gives
\(N_{\mathrm{stab}}(2)=15\): nine product-state subspaces and six Bell-pair
subspaces.  This is the universal counting input used in
Table~\ref{tab:tc_sig_distribution}; the table itself remains in the
toric-code section because the value of \(\Delta_{\mathrm{Logi.}}\) also depends on the chosen initial logical state.

The channel-level Choi branch probes a different aspect of the same random instrument.  A finite-field symplectic count for a uniform global physical Clifford gives, at fixed \(k=2\) and large \(N\),
\begin{equation}
    \begin{aligned}
        \Pr(r=0,1,2)
        &=
        \frac{1}{51}(16,30,5),\\
        \mathbb E_{\mathrm{Cl}}[I_c]
        &=
        \frac{62}{51}
        \approx1.216.
    \end{aligned}
    \label{eq:k2_global_clifford_benchmark}
\end{equation}
The value \(62/51\) is thus an independent channel-level prediction of the
global-Clifford instrument ansatz, rather than a consequence of the 15-state
count.

\paragraph*{Finite-rate random-state and Page benchmarks.}
For a finite-rate family with \(K=R_{\mathrm{code}}N\), the definition in
Eq.~\eqref{eq:normalized_coherent_information} becomes
\(i_c=I_c/K\).  In the uniform-global-Clifford ensemble, \(r\) remains \(O(1)\) even though \(K=O(N)\).  Its limiting distribution is the balanced-cut entanglement-deficit distribution of a random stabilizer state~\cite{DahlstenPlenio2006,SmithLeung2006},
\begin{equation}
    \Pr_{\mathrm{Cl}}(r)
    =
    \frac{1}{Z_{\mathrm{Cl}}}
    \frac{2^r}{\prod_{j=1}^{r}(2^j-1)^2},
    \quad
    \mu_{\mathrm{Cl}}
    \equiv
    \mathbb E_{\mathrm{Cl}}[r]
    \approx0.85018.
    \label{eq:finite_rate_clifford_rank_distribution}
\end{equation}
It follows that
\begin{align}
    \mathbb E_{\mathrm{Cl}}[i_c]
    &=
    1-\frac{\mu_{\mathrm{Cl}}}{R_{\mathrm{code}}N}+o(N^{-1})
    \notag\\
    &=
    \exp\!\left[
        -\frac{\mu_{\mathrm{Cl}}/R_{\mathrm{code}}}{N}
        +o(N^{-1})
    \right].
    \label{eq:finite_rate_clifford_benchmark}
\end{align}
For a generic global unitary, a normalized conditional Choi branch instead
approaches a Haar-random state on the balanced
reference--logical bipartition.  Page's result~\cite{Page1993} then gives
\begin{equation}
    \begin{aligned}
        K-\mathbb E_{\mathrm{Haar}}[I_c]
        &\rightarrow
        \frac{1}{2\ln 2},\\
        \mathbb E_{\mathrm{Haar}}[i_c]
        &=
        \exp\!\left[
            -\frac{1/(2R_{\mathrm{code}}\ln 2)}{N}
            +o(N^{-1})
        \right].
    \end{aligned}
    \label{eq:finite_rate_haar_benchmark}
\end{equation}
Both ensembles therefore give an \(O(1)\) total coherent-information deficit
and a \(1/N\) deficit per logical qubit.  More generally, if the local error
ensemble has
\begin{equation}
    \mu(p)
    =
    \lim_{N\to\infty}
    \bigl(
        \mathbb E_{\mathcal C,\mathcal U}
        \!\left[K(\mathcal C)-I_c(\mathcal C,\mathcal U)\right]
    \bigr)
    <\infty,
\end{equation}
then \(h(p)=\mu(p)/R_{\mathrm{code}}\) for ensembles of fixed asymptotic
rate \(R_{\mathrm{code}}\).  The global ensembles predict candidate fully
scrambled values of \(\mu\); the \(p\)-dependence and the approach to those
values remain properties of the microscopic error model.  The symplectic
counts and the Haar/Wishart calculation are derived in
App.~\ref{app:random_instrument_benchmark}.

In this precise sense, the \(p>p_c\) phase reduces to a random-state readout
problem: a globally scrambled encoded state is projected into syndrome
sectors, and recoverability is governed by the rank or entanglement of the
resulting conditional branch.  It should not be described as a random
logical unitary unless the necessary and sufficient condition
\(E_s\propto I_{\mathcal L}\) holds for every nonzero branch.

\subsection{Syndrome distribution and its information-theoretic diagnostics}
\label{sec:syndrome_distribution}

We now turn to the syndrome distribution
\begin{equation}
    \mathbb{P}[s] = \operatorname{Tr}(\Pi_s \rho_E),
    \label{eq:syndrome_distribution}
\end{equation}
which gives the probability of obtaining a particular syndrome \(s\).
The instrument formulas above use an independent generating set of
\(N-k\) checks, so that \(\{\Pi_s\}\) is an orthogonal resolution of the
identity.  For spatial syndrome-distribution diagnostics, however, we may
record \(N_s\) commuting check outcomes that include redundant local checks.

For stabilizer states, \(\mathbb{P}[s]\) admits a simple constraint form (see
App.~\ref{app:stab_measurement} for details).  Let \(\Gamma\) be the binary
syndrome-constraint matrix.  Then
\begin{equation}
    \mathbb{P}[s] = Z^{-1} \prod_{\alpha=1}^{r_\Gamma} \delta_\alpha[s],
    \label{eq:syndrome_distribution_explicit}
\end{equation}
where
\begin{equation}
    \delta_\alpha[s]
    =
    \frac{1}{2}
    \left(
        1+\epsilon_\alpha
        \prod_{i=1}^{N_s}(s_i)^{\Gamma_{i,\alpha}}
    \right),
\end{equation}
where \(\epsilon_\alpha=\pm1\) is the relative sign in the stabilizer relation
\(g_\alpha'=\epsilon_\alpha\prod_i
O_i^{\Gamma_{i,\alpha}}\), while
\(\Gamma_{i,\alpha}\in\{0,1\}\)
specifies whether syndrome bit \(s_i\) participates in the
\(\alpha\)-th parity constraint, \(N_s\) is the total number of syndrome
bits, and \(r_\Gamma=\rank_{\mathbb F_2}\Gamma\) is the number of independent
constraints generated by the Pauli measurement process.  The normalization
factor is
\begin{equation}
    Z
    =
    \sum_s\prod_{\alpha=1}^{r_\Gamma}\delta_\alpha[s]
    =
    2^{N_s-r_\Gamma}.
\end{equation}

This constraint subspace is related to the logical POVM structure discussed in
Sec.~\ref{sec:random_instrument}.  If the POVM in
Eq.~\eqref{eq:logical_povm_projector_form} measures a logical Pauli for which
the chosen input is an eigenstate, then the corresponding logical readout fixes
a syndrome parity.  That fixed parity appears as a column of \(\Gamma\).  In
the post-threshold phase such POVM-selected parities can become
system-spanning, producing nonlocal constraints in \(\mathbb P[s]\) that are
detected by the classical CMI of the syndrome distribution.

The local structure of \(\mathbb{P}[s]\) can be probed using the classical
conditional mutual information (CMI)
\begin{equation}
    I(A:B \mid C)_{\mathbb{P}}
    =
    \sum_s \mathbb{P}[s]
    \log_2
    \frac{\mathbb{P}_C[s_C]\,\mathbb{P}[s]}
         {\mathbb{P}_{AC}[s_{AC}]\,\mathbb{P}_{BC}[s_{BC}]},
\end{equation}
where \(\mathbb{P}_{\mathcal{D}}[s_{\mathcal{D}}]\) denotes the marginal over
a domain \(\mathcal{D}\). For a linear hard-constraint distribution, this
quantity depends only on the rank function of the restricted constraint
matrix.  Let \(A\) and \(B\) be disjoint sets of syndrome bits, let
\(C=(A\cup B)^c\), and let \(\Gamma_R\) denote the row restriction of
\(\Gamma\) to the bits in \(R\).  Then
\begin{equation}
 I(A:B \mid C)_{\mathbb{P}}
 =
 \rank(\Gamma_A)+\rank(\Gamma_B)-\rank(\Gamma_{AB}),
\end{equation}
with ranks taken over \(\mathbb F_2\).  Thus the CMI is insensitive to the
constraint signs \(\epsilon_\alpha\) and to the branch rank \(r\) alone; it
detects the spatial support and linear independence of the syndrome
characters selected by the POVM.  It is nonzero precisely when constraints
link \(A\) and \(B\) in a way that cannot be generated separately from
constraints restricted to the two regions.

The global structure of \(\mathbb{P}[s]\) is captured by the Shannon
entropy
\begin{equation}
    S_{\mathbb{P}}
    = - \sum_s \mathbb{P}[s] \log_2 \mathbb{P}[s]
    = \log_2 Z.
\end{equation}
This motivates the definition of the \emph{global free entropy}
\begin{equation}\label{eq:global_free_entropy}
    \phi \equiv \log_2 Z = N_s-r_\Gamma,
\end{equation}
and the corresponding \emph{reduced free entropy density}
\begin{equation}\label{eq:reduced_free_entropy}
    \varphi \equiv 1-\frac{\phi}{N_s}=\frac{r_\Gamma}{N_s}.
\end{equation}
The quantity \(\varphi\) measures the density of emergent constraints in the
syndrome distribution.  Together, the local quantity
\(I(A:B\mid C)_{\mathbb P}\) and the global Shannon/free-entropy diagnostic
\(\varphi\) characterize complementary aspects of the correlations generated
by the syndrome-resolved projection.  They need not exhibit the same singular
behavior: the CMI is sensitive to the spatial organization of constraints,
whereas \(\varphi\) records only their total density.

The general framework developed above provides four complementary diagnostics
of coherent-error-induced instability in quantum codes: the logical-group
signature \(\Delta_{\mathrm{Logi.}}\), which measures changes in the
post-measurement logical stabilizer structure; the state-level MAP return
probability \(P_{\mathrm{rec}}^{\mathrm{opt}}\), which quantifies the best
Pauli-frame return fidelity for the chosen logical-basis input; the quantum
coherent information \(I_c\), which probes reversibility
at the channel level; and the syndrome-distribution diagnostics
    \(I(A:B\mid C)_{\mathbb{P}}\) and \(\varphi\), which characterize the local
and global structure of the syndrome ensemble. In the following sections, we
apply this framework to the toric code and to finite-rate random
stabilizer-code ensembles.  The logical and channel diagnostics locate a
coherent-error-induced transition at a critical error strength \(p_c\), while
the syndrome-distribution observables reveal complementary local and global
constraint structure that can have different finite-size scaling.

\section{Toric Code}\label{sec:TC}

\begin{figure}[ht]
    \centering
    \includegraphics[width=1\linewidth]{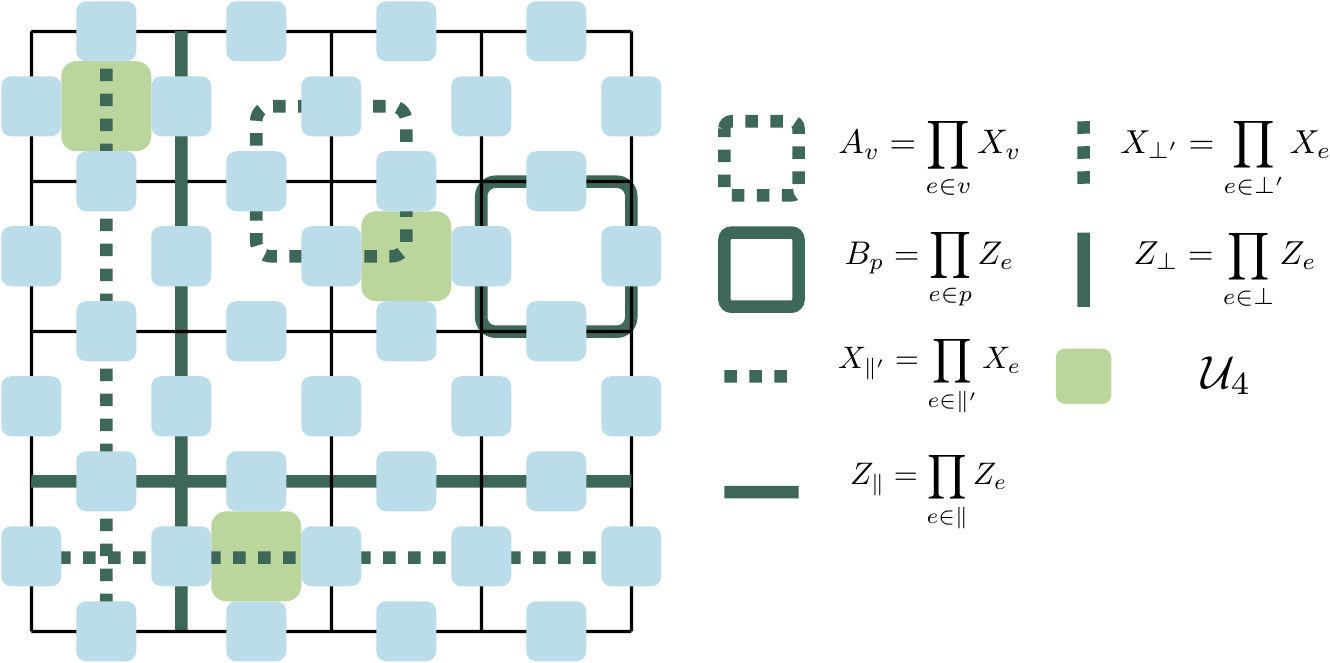}
    \caption{Illustration of toric-code stabilizer checks, logical operators, and the coherent error model. The top and bottom edges, as well as the left and right edges, are identified to form a torus. Qubits reside on lattice edges. The vertex operator \(A_v\) is denoted by a dashed square, while the plaquette operator \(B_p\) is denoted by a solid square. The logical operators \(X_{\perp'}\) and \(Z_{\perp}\) appear as vertical thick dashed and solid lines, respectively, while \(X_{\parallel'}\) and \(Z_{\parallel}\) appear as horizontal thick dashed and solid lines. The green plaquettes indicate a random realization of coherent error unitaries, namely 4-qubit Clifford gates applied with probability \(p\).}
    \label{fig:tc}
\end{figure}

The toric code~\cite{kitaev2003fault, dennis2002topological} is defined as the ground-state space of the Hamiltonian
\begin{equation}
    H = -\sum_{v} A_v - \sum_{p} B_p,
\end{equation}
with vertex and plaquette operators
\begin{equation}
    A_v = \prod_{e \in v} X_e, \quad B_p = \prod_{e \in p} Z_e.
\end{equation}
Here, qubits reside on the edges of a two-dimensional lattice. The operators \(A_v\) and \(B_p\) apply Pauli-\(X\) and Pauli-\(Z\) to the edges adjacent to vertex \(v\) and around plaquette \(p\), respectively, as illustrated in Fig.~\ref{fig:tc}. Since \([A_v,B_p]=0\) for all \(v\) and \(p\), they form a commuting stabilizer-check set that defines the code.

As shown in Fig.~\ref{fig:tc}, on a torus the toric code supports logical operators generated by non-contractible Pauli string operators:
\begin{align}
    Z_{\perp} &= \prod_{e \in \perp} Z_e, & \quad Z_{\parallel} &= \prod_{e \in \parallel} Z_e, \\
    X_{\perp'} &= \prod_{e \in \perp'} X_e, & \quad X_{\parallel'} &= \prod_{e \in \parallel'} X_e.
\end{align}
These operators commute with all stabilizer generators but act nontrivially on the code space, giving a four-fold ground-state degeneracy. Accordingly, the toric code on an \(L\times L\) torus is a \([[2L^2,2,L]]\) stabilizer code. Unless otherwise stated, all toric-code numerical results in this section are disorder-averaged over independent realizations of the coherent error model \(\mathcal U\) at fixed plaquette-gate activation probability \(p\).

For simplicity, we define the logical-\(Z\) classes as
\begin{equation}
    \bar Z_1 = Z_{\perp}\mathcal S,
    \qquad
    \bar Z_2 = X_{\perp'}\mathcal S,
    \label{eq:logical_zx}
\end{equation}
with physical representatives \(Z_{\perp}\) and \(X_{\perp'}\) corresponding
to the vertical strings in Fig.~\ref{fig:tc}.  Let
\(\underline Z_j\in\underline G_{\mathcal L}\) denote the phase-free image of
\(\bar Z_j\). We take the initial logical code state to be
\begin{equation}\label{eq:tc_logical_state}
    \ket{c} = \ket{0_1\,0_2},
\end{equation}
which is the \(+1\)-eigenstate of the representatives \(Z_{\perp}\) and
\(X_{\perp'}\), equivalently of the induced operators \(\widehat Z_1\) and
\(\widehat Z_2\). The corresponding logical stabilizer group is
\begin{equation}\label{eq:tc_logical_stabilizer}
    \mathcal{G}_\mathcal{L} = \langle \bar Z_1, \bar Z_2\rangle.
\end{equation}

Our coherent error model for the toric code consists of one round of random
4-qubit Clifford gates, denoted by \(\mathcal U_4\).  For the even values of
\(L\) used numerically, the plaquettes are divided into the two checkerboard
sublattices and the sublayers are applied in a fixed order.  On each
plaquette, independently, a uniformly sampled 4-qubit Clifford
gate~\cite{koenig2014efficiently} is applied to its four edge qubits with
probability \(p\); otherwise the identity is applied. Thus \(p\) controls the
density of coherent plaquette errors. Unless otherwise stated, all
toric-code results are averaged over independently sampled realizations of
this two-sublayer error round.

\subsection{Syndrome-state diagnostics}
\label{sec:syndrome_state}

We first specialize the general logical-group diagnostic of
Sec.~\ref{sec:recoverability} to the toric code.  Using the projective logical
subspaces defined above, the combined phase-free subspace is
\begin{equation}
    G_{\text{comb.}}^{\mathrm{sf}}
    =
    \operatorname{span}_{\mathbb F_2}
    \left(
        \underline{\mathcal G}_{\mathcal L},
        \underline{\mathcal G}'_{\mathcal L}
    \right),
    \label{eq:combined_logical_group}
\end{equation}
where \(\underline{\mathcal G}_{\mathcal L}\) and
\(\underline{\mathcal G}'_{\mathcal L}\) are the projective images of the
initial and post-measurement logical stabilizer groups.  The physical
toric-code stabilizer redundancy has already been removed by passing to
\(\underline{\mathcal T}_s/\underline{\mathcal S}\), and the global Pauli
phases have already been discarded.  The
corresponding group-difference signature is
\begin{equation}
    \Delta_{\text{Logi.}}^{\underline{\mathcal G}'_{\mathcal L}}
    =
    \dim_{\mathbb F_2}G_{\text{comb.}}^{\mathrm{sf}}-2.
    \label{eq:delta_logical}
\end{equation}

In the toric-code Clifford setting studied here, the sign-free logical stabilizer structure is the same for all syndrome outcomes associated with a fixed realization of the coherent error layer \(\mathcal{U}\). Therefore \(\Delta_{\text{Logi.}}^{\underline{\mathcal G}'_{\mathcal L}}\) depends only on the disorder realization \(\mathcal{U}\), not on the syndrome branch itself.

\begin{figure}[ht]
    \centering
    \subfloat[\label{fig:tc_delta_L}]{
        \includegraphics[width=0.2\textwidth]{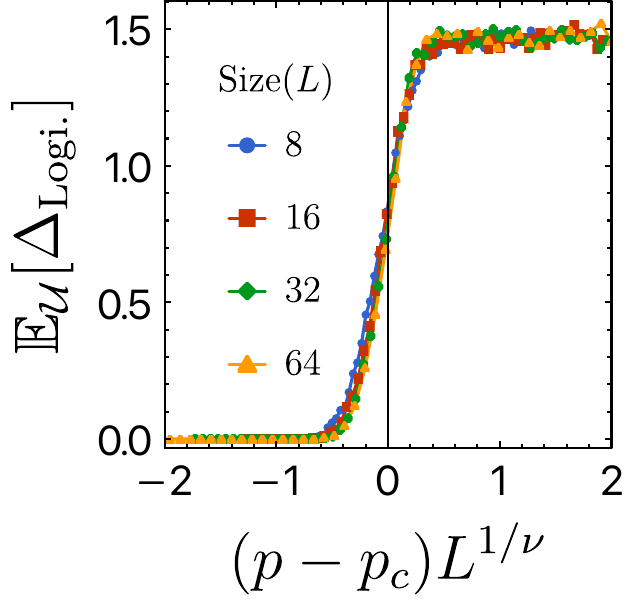}
    }\quad
    \subfloat[\label{fig:tc_stab_change}]{
        \includegraphics[width=0.2\textwidth]{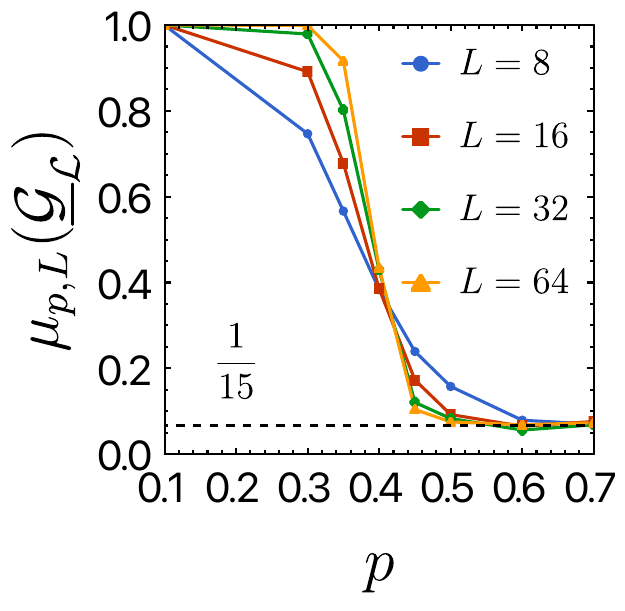}
    }\quad
    \subfloat[\label{fig:tc_succ}]{
        \includegraphics[width=0.2\textwidth]{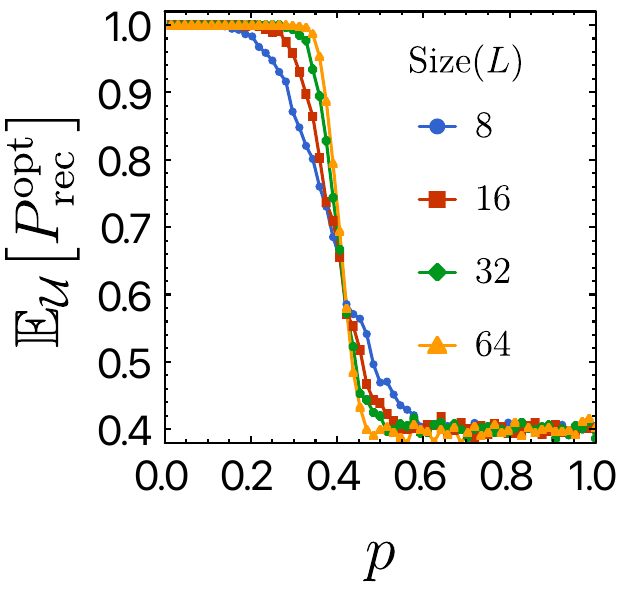}}
    \caption{\label{tc:delta_stab}
    (a) The expectation \(\mathbb E_{\mathcal U}[\Delta_{\mathrm{Logi.}}](p,L)\) as a function of the plaquette-gate activation probability \(p\) for system sizes \(L=8,16,32,64\). For each sample, plaquette locations are activated independently with probability \(p\), a random 4-qubit Clifford is sampled at each active location, a stabilizer measurement is performed, and the phase-free logical stabilizer subspace \(\underline{\mathcal G}'_{\mathcal L}\) of the resulting post-measurement state is determined. The quoted \(p_c\) and \(\nu\) are obtained from the finite-size scaling collapse discussed in the text.
    (b) Retention probability
    \(\mu_{p,L}(\underline{\mathcal G}_{\mathcal L})\equiv
    \Pr_{\mathcal U}[\underline{\mathcal G}'_{\mathcal L}(\mathcal U)=\underline{\mathcal G}_{\mathcal L}]\),
    shown versus \(p\) for \(L=8,16,32,64\).
    (c) Disorder-averaged state-level MAP return probability,
    \(\mathbb E_{\mathcal U}\!\left[P_{\mathrm{rec}}^{\mathrm{opt}}\right]\),
    shown versus \(p\) for \(L=8,16,32,64\). Data are obtained from 1024
    independent realizations of the coherent error layer.}
\end{figure}

As shown in Fig.~\ref{tc:delta_stab}(a), the expectation
\(\mathbb E_{\mathcal U}[\Delta_{\text{Logi.}}^{\underline{\mathcal G}'_{\mathcal L}}]\)
exhibits a phase transition near \(p_c \approx 0.41\). In the large-\(L\) limit, it approaches
\begin{equation}
   \mathbb E_{\mathcal U}[\Delta_{\text{Logi.}}] \to
    \begin{cases}
        0, & p < p_c, \\[4pt]
        22/15 \approx 1.46, & p > p_c,
    \end{cases}
    \quad (L\to\infty),
\end{equation}
where the average is taken over random realizations of the coherent error layer.

Furthermore, the data collapse onto a universal scaling function
\begin{equation}
    \mathbb E_{\mathcal U}[\Delta_{\text{Logi.}}]
    = f_{\Delta}\left(L^{1/\nu} (p - p_c)\right),
    \label{eq:logical_collapse}
\end{equation}
with critical exponent \(\nu \approx 2.34\).

This transition can be understood by examining the disorder-induced
distribution of phase-free logical stabilizer subspaces.  Let
\begin{equation}
    \mu_{p,L}(\underline{\mathcal H})
    =
    \Pr_{\mathcal U}\!\left[
        \underline{\mathcal G}'_{\mathcal L}(\mathcal U)=\underline{\mathcal H}
    \right]
\end{equation}
denote the probability mass function over phase-free logical stabilizer
subspaces \(\underline{\mathcal H}\), where the probability is over random
coherent-error realizations at fixed \(p\) and \(L\).
Figure~\ref{tc:delta_stab}(b) shows the mass assigned to the original
subspace, \(\mu_{p,L}(\underline{\mathcal G}_{\mathcal L})\).
For \(p < p_c\), the distribution is sharply concentrated on the original
phase-free logical stabilizer subspace:
\begin{equation}
\mu_{p,L}(\underline{\mathcal H}) \approx
\begin{cases}
1, & \text{if } \underline{\mathcal H} = \underline{\mathcal G}_{\mathcal L}, \\[4pt]
0, & \text{otherwise.}
\end{cases}
\end{equation}
Above threshold (\(p > p_c\)), the distribution broadens and becomes effectively uniform over all permissible phase-free logical stabilizer subspaces:
\begin{equation}\label{eq:uniform_distr}
    \mu_{p,L}(\underline{\mathcal H}) \approx \frac{1}{|\underline{\mathfrak S}_{\mathcal L}|},
    \quad
    \underline{\mathcal H}\in\underline{\mathfrak S}_{\mathcal L},
\end{equation}
where \(\underline{\mathfrak S}_{\mathcal L}\) is the set of distinct
phase-free two-logical-qubit stabilizer subspaces.
Equation~\eqref{eq:number_stabilizer_groups}
gives \(|\underline{\mathfrak S}_{\mathcal L}|=15\), comprising nine
product-state subspaces and six Bell-pair subspaces.  Their contributions to
\(\Delta_{\mathrm{Logi.}}\) for the chosen input are listed in
Table~\ref{tab:tc_sig_distribution}.  The near-uniform numerical
distribution is thus consistent with the fixed-input random-stabilizer
benchmark of Sec.~\ref{sec:random_instrument}; it is not assumed merely from
the condition \(p>p_c\).

\begin{table}[ht]
    \centering
    \begin{tabular}{|l|c|c|}
    \hline
     \(\Delta_{\text{Logi.}}^{\underline{\mathcal H}}\) & \(\langle O_1 I_2,\, I_1 O_2 \rangle\) & \(\langle O_1 O_2,\, O_1' O_2'\rangle\) \\
    \hline
    0 & 1/15 & 0 \\
    \hline
    1 & 4/15 & 2/15 \\
    \hline
    2 & 4/15 & 4/15 \\
    \hline
    \end{tabular}
    \caption{\label{tab:tc_sig_distribution}
    Contribution to the probability mass of
    \(\Delta_{\mathrm{Logi.}}^{\underline{\mathcal H}}=0,1,2\) under the
    uniform measure on the 15 phase-free two-logical-qubit stabilizer
    subspaces \(\underline{\mathcal H}\in\underline{\mathfrak S}_{\mathcal L}\),
    for the fixed initial subspace
    \(\underline{\mathcal G}_{\mathcal L}
    =\operatorname{span}_{\mathbb F_2}
    \{\underline Z_1,\underline Z_2\}\).
    The columns decompose this mass into product-state subspaces
    \(\langle O_1 I_2,\, I_1 O_2\rangle\) and Bell-pair subspaces
    \(\langle O_1 O_2,\, O_1' O_2'\rangle\).  Here \(O_i\) and \(O_i'\)
    denote phase-free nonidentity single-qubit logical Pauli classes acting on
    logical qubit \(i\).  In the Bell-pair column \(O_i\neq O_i'\) for each \(i\), so
    the two generators anticommute on each logical qubit separately and commute
    as two-qubit logical Paulis.}
\end{table}

The expectation value of the above-threshold distribution is therefore
\begin{equation}
\mathbb E_{\mathcal U}[\Delta_{\text{Logi.}}]
= \sum_{\underline{\mathcal H}\in\underline{\mathfrak S}_{\mathcal L}}
\mu_{p,L}(\underline{\mathcal H})\,\Delta_{\text{Logi.}}^{\underline{\mathcal H}}
\approx \frac{22}{15} \approx 1.46,
\end{equation}
in agreement with the numerical data.

By the general result of Sec.~\ref{sec:recoverability}, the state-level MAP
return probability for the fixed toric-code input is related to the
group-difference signature by
\begin{equation}
P_{\mathrm{rec}}^{\mathrm{opt}}(\mathcal U)=2^{-\Delta_{\mathrm{Logi.}}^{\underline{\mathcal G}'_{\mathcal L}}(\mathcal U)}.
\label{eq:P_from_delta_branch_toric}
\end{equation}
Since, for the toric-code Clifford setting considered here, \(\Delta_{\mathrm{Logi.}}^{\underline{\mathcal G}'_{\mathcal L}}\) is syndrome-independent at fixed disorder realization, we may average directly over disorder. Using Eq.~\eqref{eq:P_from_delta_branch_toric} together with Table~\ref{tab:tc_sig_distribution}, we obtain the post-threshold value
\begin{equation}
\mathbb E_{\mathcal U}\!\left[P_{\mathrm{rec}}^{\mathrm{opt}}\right]
\approx
\frac{1}{15}\cdot 1
+\frac{6}{15}\cdot \frac{1}{2}
+\frac{8}{15}\cdot \frac{1}{4}
=
\frac{6}{15}
=
0.4,
\end{equation}
which matches the numerical data in Fig.~\ref{tc:delta_stab}(c). In the large-\(L\) limit,
\begin{equation}
\mathbb E_{\mathcal U}\!\left[P_{\mathrm{rec}}^{\mathrm{opt}}\right]\to
\begin{cases}
1, & p < p_c, \\[4pt]
0.4, & p > p_c,
\end{cases}
\quad (L\to\infty).
\end{equation}

\subsection{Diagnostics of the uncorrected error-and-syndrome channel}

We next analyze the uncorrected error-and-syndrome channel \(\mathcal E_e\)
introduced in Sec.~\ref{sec:effective_channel}. For the toric code, we focus
on two complementary diagnostics: the local structure of the averaged
syndrome state \(\rho_{\mathrm{ns}}\), as measured by the quantum conditional
mutual information (qCMI), and the global channel performance, as measured by
the quantum coherent information (qCI). The state-specific MAP correction is
not applied in these channel-level numerics.

\subsubsection*{Quantum conditional mutual information}

\begin{figure}[ht]
    \centering
    \subfloat[\label{fig:qcmi_geometry}Geometry of the qCMI domains]{%
        \includegraphics[width=0.2\textwidth]{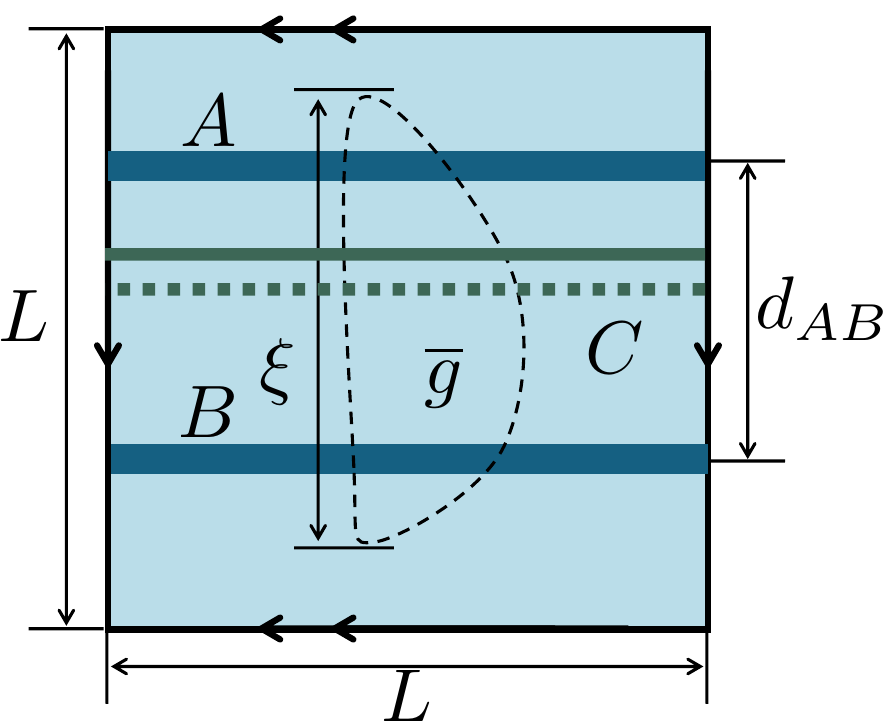}
    }\quad
    \subfloat[\label{fig:qcmi_data_collapse_d}Critical-point qCMI scaling with separation \(u = \sin (\pi d_{AB}/L)\)]{%
        \includegraphics[width=0.2\textwidth]{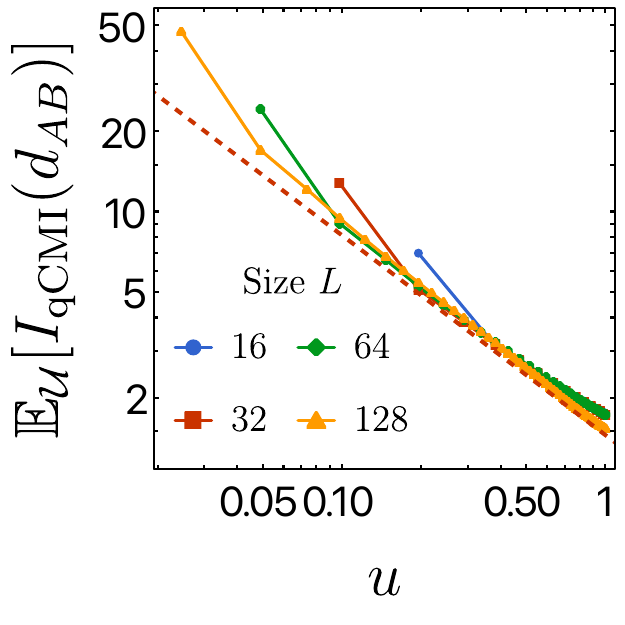}
    }\quad
    \subfloat[\label{fig:qcmi_data_collapse}Universal data collapse of qCMI]{%
        \includegraphics[width=0.2\textwidth]{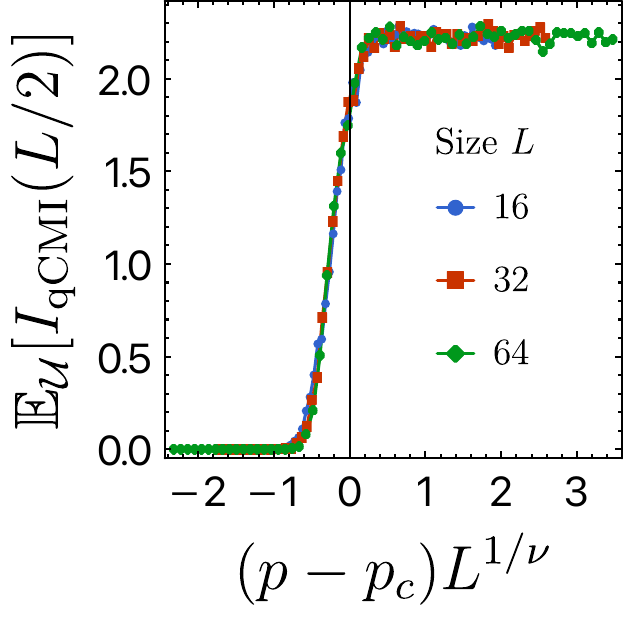}
    }
    \caption{\label{fig:qcmi_figure}
    (a) Geometry of the regions \(A\), \(B\), and \(C\) used to define the qCMI. Dark blue strips denote regions \(A\) and \(B\), while the light blue area indicates \(C\). A nonlocal stabilizer \(g_{\mathrm{nl}}\) (dashed lines) has characteristic length scale \(\xi\). The dashed lines parallel to \(A\) and \(B\) represent the two logical stabilizers of the toric-code state \(\rho_0\). Identical arrows on the boundaries indicate periodic boundary conditions.
    (b) Numerical data showing the disorder-averaged qCMI
    decay at \(p = 0.41\):
    \(\mathbb E_{\mathcal U}[I_{\mathrm{qCMI}}(d_{AB})] \approx u^{-0.75}\), where
    \(u = \sin(\pi d_{AB} / L)\).
    (c) Universal data collapse of
    \(\mathbb E_{\mathcal U}[I_{\mathrm{qCMI}}(L/2)]\) as a function of
    \(L^{1/\nu}(p - p_c)\) with \(\nu \approx 2.34\) near the critical point.
    }
\end{figure}

For each coherent-error realization, we investigate the qCMI of the
nonselective syndrome state
\begin{equation}\label{eq:errored_state}
    \rho_{\mathrm{ns}}(\mathcal U;\rho_0) = \mathcal E_e^{(\mathcal U)}(\rho_0),
\end{equation}
where \(\rho_0=\ket{0_1,0_2}\bra{0_1,0_2}\) is the toric-code logical
state from Eq.~\eqref{eq:tc_logical_state}. We consider a torus partitioned
into two disjoint ring-shaped physical-qubit regions \(A\) and \(B\), with
\(C=(A\cup B)^c\), as shown in Fig.~\ref{fig:qcmi_geometry}.

For three regions \(A\), \(B\), and \(C\), the qCMI is
\begin{equation}
\begin{aligned}
    I(A:B\mid C)_{\rho_{\mathrm{ns}}}
    &=
    S(\rho_{\mathrm{ns}}^{AC})
    +S(\rho_{\mathrm{ns}}^{BC})\\
    &\quad
    -S(\rho_{\mathrm{ns}}^{C})
    -S(\rho_{\mathrm{ns}}^{ABC}),
\end{aligned}
\end{equation}
where \(S(\cdot)\) denotes the von Neumann entropy.
Let \(A(d_{AB})\) and \(B(d_{AB})\) denote the two strips at separation
\(d_{AB}\), and let \(C=(A\cup B)^c\).  We denote the plotted average simply
by \(\mathbb E_{\mathcal U}[I_{\mathrm{qCMI}}]\).  This is the average of
the qCMI evaluated for each realization, not the qCMI of the density matrix
averaged over \(\mathcal U\).

In the toric-code geometry, the qCMI probes the spatial structure of the
stabilizers generated in the nonselective syndrome state. For stabilizer
systems, it effectively counts the number of stabilizers connecting regions
\(A\) and \(B\), and therefore diagnoses the growth of nonlocal structure in
\(\rho_{\mathrm{ns}}\).

As shown in Fig.~\ref{fig:qcmi_figure}, the qCMI is zero when \(p < p_c\) and finite when \(p > p_c\). At the critical point \(p=p_c\), the qCMI exhibits a power-law decay
\begin{equation}
    \mathbb E_{\mathcal U}[I_{\mathrm{qCMI}}(d_{AB})] \approx u^{-\alpha},
    \quad
    \alpha = 0.75,
\end{equation}
where \(u=\sin(\pi d_{AB}/L)\) and \(d_{AB}\) is the separation between regions \(A\) and \(B\). Fixing \(d_{AB}=L/2\), the data collapse onto the scaling form
\begin{equation}
    \mathbb E_{\mathcal U}[I_{\mathrm{qCMI}}(L/2)]
    \approx f_{\mathrm{qCMI}}\Bigl(L^{1/\nu}(p-p_c)\Bigr),
\end{equation}
with \(p_c\approx0.41\) and \(\nu\approx2.34\), consistent with Eq.~\eqref{eq:logical_collapse}. This implies a diverging length scale
\begin{equation}
    \xi \approx |p-p_c|^{-\nu},
\end{equation}
so that above threshold the stabilizers induced by syndrome measurement become arbitrarily large and directly influence the global logical structure.

\subsubsection*{Coherent information}

To confirm that the transition in logical stabilizer structure corresponds to
an error-correction phase transition, we now evaluate the quantum coherent
information defined in Sec.~\ref{sec:effective_channel} for the uncorrected
error-and-syndrome channel \(\mathcal E_e\). For the toric code on a torus,
the code encodes \(k=2\) logical qubits, and we choose \(\rho_Q\) to be the
maximally mixed state on the toric-code space, so that
\begin{equation}
    S(\rho_Q)=2.
\end{equation}

\begin{figure}[ht]
    \centering
    \includegraphics[width=0.45\linewidth]{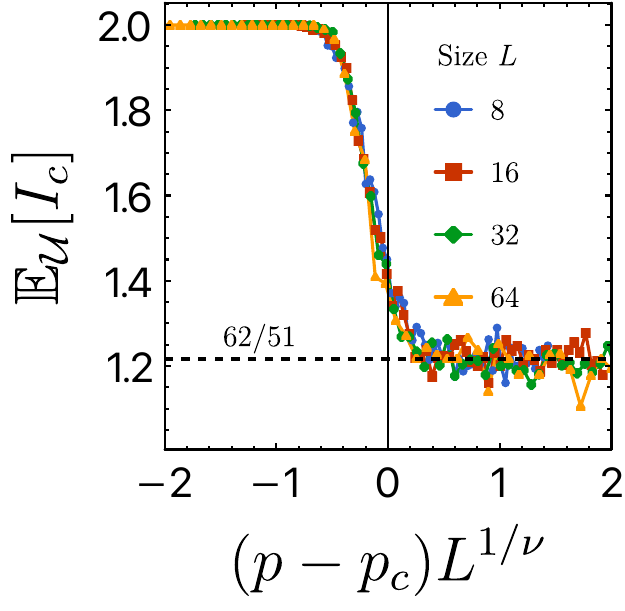}
    \caption{\label{fig:tc_co_info}
    Disorder-averaged coherent information
    \(\mathbb E_{\mathcal U}[I_c]\) as a function of the gate-activation probability
    \(p\) for different system sizes \(L=8,16,32,64\).}
\end{figure}

As shown in Fig.~\ref{fig:tc_co_info}, the coherent information remains
maximal below threshold and approaches a post-threshold plateau consistent
with the fixed-\(k=2\) global-Clifford benchmark derived in
Sec.~\ref{sec:random_instrument}:
\begin{equation}
    \mathbb E_{\mathcal U}[I_c]
    \xrightarrow[L\to\infty]{}
    \begin{cases}
        2, & p < p_c, \\[5pt]
        62/51 \approx 1.216, & p > p_c,
    \end{cases}
\end{equation}
with critical gate-activation probability \(p_c \approx 0.41\). The data collapse onto the scaling form
\begin{equation}
    \mathbb E_{\mathcal U}[I_c]
    =
    f_{I_c}\bigl(L^{1/\nu}(p-p_c)\bigr),
    \label{eq:coherent_collapse}
\end{equation}
with \(\nu \approx 2.34\), consistent with the other observables.

This plateau agrees with the global-Clifford instrument benchmark of
Sec.~\ref{sec:random_instrument}.  In the \(k=2\) global-Clifford calculation,
the measured-logical-rank distribution is
\(\Pr(r=0,1,2)=(16,30,5)/51\), and
Eq.~\eqref{eq:rmt_rank_to_coherent_information} gives \(I_c=k-r\).  Thus the
same global-Clifford picture of the \(p>p_c\) phase accounts for both
diagnostics: the fixed-input logical stabilizer distribution becomes nearly
uniform over the 15 phase-free subspaces, while the disorder-averaged
channel-level coherent information approaches the corresponding
global-Clifford ensemble average.  Since this average is below the reversible
value \(I_c=2\), the dense-error toric-code phase retains a nonunitary
channel-level component.

\subsection{Syndrome-distribution diagnostics}

We now turn to the syndrome distribution \(\mathbb{P}[s]\). Rather than
repeating the general theory from Sec.~\ref{sec:syndrome_distribution}, we
directly apply the two diagnostics introduced there: the classical conditional
mutual information \(I(A:B\mid C)_{\mathbb P}\), which probes the local
structure of the syndrome constraints, and the reduced free-entropy density
\(\varphi\), which probes their global density.

\subsubsection*{Classical conditional mutual information}

\begin{figure}[ht]
    \centering
    \subfloat[\label{fig:cmi_geometry}Geometry of the CMI domains]{
        \includegraphics[width=0.2\textwidth]{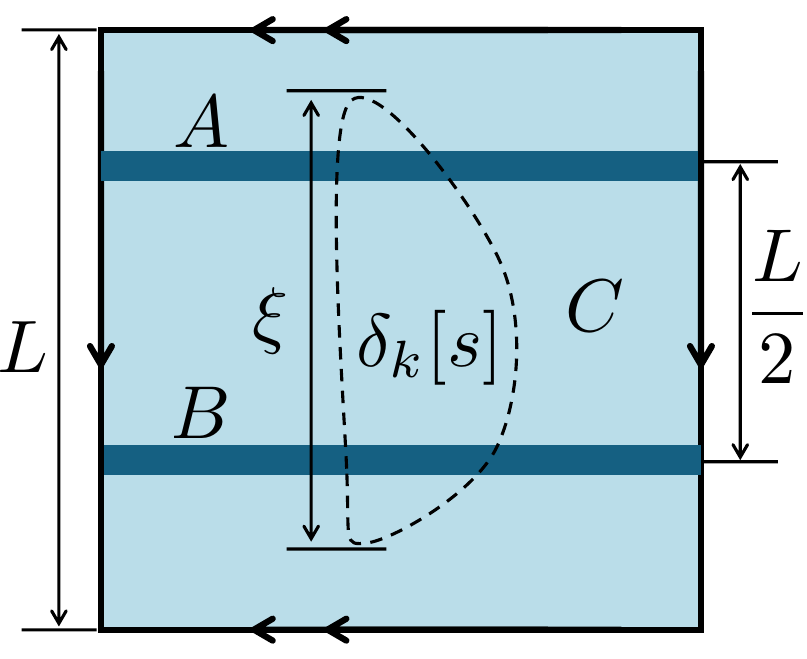}
    }\quad
    \subfloat[\label{fig:cmi_data_collapse}Data collapse of the CMI]{
        \includegraphics[width=0.2\textwidth]{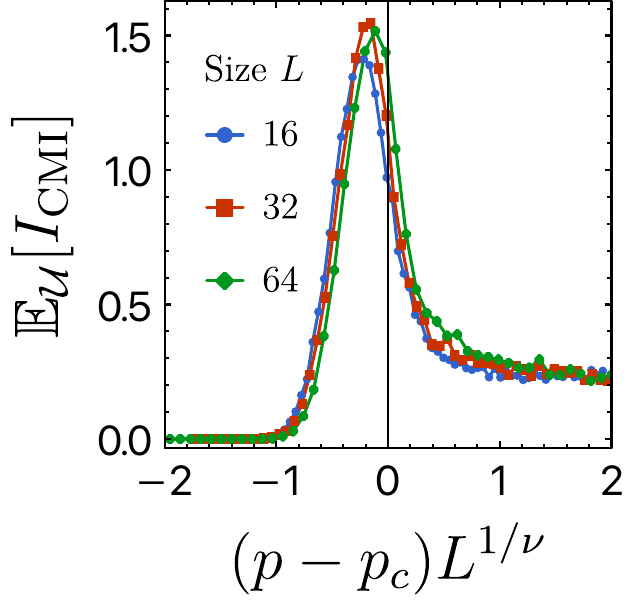}
    }
    \caption{\label{fig:cmi_figure}
    (a) Geometry of regions \(A\), \(B\), and \(C\). The dark blue strips represent \(A\) and \(B\), while the light blue region represents \(C\). A nonlocal constraint \(\delta_\alpha[s]\) is indicated by dashed lines, with characteristic length scale \(\xi\). Identical arrows on the boundaries indicate periodic boundary conditions.
    (b) Numerical data collapse of
    \(\mathbb E_{\mathcal U}[I_{\mathrm{CMI}}]\), demonstrating universal
    scaling near the critical point.
    }
\end{figure}

For the toric-code syndrome distribution, we use strips with the same spatial
boundaries as in the qCMI analysis, shown in Fig.~\ref{fig:cmi_geometry}, but
the random variables are now syndrome coordinates rather than physical
qubits.  Before forming spatial regions, the two dependent check rows are
row-reduced into fixed global-parity coordinates; the remaining coordinates
are the outcomes of \(2L^2-2\) local star and plaquette checks.  This convention
removes the two always-present toric-code relations from the spatial CMI.  A
strip contains both star- and plaquette-check coordinates anchored in that
strip, while \(C\) contains the remainder.  For one realization,
\(I(A:B\mid C)_{\mathbb P_{\mathcal U}}\) measures the number of independent
syndrome constraints shared between the strips.  We denote the plotted
average simply by \(\mathbb E_{\mathcal U}[I_{\mathrm{CMI}}]\).  This is the
average of the CMI evaluated for each realization, not the CMI of the syndrome
distribution averaged over \(\mathcal U\).

For the geometry shown in Fig.~\ref{fig:cmi_geometry}, the typical constraint length scale \(\xi\) diverges as
\begin{equation}
    \xi \approx |p - p_c|^{-\nu},
\end{equation}
leading to the scaling form
\begin{equation}
    \mathbb E_{\mathcal U}[I_{\mathrm{CMI}}]
    = f_{\mathrm{CMI}}\bigl(L^{1/\nu}(p - p_c)\bigr).
\end{equation}
The data collapse in Fig.~\ref{fig:cmi_data_collapse} confirms universal scaling behavior and yields the same critical exponent, \(\nu \approx 2.34\), extracted from the logical-group and qCI analyses.

\subsubsection*{Reduced free entropy density}

To probe the global structure of the syndrome distribution, we study the
reduced free-entropy density \(\varphi\) introduced in
Eq.~\eqref{eq:reduced_free_entropy}. This quantity measures the density of
independent constraints in one realization; the plotted observable is
\(\mathbb E_{\mathcal U}[\varphi]\).  Here all
\(L^2\) star and \(L^2\) plaquette outcomes are recorded, so
\(N_s=2L^2\); the two global check dependencies are included in
\(\Gamma\).

\begin{figure}[ht]
    \centering
    \includegraphics[width=0.45\linewidth]{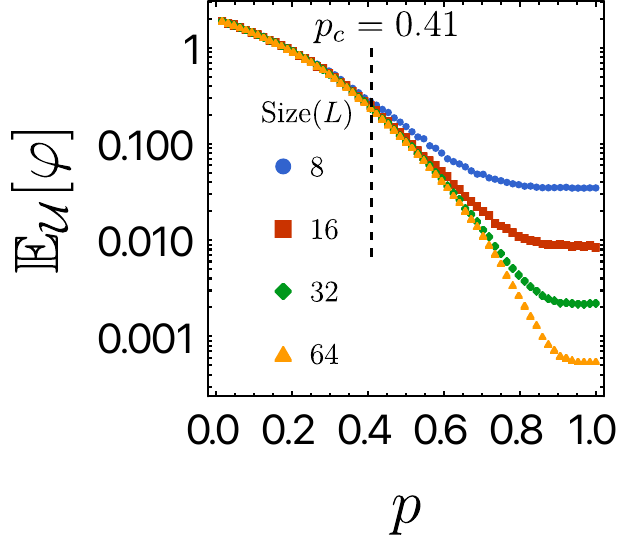}
    \caption{\label{fig:tc_free_entropy}
    Reduced free-entropy density \(\mathbb E_{\mathcal U}[\varphi]\) versus
    gate-activation probability \(p\) for \(L=8,16,32,64\).}
\end{figure}

Figure~\ref{fig:tc_free_entropy} shows that the disorder-averaged constraint
density decreases smoothly with \(p\).  Over the range where convergence with
\(L\) is resolved, the curves are consistent with a nonzero thermodynamic
density at fixed \(p<1\), although that density becomes rapidly smaller as
\(p\) approaches one.  Near \(p=1\), the finite-size data are dominated by an
\(O(1)\) number of surviving constraints---including the two exact toric-code
check dependencies---and hence by an \(O(L^{-2})\) contribution to
\(\mathbb E_{\mathcal U}[\varphi]\).  No singular feature is resolved at
\(p_c\approx0.41\), so this observable quantifies continuous depletion of the
total constraint density rather than serving as an independent order
parameter for the logical transition.  By contrast, the classical CMI probes
the spatial organization of the constraints and exhibits critical behavior
near the transition.

\section{The Random Stabilizer Code Ensemble}\label{sec:rsce}

In this section, we study two examples of finite-rate \emph{random stabilizer code ensembles} (RSCEs). This contrasts with topological codes such as the toric code, where the code space remains \(O(1)\) and the code rate therefore vanishes in the thermodynamic limit.

We begin with a structured example: quantum low-density parity-check (qLDPC) codes derived from the hypergraph product of two classical LDPC codes, namely hypergraph-product (HGP) codes~\cite{mackay1997near, tillich2013quantum, leverrier2015quantum}. These codes inherit sparse-check structure from their classical counterparts and support an extensive number of logical qubits.

We then consider a simpler but highly scrambled example: \emph{random Clifford codes} (RCCs)~\cite{brown2013short, nelson2023fault, PhysRevResearch.6.023055, PhysRevX.11.031066, turkeshi2024error}, generated by random Clifford circuits. Despite their simplicity, these codes display strong scrambling properties together with a finite encoding rate.

In both cases, the code family is random but the logical rate remains finite in the thermodynamic limit. We study both under coherent errors and show that they exhibit coherent-error-induced phase transitions qualitatively distinct from that of the toric code.

Unless otherwise stated, all numerical quantities reported in this section are
averaged over both the random code realization and the random coherent-error
realization at fixed gate-activation probability \(p\). We denote such joint
averages by \(\mathbb E_{\mathcal C,\mathcal U}[\cdots]\), where
\(\mathcal C\) labels the code realization and \(\mathcal U\) labels the
coherent-error realization.

In this finite-rate section we write \(K(\mathcal C)\) for the number of
logical qubits of a code instance and
\(R_{\mathrm{code}}(\mathcal C)=K(\mathcal C)/N\) for its rate.  We suppress
the argument \(\mathcal C\) inside per-instance formulas.  Thus \(K\) plays
the role of the generic \(k\) in the \([[N,k,d]]\) notation of
Sec.~\ref{sec:prelim}; lower-case \(k\) is retained in the general and
fixed-\(k\) benchmark formulas.
For both ensembles,
\(\Delta_{\mathrm{Logi.}}(\mathcal C,\mathcal U)\),
\(\delta_{\mathrm{Logi.}}(\mathcal C,\mathcal U)\), and
\(P_{\mathrm{rec}}^{\mathrm{opt}}(\mathcal C,\mathcal U)\) are
syndrome-independent at fixed \((\mathcal C,\mathcal U)\); these arguments are
suppressed when no ambiguity arises.

\subsection{Coherent random Clifford error models}\label{sec:rcc_error_model}

For the RSCEs, we consider two \(q\)-parameterized coherent Clifford error
models.

The first is a \emph{long-range} \(q\)-unitary error model, in which a
Clifford unitary acts on \(q\) qubits chosen without regard to the geometric
or graph structure of the code. Concretely, the physical qubits are scanned
in increasing index order.  Independently for each anchor qubit \(i\), with
probability \(p\), \(q-1\) distinct partners are sampled uniformly without
replacement from the other physical qubits and a uniformly random
\(q\)-qubit Clifford is applied to the resulting support.  Overlapping gates
are composed in this deterministic anchor order.

The second is a \emph{short-range}, or \emph{\(q\)-local}, error model.  For
HGP codes, every physical qubit again serves as an anchor; conditional on
activation with probability \(p\), a walk of length \(q-1\) is sampled on the
qubit adjacency graph induced by shared checks.  The neighbor list includes
the current vertex, so self-steps and other revisits are possible.  A uniformly
random Clifford is applied to the distinct vertices visited by the walk, so
its support has size at most \(q\).  Here the term \(q\)-local denotes the
maximum walk arity, not an exactly \(q\)-body gate.  Anchors are processed in
increasing index order.  For RCCs, the physical qubits are placed in a
periodic one-dimensional order.  The circuit consists of \(q\) shifted layers
of contiguous \(q\)-site blocks; within each layer the blocks are disjoint,
each block is independently activated with probability \(p\), and the layers
are applied in a fixed order.  Each layer contains \(\lfloor N/q\rfloor\)
blocks, with all site indices understood modulo \(N\).  Thus, when \(N\) is
not divisible by \(q\), blocks crossing the boundary wrap around periodically
and the remainder shifts with the layer offset.

When \(q=1\), both constructions reduce to independent single-qubit Clifford
rotations and therefore coincide.  All thresholds and scaling functions below
are understood separately for each fixed \(q\) and schedule
\(\chi\in\{\mathrm{long},\mathrm{local}\}\).  Thus, for example,
\(p_c\), \(h(p)\), and \(A(p)\) abbreviate
\(p_c^{(q,\chi)}\), \(h_{q,\chi}(p)\), and \(A_{q,\chi}(p)\).  In the
finite-size scaling plots below, \(\log\) denotes the natural logarithm.

\subsection{HGP code}\label{sec:hgp}

We first discuss the coherent-error-induced phase transition in the hypergraph-product (HGP) code~\cite{tillich2013quantum, leverrier2015quantum}. The HGP code is a Calderbank-Shor-Steane (CSS) code constructed from two classical linear codes \(C_1\) and \(C_2\) over \(\mathbb{F}_2\), with parity-check matrices \(H_1\) and \(H_2\).

\subsubsection*{Code construction}

Let
\(H_i\in\mathbb F_2^{m_i\times n_i}\), \(i=1,2\), where \(n_i\) is the
block length and \(m_i\) is the number of parity-check rows.  Write
\begin{align}
    k_i&=\dim\ker H_i=n_i-\rank H_i,\notag\\
    k_i^\top&=\dim\ker H_i^\top=m_i-\rank H_i .
\end{align}
The \(X\)- and \(Z\)-check rows of the HGP code are specified by
\begin{equation}
    \begin{aligned}
        H_X &=
        \begin{pmatrix}
            H_1 \otimes I_{n_2} \quad I_{m_1} \otimes H_2^\top
        \end{pmatrix},\\[6pt]
        H_Z &=
        \begin{pmatrix}
            I_{n_1} \otimes H_2 \quad H_1^\top \otimes I_{m_2}
        \end{pmatrix},
    \end{aligned}
\end{equation}
which obey \(H_XH_Z^\top=0\).  The row sets may be redundant.  The total
number of physical qubits is
\begin{equation}
    N = n_1 n_2 + m_1 m_2.
\end{equation}

Logical operators arise from nontrivial homology classes.  In binary
\(X\)-support notation, if
\([v_1]\in\mathbb F_2^{n_1}/\operatorname{im}H_1^\top\) is nonzero and
\(0\ne v_2\in\ker H_2\), then
\begin{equation}\label{eq:hgp_xi}
      X^{\mathrm{I}} = (\,v_1 \otimes v_2 \,\vert\, 0\,)
\end{equation}
defines a type-I logical \(X\) representative. Similarly, if
\(0\ne v_1'\in\ker H_1^\top\) and
\([v_2']\in\mathbb F_2^{m_2}/\operatorname{im}H_2\) is nonzero, then
\begin{equation}\label{eq:hgp_xii}
      X^{\mathrm{II}} = (\,0 \,\vert\, v_1' \otimes v_2'\,)
\end{equation}
defines a type-II logical \(X\) representative. Logical \(Z\) operators arise
from the dual construction. The total number of logical qubits is
\begin{equation}
    K = k_1 k_2 + k_1^\top k_2^\top,
\end{equation}
Define \(d_i\) as the minimum nonzero Hamming weight in \(\ker H_i\) and
\(d_i^\top\) as the minimum nonzero Hamming weight in \(\ker H_i^\top\),
using \(+\infty\) when the corresponding kernel is trivial.  In general the
standard HGP parameter theorem gives
\begin{equation}
    d_{\mathrm{HGP}}
    \ge \min\{d_1,d_2,d_1^\top,d_2^\top\}.
\end{equation}
For the positive-rate case in which both \(H_i\) have full row rank, the
transpose kernels are trivial and the familiar equality
\(d_{\mathrm{HGP}}=\min\{d_1,d_2\}\) applies.

As a concrete example, we take \(H_1\) and \(H_2\) to be random
\(\mathrm{LDPC}_n(3,6)\) matrices with \(m_i=n/2\), hence design rate
\(1/2\)~\cite{mezard2009information}; their actual dimensions are determined
by the sampled matrix ranks. In that case,
\begin{equation}
    N = \frac{5}{4} n^2,
    \quad
    K \approx \frac{1}{4} n^2,
    \quad
    R_{\mathrm{code}}\approx\frac{1}{5}.
\end{equation}

\subsubsection*{Syndrome state logicals}

We now analyze how coherent errors modify the logical stabilizer structure of the post-measurement syndrome state. Following the general definitions in Sec.~\ref{sec:recoverability}, we quantify this change using the group-difference signature.

For each code realization \(\mathcal C\), the simulation chooses a complete
independent set of \(K(\mathcal C)\) logical \(X\) representatives of the two
types in Eqs.~\eqref{eq:hgp_xi} and \eqref{eq:hgp_xii}.  The fixed encoded
input \(\rho_{\mathrm{enc}}^{(\mathcal C)}\) is the joint \(+1\) eigenstate of
all physical checks and these logical \(X\) operators--equivalently, the
encoded logical state \(\ket{+}^{\otimes K(\mathcal C)}\) in the conjugate
logical-\(Z\) basis.  Its logical stabilizer group
\(\mathcal G_{\mathcal L}(\mathcal C)\) is generated by the corresponding
logical classes. For a fixed pair \((\mathcal C,\mathcal U)\), let
\(\mathcal G_{\mathcal L}'(\mathcal C,\mathcal U)\) denote the logical
stabilizer group of the post-measurement syndrome state. We define the
combined phase-free logical subspace
\begin{equation}
    G_{\mathrm{comb.}}^{\mathrm{sf}}
    =
    \operatorname{span}_{\mathbb F_2}
    \left(
        \underline{\mathcal G}_{\mathcal L},
        \underline{\mathcal G}'_{\mathcal L}
    \right),
\end{equation}
and the corresponding group-difference signature
\begin{equation}
    \Delta_{\mathrm{Logi.}}
    =
    \dim_{\mathbb F_2} G_{\mathrm{comb.}}^{\mathrm{sf}}
    -
    \dim_{\mathbb F_2}\underline{\mathcal{G}}_{\mathcal{L}}.
\end{equation}
Since the HGP code has a finite logical rate, it is convenient to normalize by the number of logical qubits:
\begin{equation}
    \delta_{\mathrm{Logi.}} = \frac{1}{K}\,\Delta_{\mathrm{Logi.}}.
\end{equation}

We report the expectation
\(\mathbb E_{\mathcal C,\mathcal U}[\delta_{\mathrm{Logi.}}]\) over the
random code and coherent-error realizations.

\begin{figure}[ht]
    \centering
    \subfloat[\label{fig:hgp_delta_L1} \(q = 1\)]{
        \includegraphics[width=0.2\textwidth]{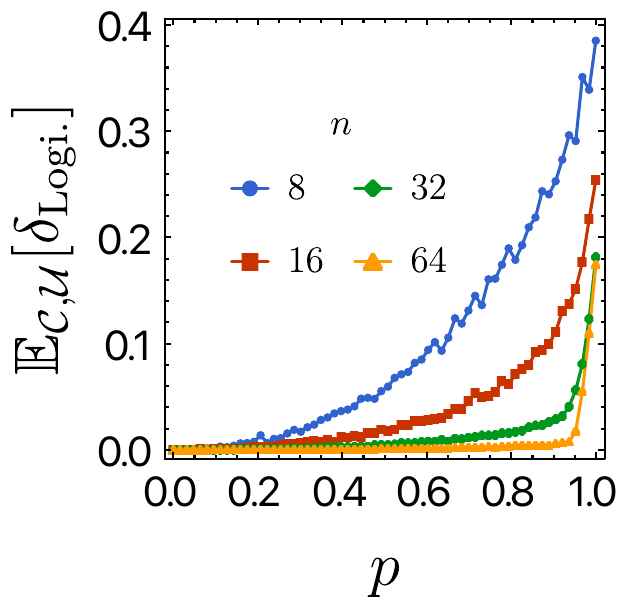}
    }\quad
    \subfloat[\label{fig:hgp_delta_L2} long-range \(q = 2\)]{
        \includegraphics[width=0.2\textwidth]{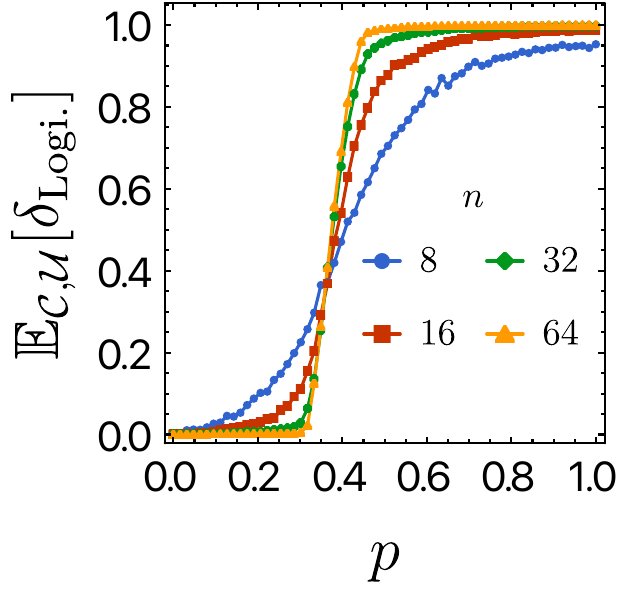}
    }\quad
    \subfloat[\label{fig:hgp_delta_L3} long-range \(q = 3\)]{
        \includegraphics[width=0.2\textwidth]{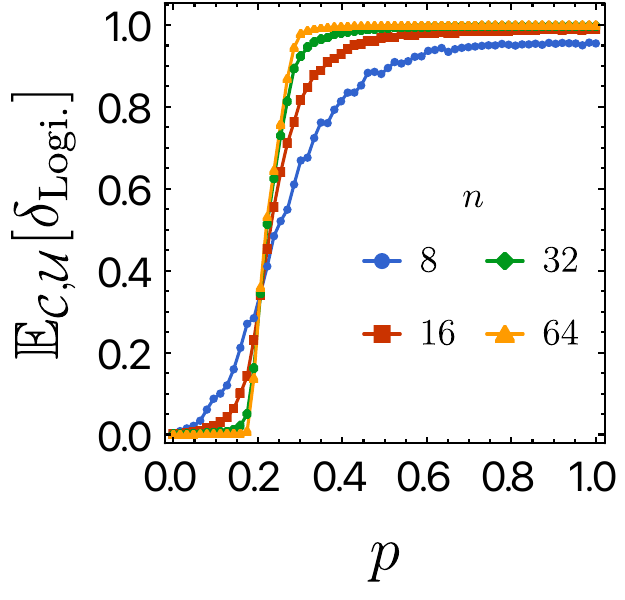}
    }\quad
    \subfloat[\label{fig:hgp_delta_S2} local \(q = 2\)]{
        \includegraphics[width=0.2\textwidth]{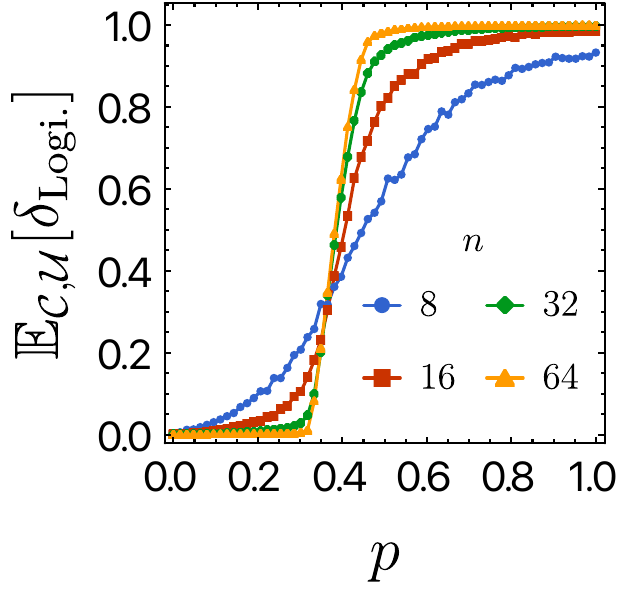}
    }\quad
    \subfloat[\label{fig:hgp_delta_S3} local \(q = 3\)]{
        \includegraphics[width=0.2\textwidth]{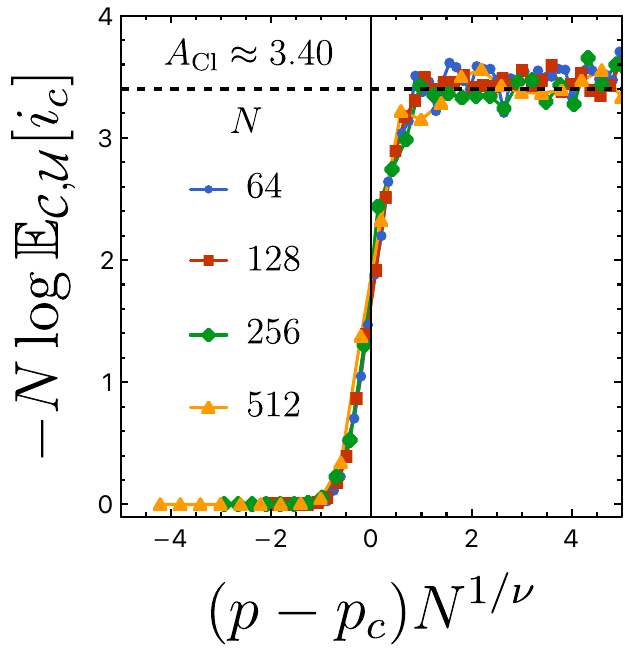}
    }
    \caption{\label{fig:hgp_delta}
    The expectation \(\mathbb E_{\mathcal C,\mathcal U}[\delta_{\mathrm{Logi.}}]\) as a function of the gate-activation probability \(p\) for HGP codes of size \(n=8,16,32,64\). Panels (a)--(c) show the \(q=1\) and long-range \(q\)-unitary models, while panels (d)--(e) show the at-most-\(q\) graph-walk local model. Each data point is averaged over both random HGP-code realizations and random coherent-error realizations.}
\end{figure}

Figure~\ref{fig:hgp_delta} shows that for HGP codes built from \(\mathrm{LDPC}_n(3,6)\times \mathrm{LDPC}_n(3,6)\), a phase transition appears at a threshold \(p_c\) for both the long-range and \(q\)-local models. For \(q>1\), the data are consistent with
\begin{equation}
    \mathbb E_{\mathcal C,\mathcal U}[\delta_{\mathrm{Logi.}}] \to
    \begin{cases}
        0, & p < p_c, \\[4pt]
        1, & p > p_c,
    \end{cases}
    \quad (n\to\infty),
\end{equation}
while for \(q=1\) we again find a transition from
\(\mathbb E_{\mathcal C,\mathcal U}[\delta_{\mathrm{Logi.}}]=0\) below
threshold to a nonzero value above threshold.

Thus, once the gate-activation probability exceeds \(p_c\), the ensemble-averaged logical
stabilizer change is extensive. By the state-level relation established in
Sec.~\ref{sec:recoverability}, each realization obeys
\begin{equation}
    P_{\mathrm{rec}}^{\mathrm{opt}}
    =
    2^{-\Delta_{\mathrm{Logi.}}},
\end{equation}
for a fixed pair of realizations \((\mathcal{C},\mathcal{U})\). For \(q>1\),
the convergence of the bounded variable
\(\delta_{\mathrm{Logi.}}\) to one in mean implies convergence in probability,
so a typical post-threshold sample has
\(\Delta_{\mathrm{Logi.}}=\Theta(K)\) and an exponentially small return
probability.  For \(q=1\), the nonzero limiting mean establishes an extensive
change on average; exponential suppression applies to those samples for which
\(\delta_{\mathrm{Logi.}}\) remains bounded away from zero, while a
concentration statement would require additional distributional data.
Channel-level recoverability is assessed separately below.

\subsubsection*{Coherent information}

We next examine \(I_c\) for the uncorrected error-and-syndrome channel.  We use
the same maximally mixed encoded input \(\rho_Q\) and reference
purification \(\ket{\Phi_D}_{\mathsf R\mathcal L}\) defined in
Eqs.~\eqref{eq:encoded_maximally_mixed_state} and
\eqref{eq:maximally_mixed_purification}, now with \(D=2^K\).
Because the HGP code supports an extensive number \(K\) of logical qubits,
we focus on the coherent-information density defined in
Eq.~\eqref{eq:normalized_coherent_information},
\begin{equation}
    i_c
    =
    \frac{I_c}{K},
\end{equation}
and report its ensemble average
\begin{equation}
    \mathbb E_{\mathcal C,\mathcal U}\!\left[i_c\right].
\end{equation}

\begin{figure}[ht]
    \centering
    \subfloat[\label{fig:hgp_delta_1} \(q = 1\)]{
        \includegraphics[width=0.2\textwidth]{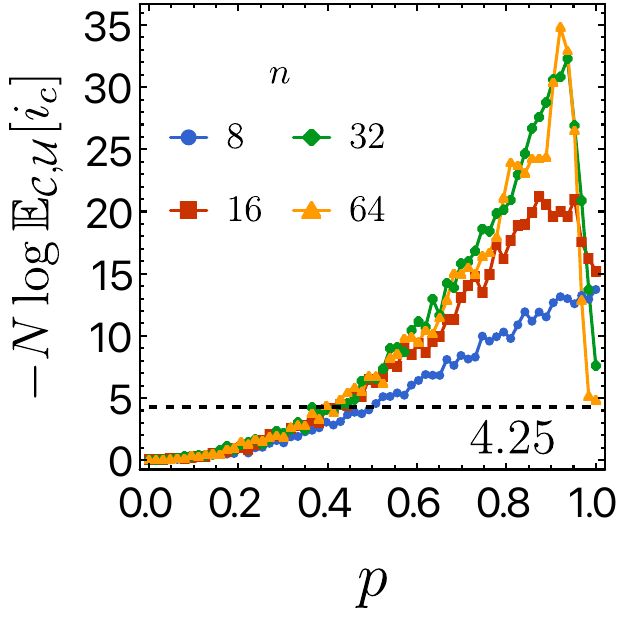}
    }\quad
    \subfloat[\label{fig:hgp_coInfo_L2} long-range \(q = 2\)]{
        \includegraphics[width=0.2\textwidth]{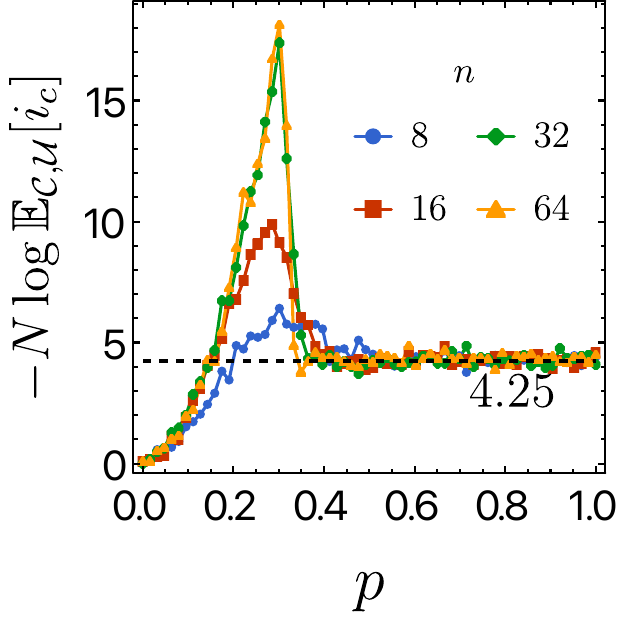}
    }\quad
    \subfloat[\label{fig:hgp_coInfo_L3} long-range \(q = 3\)]{
        \includegraphics[width=0.2\textwidth]{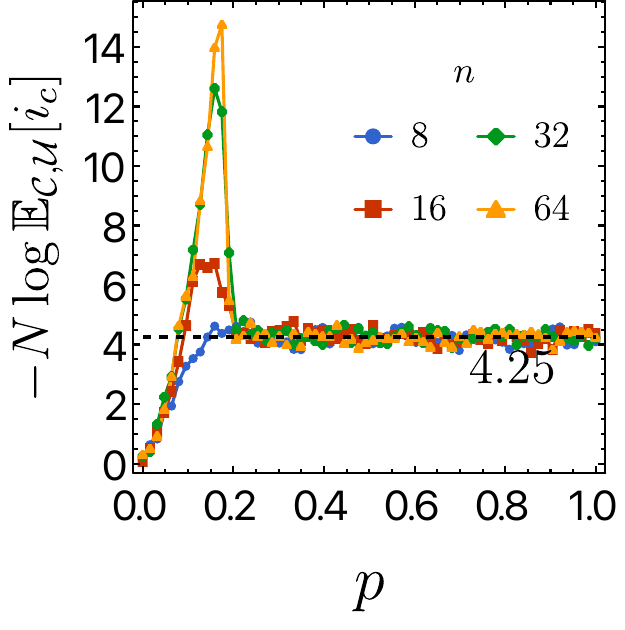}
    }\quad
    \subfloat[\label{fig:hgp_coInfo_2} local \(q = 2\)]{
        \includegraphics[width=0.2\textwidth]{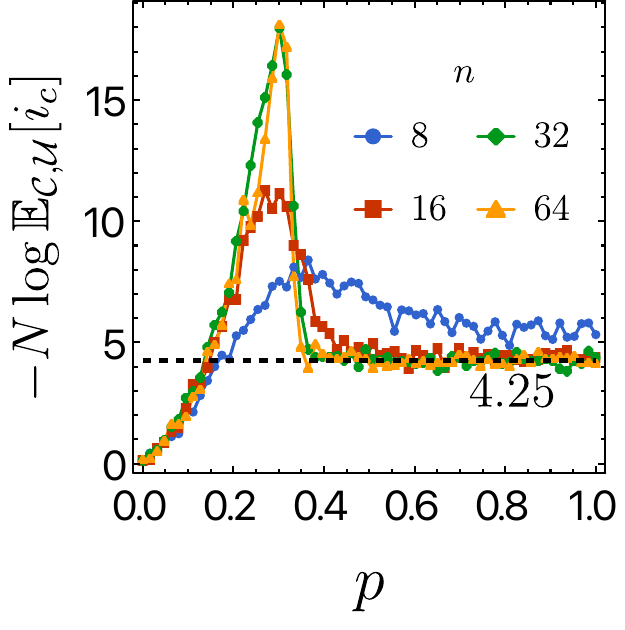}
    }\quad
    \subfloat[\label{fig:hgp_coInfo_3} local \(q = 3\)]{
        \includegraphics[width=0.2\textwidth]{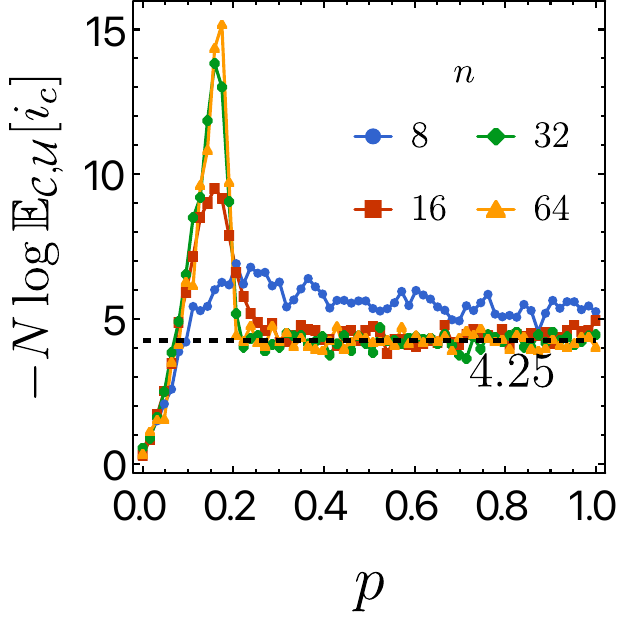}
    }
    \caption{\label{fig:hgp_coInfo}
    Scaled coherent-information-density deficit
    \(-N\log\mathbb E_{\mathcal C,\mathcal U}[i_c]\) under
    \(q\)-parameterized coherent errors for HGP codes. Panels (a)--(c) show the
    \(q=1\) and long-range \(q\)-unitary models, while panels (d)--(e) show
    the at-most-\(q\) graph-walk local model. Each data point is averaged over both random
    HGP-code realizations and random coherent-error realizations. For
    \(p>p_c\), \(h(p)=\mu(p)/R_{\mathrm{code}}\), where \(\mu(p)\) is the
    asymptotic total coherent-information deficit. The fully scrambled
    uniform-global-Clifford benchmark is
    \(h_{\mathrm{Cl}}=\mu_{\mathrm{Cl}}/R_{\mathrm{code}}\approx4.25\), using
    \(R_{\mathrm{code}}\approx1/5\) and \(\mu_{\mathrm{Cl}}\approx0.85018\).}
\end{figure}

As shown in Fig.~\ref{fig:hgp_coInfo}, the per-logical-qubit coherent information is well described by
\begin{equation}\label{eq:hgp_coInfo}
    \mathbb E_{\mathcal C,\mathcal U}\!\left[i_c\right]
    \approx
    \exp\!\Bigl(-\frac{h(p)}{N}\Bigr),
\end{equation}
where \(h(p)\ge 0\) is a function of the gate-activation probability \(p\).  This scaling is
consistent with an \(O(1)\) ensemble-mean total deficit
\begin{equation}
    \mu_N(p)
    \equiv
    \mathbb E_{\mathcal C,\mathcal U}
    \!\left[K(\mathcal C)-I_c(\mathcal C,\mathcal U)\right].
\end{equation}
For a fixed-rate ensemble with \(K=R_{\mathrm{code}}N\), the displayed
normalized form gives
\(\mu_N(p)=R_{\mathrm{code}}h(p)+o(1)\).  When \(K\) fluctuates with
\(\mathcal C\), this conclusion additionally uses concentration of the rate;
the unnormalized \(\mu_N\) is the quantity that tests it directly.  The
uniform-global-Clifford value
\(\mu_{\mathrm{Cl}}\approx0.85018\) provides a candidate fully scrambled
limit, but does not determine the observed \(p\)-dependence.

Combining this observation with the extensive change in logical stabilizer
structure above threshold, the HGP data are consistent with a regime dominated
not by complete erasure of logical information, but by \emph{logical
scrambling}: the ensemble-mean total coherent-information deficit remains
\(O(1)\) while the logical structure is extensively rearranged.  This
asymptotically maximal coherent-information density does not by itself imply
that the full uncorrected error-and-syndrome channel is close to a unitary
channel.

\subsubsection*{Syndrome distribution}

The phase transition also appears in the syndrome distribution.  Let
\(s=(s^X,s^Z)\) collect the outcomes of the measured rows of \(H_X\) and
\(H_Z\), including any redundant rows.  Thus the raw recorded-bit count is
\begin{equation}
    N_s=m_1n_2+n_1m_2 .
\end{equation}
Let
\(\Pi_X^{s^X}\) and \(\Pi_Z^{s^Z}\) be the corresponding products of commuting
single-check projectors.  For a code/error pair \((\mathcal C,\mathcal U)\),
\begin{equation}
    \mathbb{P}_{\mathcal C,\mathcal U}[s]
    =
    \Tr\!\left[
        \Pi_X^{s^X}\Pi_Z^{s^Z}
        \rho_E^{(\mathcal C,\mathcal U)}
    \right],
\end{equation}
where
\(\rho_E^{(\mathcal C,\mathcal U)}
=\mathcal U\rho_{\mathrm{enc}}^{(\mathcal C)}\mathcal U^\dagger\).
Following Sec.~\ref{sec:syndrome_distribution}, we characterize its global
structure by the reduced free-entropy density \(\varphi\) and report
\(\mathbb E_{\mathcal C,\mathcal U}[\varphi]\).

\begin{figure}[ht]
    \centering
    \subfloat[\label{fig:hgp_free_entropy_L1} \(q = 1\)]{
        \includegraphics[width=0.2\textwidth]{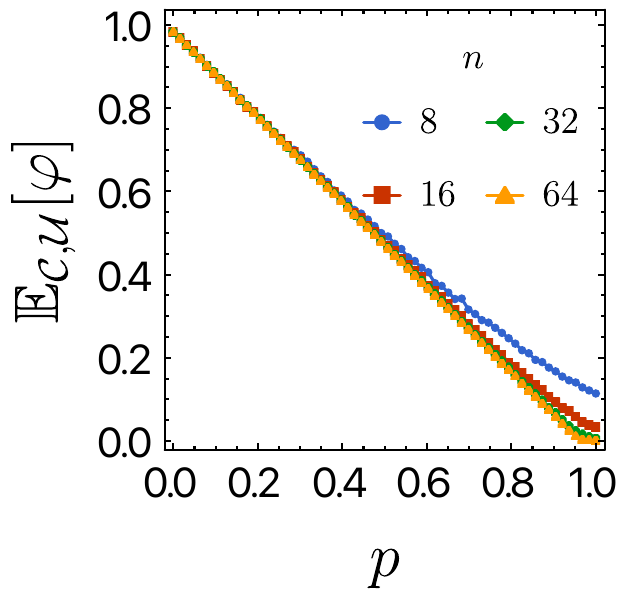}
    }\quad
    \subfloat[\label{fig:hgp_free_entropy_L2} long-range \(q = 2\)]{
        \includegraphics[width=0.2\textwidth]{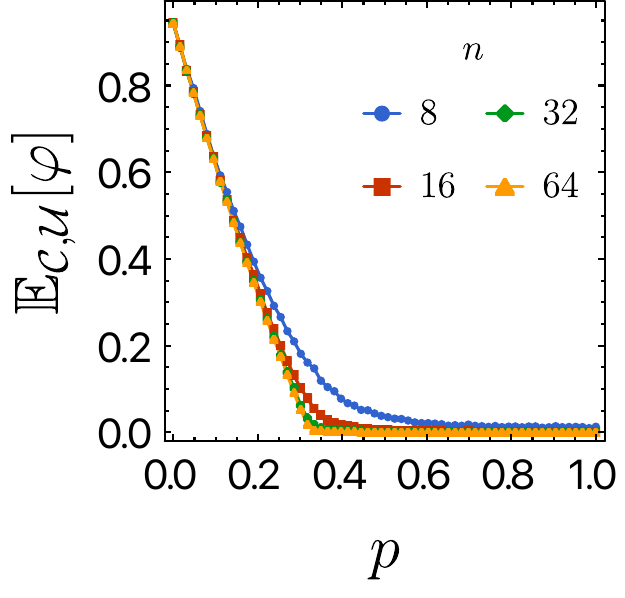}
    }\quad
    \subfloat[\label{fig:hgp_free_entropy_L3} long-range \(q = 3\)]{
        \includegraphics[width=0.2\textwidth]{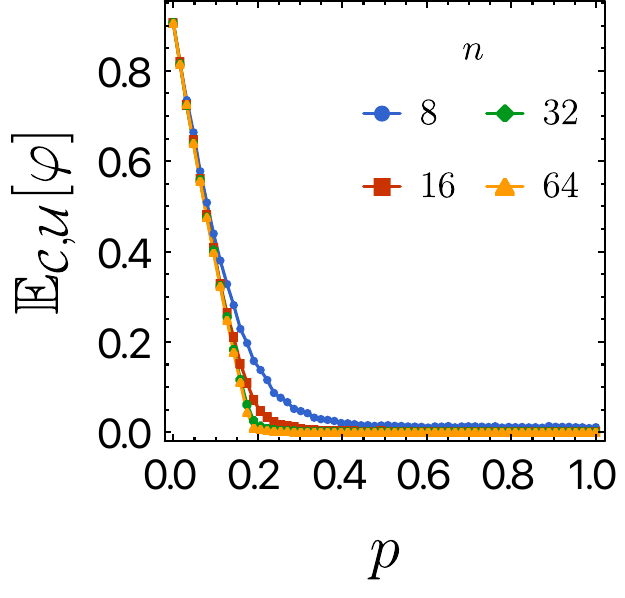}
    }\quad
    \subfloat[\label{fig:hgp_free_entropy_S2} local \(q = 2\)]{
        \includegraphics[width=0.2\textwidth]{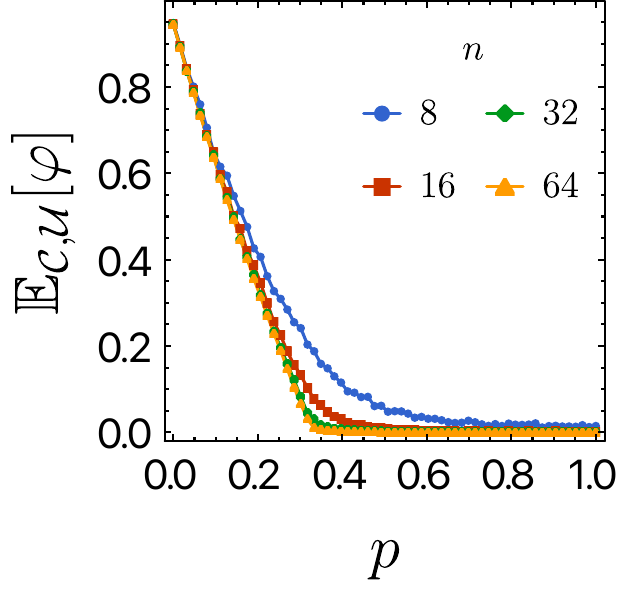}
    }\quad
    \subfloat[\label{fig:hgp_free_entropy_S3} local \(q = 3\)]{
        \includegraphics[width=0.2\textwidth]{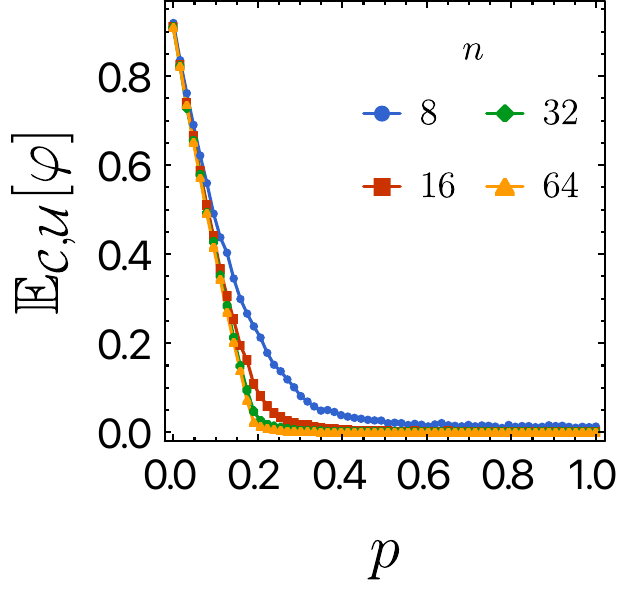}
    }
    \caption{\label{fig:hgp_free_entropy}
    The expectation \(\mathbb E_{\mathcal C,\mathcal U}[\varphi]\) versus the gate-activation probability \(p\) for HGP codes of size \(n=8,16,32,64\). Panels (a)--(c) show the \(q=1\) and long-range \(q\)-unitary models, while panels (d)--(e) show the at-most-\(q\) graph-walk local model. Each data point is averaged over both random HGP-code realizations and random coherent-error realizations.}
\end{figure}

Figure~\ref{fig:hgp_free_entropy} shows that in the large-system limit,
\begin{equation}
    \mathbb E_{\mathcal C,\mathcal U}[\varphi]
    \to
    \begin{cases}
        F_{\mathrm{HGP}}(p), & p < p_c, \\[3pt]
        0, & p > p_c,
    \end{cases}
\end{equation}
where \(F_{\mathrm{HGP}}(p)>0\). The vanishing of
\(\mathbb E_{\mathcal C,\mathcal U}[\varphi]\) above threshold means that the number of independent
syndrome constraints is subextensive and that the entropy per recorded
syndrome bit approaches one.  It does not imply pointwise convergence to
\(2^{-N_s}\): a subextensive set of constraints, potentially including
nonlocal logical information, may remain.  The data therefore show the loss of
an extensive density of constraint information, not the absence of every
decoding-relevant correlation.

Taken together, the HGP results reveal a post-threshold regime qualitatively
different from that of the toric code. In the toric code, coherent errors
eventually lead to a finite probability of genuine logical-information loss.
In the HGP code, by contrast, the logical subspace remains nearly intact at
the channel level, but its logical structure changes extensively in ensemble
mean.  For \(q>1\), this change is typical and the state-level Pauli-frame
return probability is exponentially small; for \(q=1\), the present mean data
do not establish concentration. We do not infer from this state-level result
that all state-independent recovery channels must fail.

\subsection{Random Clifford code}\label{sec:rcc}

\begin{figure}[ht]
    \centering
    \subfloat[\label{fig:random_cliff_code_enc} Clifford-code encoding]{
        \includegraphics[width=0.22\textwidth]{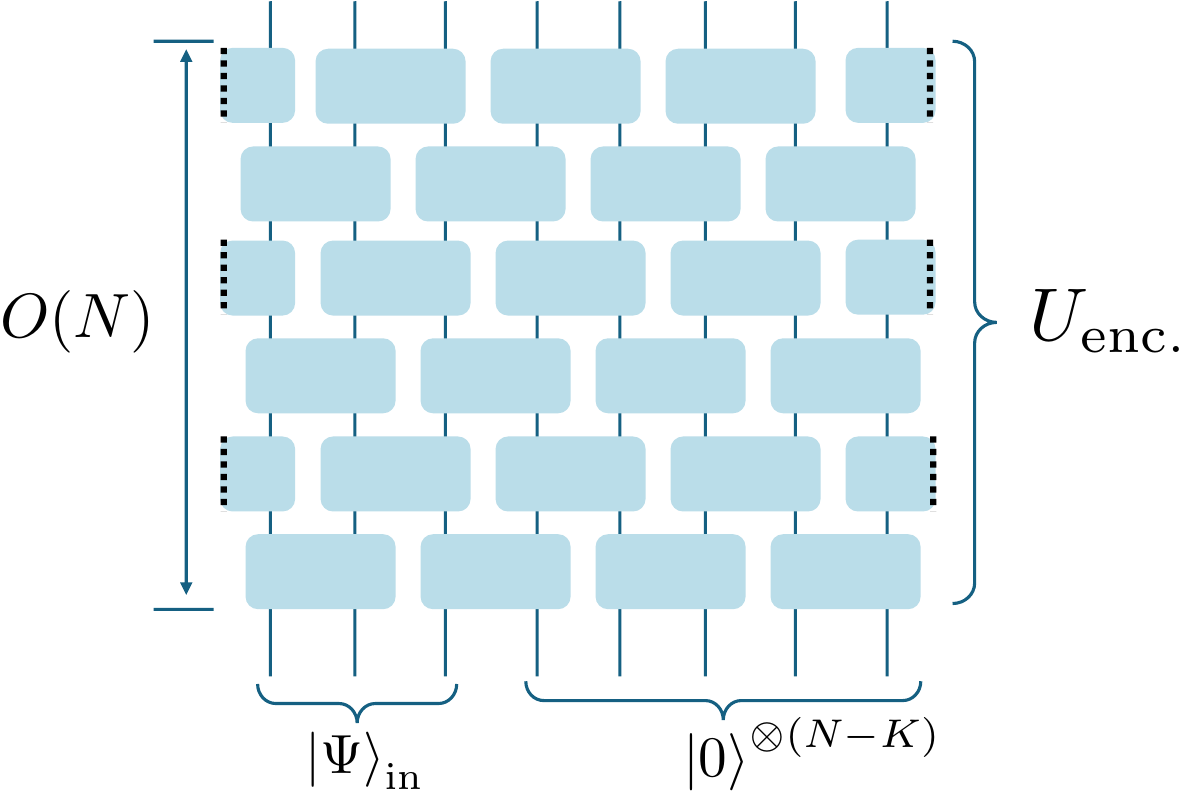}
    }\quad
    \subfloat[\label{fig:random_cliff_code_dec} Decoding and syndrome measurement]{
        \includegraphics[width=0.12\textwidth]{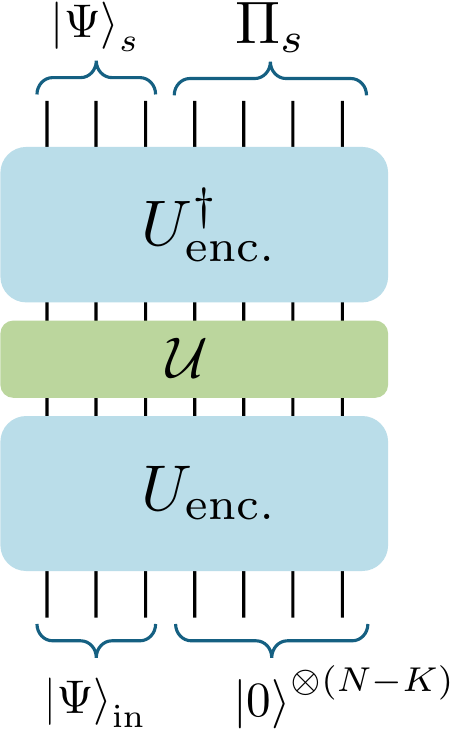}
    }
    \caption{\label{fig:random_cliff_code}
    (a) Encoding by a random Clifford unitary \(U_{\mathrm{enc}}\), implemented via a two-qubit brickwork circuit of depth \(O(N)\). Of the \(N\) qubits, the first \(K\) store the logical information, while the remaining qubits act as check qubits.
    (b) Decoding by the inverse circuit \(U_{\mathrm{enc}}^\dagger\), followed by syndrome measurement on the check qubits.}
\end{figure}

We now turn to another example of the RSCE: the \emph{random Clifford code}
(RCC)~\cite{brown2013short, nelson2023fault, PhysRevResearch.6.023055,
PhysRevX.11.031066, turkeshi2024error}. Concretely, we consider an \(N\)-qubit
code generated by a random two-qubit Clifford brickwork circuit of depth
\(2N\). Of the total \(N\) input qubits, the first \(K=N/4\) serve as logical
qubits, so \(R_{\mathrm{code}}=1/4\), as illustrated in
Fig.~\ref{fig:random_cliff_code}. We study the coherent-error-induced phase
transition under the \(q\)-unitary noise model introduced in
Sec.~\ref{sec:rcc_error_model}.

More explicitly, for a sampled encoding circuit \(U_{\mathrm{enc}}^{(\mathcal
C)}\), the code isometry is
\begin{equation}
    V_{\mathcal C}\ket{\psi}_{\mathcal L}
    =
    U_{\mathrm{enc}}^{(\mathcal C)}
    \left(
        \ket{\psi}_{1:K}\otimes
        \ket{0}^{\otimes(N-K)}_{K+1:N}
    \right).
\end{equation}
The circuit has \(2N\) alternating periodic nearest-neighbor brickwork layers,
with an independently sampled uniform two-qubit Clifford (modulo Pauli phase)
on every pair.  Physical check representatives and logical-\(Z\)
representatives are, respectively,
\begin{equation}
    h_j^{(\mathcal C)}
    =
    U_{\mathrm{enc}}^{(\mathcal C)}Z_j
    U_{\mathrm{enc}}^{(\mathcal C)\dagger},
    \quad j=K+1,\ldots,N,
\end{equation}
and
\begin{equation}
    Z_{i,\mathrm{phys}}^{(\mathcal C)}
    =
    U_{\mathrm{enc}}^{(\mathcal C)}Z_i
    U_{\mathrm{enc}}^{(\mathcal C)\dagger},
    \quad i=1,\ldots,K.
\end{equation}
Thus \(\mathcal S(\mathcal C)=\langle h_{K+1}^{(\mathcal C)},\ldots,
h_N^{(\mathcal C)}\rangle\),
\(\bar Z_i=Z_{i,\mathrm{phys}}^{(\mathcal C)}\mathcal S(\mathcal C)\), and
the recorded syndrome has \(N_s=N-K\) bits.

\subsubsection*{Syndrome state logicals}

We take the input logical state to be
\begin{equation}
    \ket{\Psi_{\mathrm{in}}}_{\mathcal L}=\ket{0}^{\otimes K},
\end{equation}
with physical encoded state
\(\rho_{\mathrm{enc}}^{(\mathcal C)}
=V_{\mathcal C}\ket{\Psi_{\mathrm{in}}}\!\bra{\Psi_{\mathrm{in}}}
V_{\mathcal C}^\dagger\),
which is stabilized by the logical stabilizer group
\begin{equation}
    \mathcal{G}_{\mathcal{L}}
    =\langle \bar Z_1,\bar Z_2,\ldots,\bar Z_K\rangle.
\end{equation}
For a fixed pair of realizations \((\mathcal{C},\mathcal{U})\), let \(\mathcal{G}_{\mathcal{L}}'\) denote the logical stabilizer group of the post-measurement syndrome state. We again define the group-difference signature through
\begin{equation}
    \Delta_{\mathrm{Logi.}}
    =
    \dim_{\mathbb F_2}G_{\mathrm{comb.}}^{\mathrm{sf}}
    -
    \dim_{\mathbb F_2}\underline{\mathcal{G}}_{\mathcal{L}},
\end{equation}
with
\begin{equation}
    G_{\mathrm{comb.}}^{\mathrm{sf}}
    =
    \operatorname{span}_{\mathbb F_2}
    \left(
        \underline{\mathcal G}_{\mathcal L},
        \underline{\mathcal G}'_{\mathcal L}
    \right),
\end{equation}
and normalize by the number of logical qubits,
\begin{equation}
    \delta_{\mathrm{Logi.}}=\frac{1}{K}\,\Delta_{\mathrm{Logi.}}.
\end{equation}
The plotted quantity is
\(\mathbb E_{\mathcal C,\mathcal U}[\delta_{\mathrm{Logi.}}]\).

\begin{figure}[ht]
    \centering
    \subfloat[\label{fig:rcc_delta_L1} \(q = 1\)]{
        \includegraphics[width=0.2\textwidth]{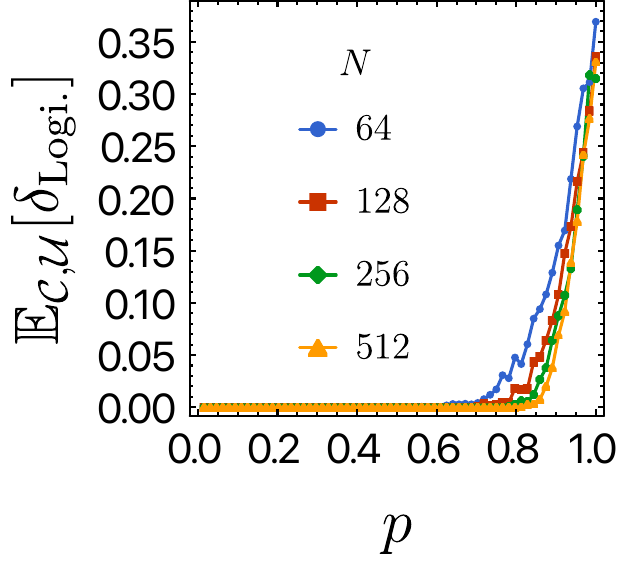}
    }\quad
    \subfloat[\label{fig:rcc_delta_L2} long-range \(q = 2\)]{
        \includegraphics[width=0.2\textwidth]{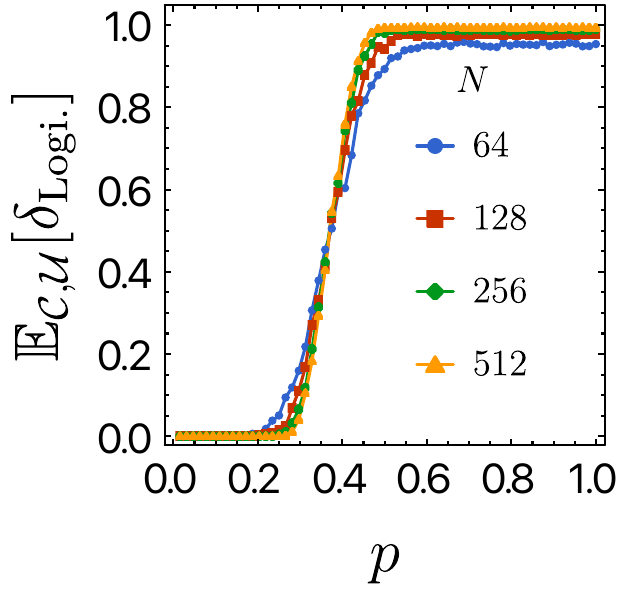}
    }\quad
    \subfloat[\label{fig:rcc_delta_L3} long-range \(q = 3\)]{
        \includegraphics[width=0.2\textwidth]{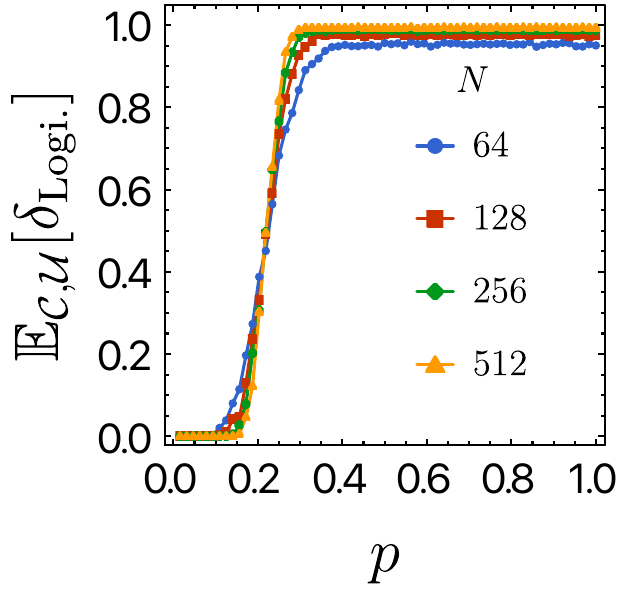}
    }\quad
    \subfloat[\label{fig:rcc_delta_S2} local \(q = 2\)]{
        \includegraphics[width=0.2\textwidth]{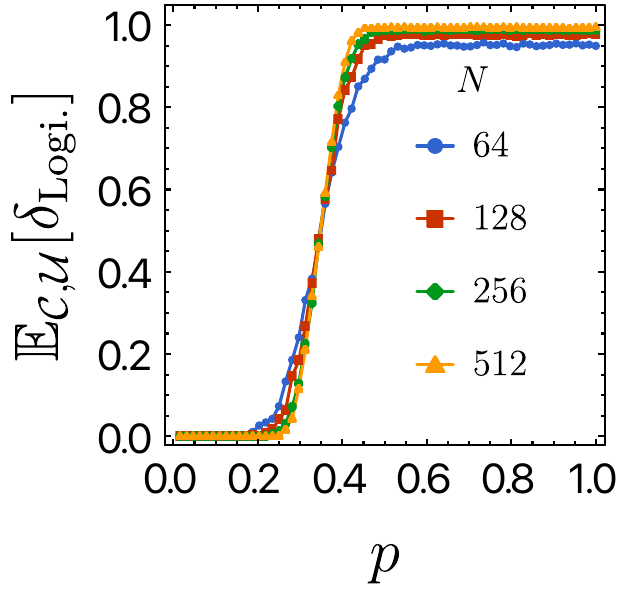}
    }\quad
    \subfloat[\label{fig:rcc_delta_S3} local \(q = 3\)]{
        \includegraphics[width=0.2\textwidth]{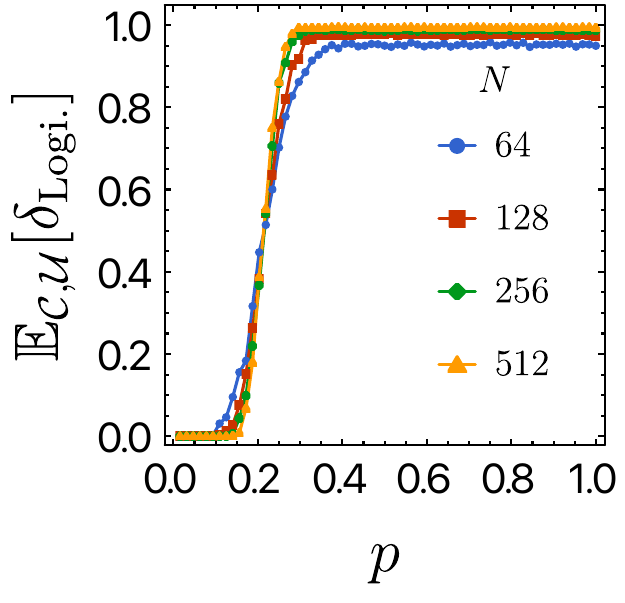}
    }
    \caption{\label{fig:rcc_delta}
    The expectation \(\mathbb E_{\mathcal C,\mathcal U}[\delta_{\mathrm{Logi.}}]\) as a function of the gate-activation probability \(p\) for RCCs with \(N=64,128,256,512\). Panels (a)--(c) show the \(q=1\) and long-range \(q\)-unitary models, while panels (d)--(e) show the \(q\)-local model. Each RCC has finite code rate with \(K=N/4\), and each data point is averaged over both random code realizations and random coherent-error realizations.}
\end{figure}

Figure~\ref{fig:rcc_delta} shows a phase transition in
\(\mathbb E_{\mathcal C,\mathcal U}[\delta_{\mathrm{Logi.}}]\) as a function
of the gate-activation probability \(p\), for both long-range and \(q\)-local
error models. For \(q>1\), the data are consistent with
\begin{equation}
    \mathbb E_{\mathcal C,\mathcal U}[\delta_{\mathrm{Logi.}}]
    \to
    \begin{cases}
        0, & p < p_c, \\[5pt]
        1, & p > p_c,
    \end{cases}
    \quad (N\to\infty),
\end{equation}
while for \(q=1\),
\(\mathbb E_{\mathcal C,\mathcal U}[\delta_{\mathrm{Logi.}}]=0\) below
threshold and becomes nonzero above threshold.

Thus, as in the HGP case, once \(p\) exceeds \(p_c\), the ensemble-averaged
logical stabilizer change is extensive. By the state-level MAP relation from
Sec.~\ref{sec:recoverability}, each realization obeys
\begin{equation}
    P_{\mathrm{rec}}^{\mathrm{opt}}=2^{-\Delta_{\mathrm{Logi.}}},
\end{equation}
for a fixed pair \((\mathcal{C},\mathcal{U})\).  For \(q>1\), the observed
convergence of \(\mathbb E_{\mathcal C,\mathcal U}[\delta_{\mathrm{Logi.}}]\) to one implies that a
typical sample has an exponentially small return probability.  For \(q=1\),
the nonzero limiting mean establishes an extensive change on average, while
exponential suppression applies samplewise whenever
\(\delta_{\mathrm{Logi.}}\) remains bounded away from zero. This diagnoses
scrambling relative to the chosen logical basis, rather than by itself proving
channel-level irreversibility.

\subsubsection*{Coherent information}

To characterize the channel-level behavior, we again use the maximally mixed
encoded input \(\rho_Q\) and its reference purification
\(\ket{\Phi_D}_{\mathsf R\mathcal L}\), now with \(D=2^K\).  We then study the
coherent-information density
\begin{equation}
    i_c
    =
    \frac{I_c}{K},
\end{equation}
and report the ensemble average
\begin{equation}
    \mathbb E_{\mathcal C,\mathcal U}\!\left[i_c\right].
\end{equation}

\begin{figure}[ht]
 \centering
    \subfloat[\label{fig:rcc_coInfo_L1} \(q = 1\)]{
        \includegraphics[width=0.2\textwidth]{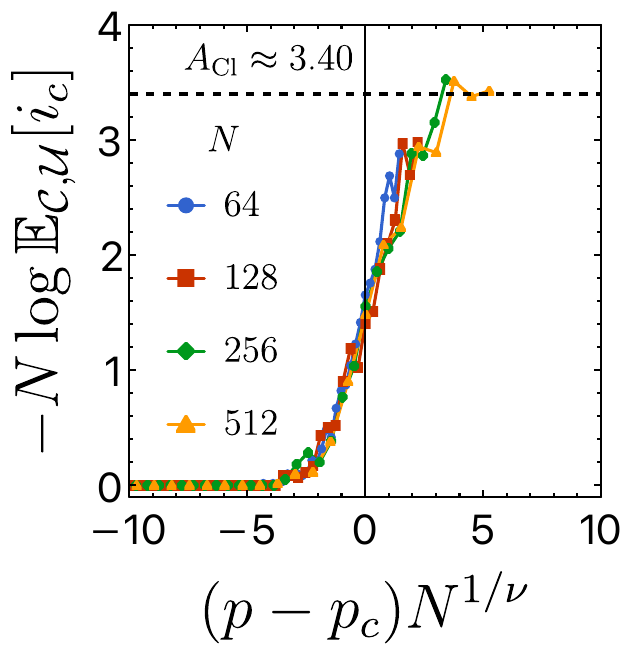}
    }\quad
    \subfloat[\label{fig:rcc_coInfo_L2} long-range \(q = 2\)]{
        \includegraphics[width=0.2\textwidth]{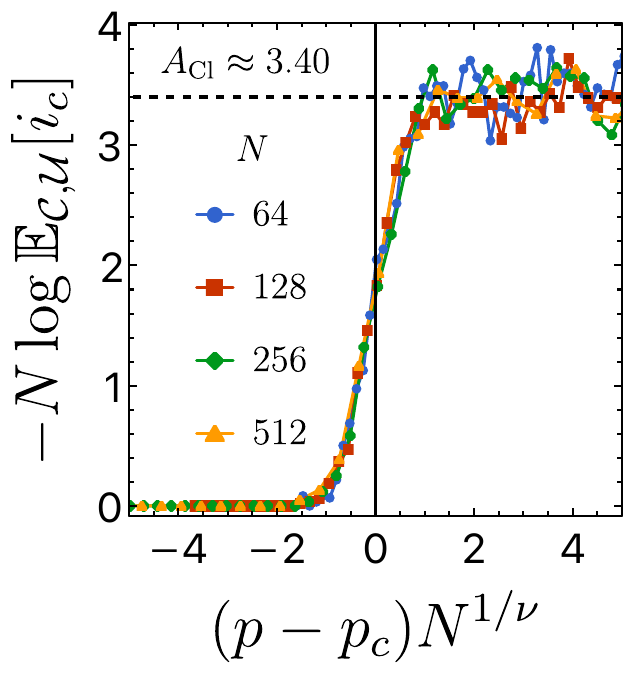}
    }\quad
    \subfloat[\label{fig:rcc_coInfo_L3} long-range \(q = 3\)]{
        \includegraphics[width=0.2\textwidth]{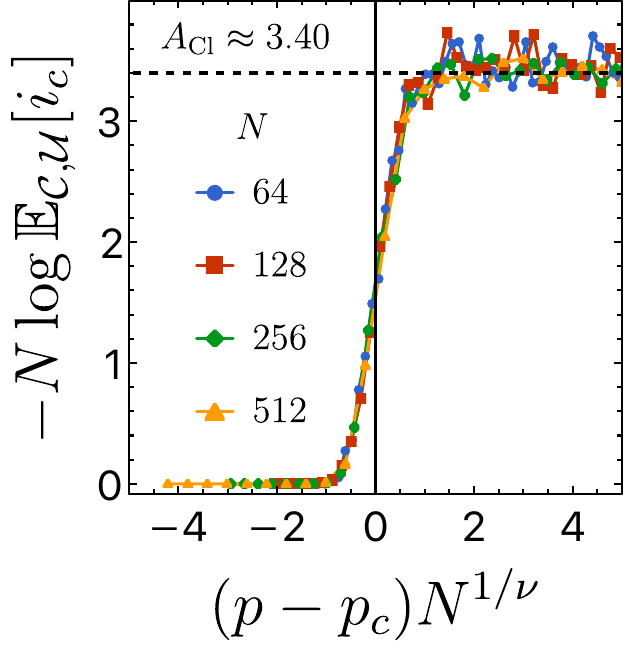}
    }\quad
    \subfloat[\label{fig:rcc_coInfo_S2} local \(q = 2\)]{
        \includegraphics[width=0.2\textwidth]{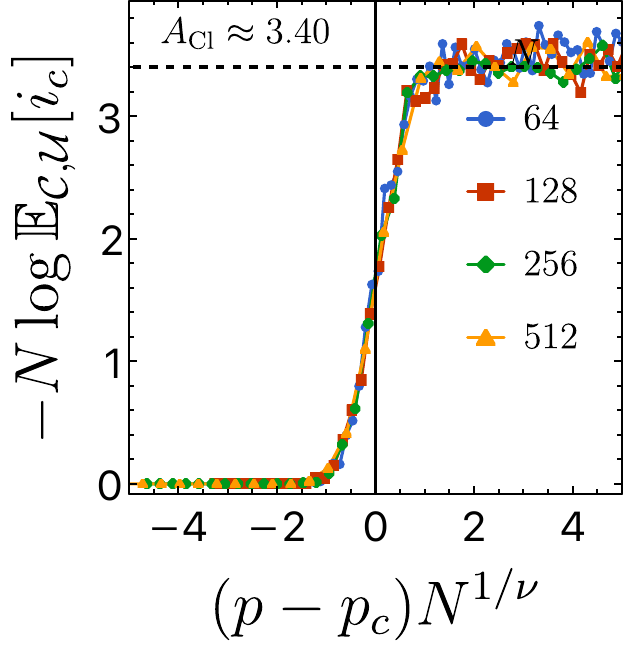}
    }\quad
    \subfloat[\label{fig:rcc_coInfo_S3} local \(q = 3\)]{
        \includegraphics[width=0.2\textwidth]{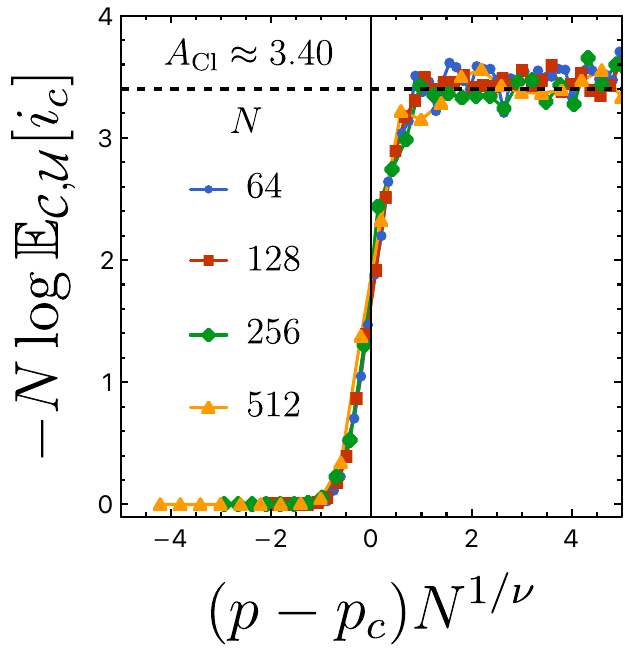}
    }
    \caption{\label{fig:rcc_coInfo}
    Scaled coherent-information-density deficit
    \(-N\log\mathbb E_{\mathcal C,\mathcal U}[i_c]\) under
    \(q\)-unitary coherent errors for RCCs. Panel (a) corresponds to the
    \(q=1\) model, panels (b)--(c) to the long-range \(q\)-unitary model,
    and panels (d)--(e) to the \(q\)-local model. Each data point is averaged
    over both random code realizations and random coherent-error
    realizations. For \(p>p_c\), \(A(p)=\mu(p)/R_{\mathrm{code}}\), where
    \(\mu(p)\) is the asymptotic total coherent-information deficit. The
    fully scrambled uniform-global-Clifford benchmark is
    \(A_{\mathrm{Cl}}=\mu_{\mathrm{Cl}}/R_{\mathrm{code}}\approx3.40\), using
    \(R_{\mathrm{code}}=1/4\) and \(\mu_{\mathrm{Cl}}\approx0.85018\).}
\end{figure}

Figure~\ref{fig:rcc_coInfo} reveals that for \(p<p_c\), the per-logical-qubit coherent information remains close to \(1\), indicating that the encoded logical information is essentially preserved at the channel level. Above threshold, it decreases but remains close to unity in a way consistent with
\begin{equation}
    \mathbb E_{\mathcal C,\mathcal U}\!\left[i_c\right]
    \approx
    \exp\!\left(-\frac{A(p)}{N}\right),
\end{equation}
where \(A(p)>0\) is an \(O(1)\) function at fixed \(p\). We further observe a data collapse of the form
\begin{equation}
    \log \mathbb E_{\mathcal C,\mathcal U}\!\left[i_c\right]
    =
    N^{-1} f_{\mathrm{RCC}}\bigl((p-p_c)N^{1/\nu}\bigr),
\end{equation}
with \(\nu \approx 1.61\) for \(q=1\) and \(\nu \approx 1.92\) for \(q>1\).
This one-variable collapse refers to the critical window
\(p-p_c=O(N^{-1/\nu})\); \(A(p)\) describes the separate fixed-\(p\)
large-\(N\) asymptotics away from that window.

As in the HGP case, the exponential fit is consistent with an \(O(1)\) total
coherent-information deficit.  If the dense-error ensemble reaches the
uniform-global-Clifford fixed point, the stronger prediction is
\(A(p)\to\mu_{\mathrm{Cl}}/R_{\mathrm{code}}\), with
\(\mu_{\mathrm{Cl}}\approx0.85018\); testing this requires the unnormalized
deficit \(K-\mathbb E_{\mathcal C,\mathcal U}[I_c]\) and the full distribution of the integer
logical-measurement rank \(r\).

As in the HGP case, the data are consistent with an RCC post-threshold phase
characterized predominantly by logical scrambling rather than complete
logical erasure: the coherent-information deficit is subextensive, while the
syndrome-resolved logical structure becomes extensively altered.  This is a
statement about coherent-information density, not convergence of the full
channel to a unitary channel.

\subsubsection*{Syndrome distribution}

We finally examine the syndrome distribution \(\mathbb{P}[s]\) through the
reduced free-entropy density \(\varphi\). The plotted quantity is
\(\mathbb E_{\mathcal C,\mathcal U}[\varphi]\).

\begin{figure}[ht]
    \centering
    \subfloat[\label{fig:rcc_free_entropy_L1} \(q = 1\)]{
        \includegraphics[width=0.2\textwidth]{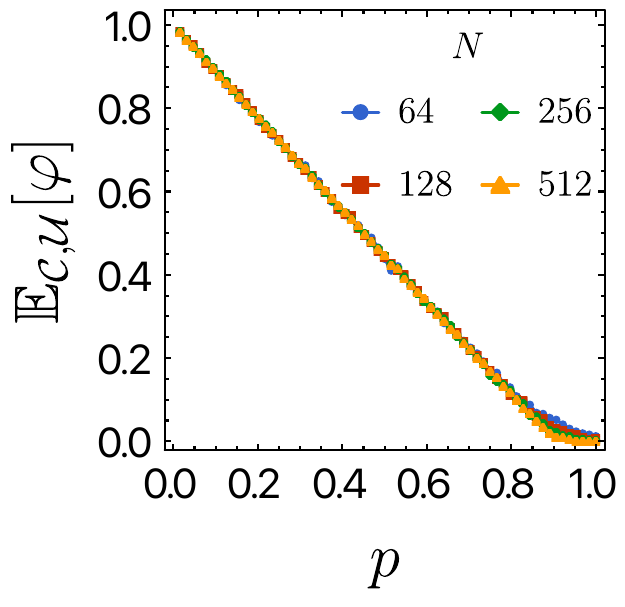}
    }\quad
    \subfloat[\label{fig:rcc_free_entropy_L2} long-range \(q = 2\)]{
        \includegraphics[width=0.2\textwidth]{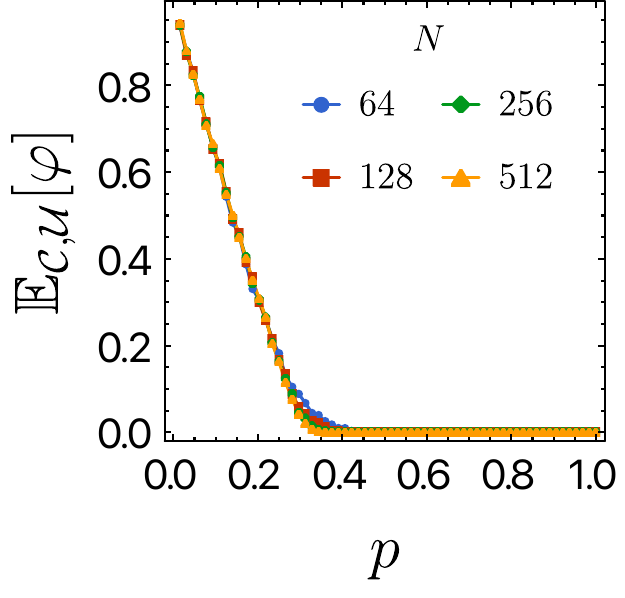}
    }\quad
    \subfloat[\label{fig:rcc_free_entropy_L3} long-range \(q = 3\)]{
        \includegraphics[width=0.2\textwidth]{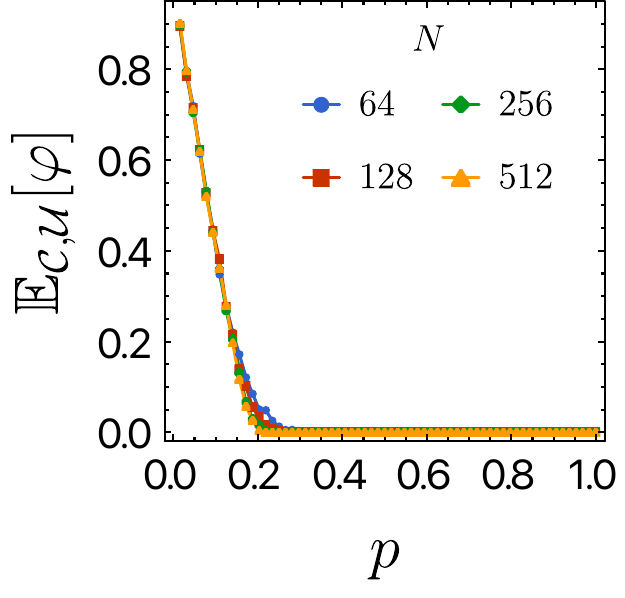}
    }\quad
    \subfloat[\label{fig:rcc_free_entropy_S2} local \(q = 2\)]{
        \includegraphics[width=0.2\textwidth]{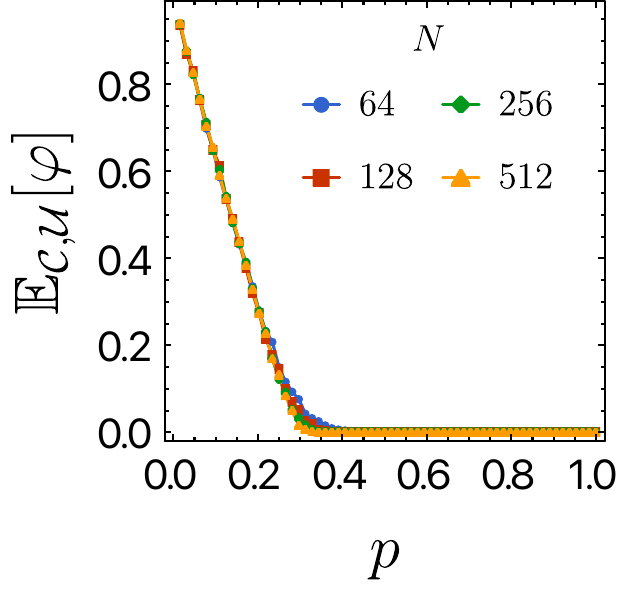}
    }\quad
    \subfloat[\label{fig:rcc_free_entropy_S3} local \(q = 3\)]{
        \includegraphics[width=0.2\textwidth]{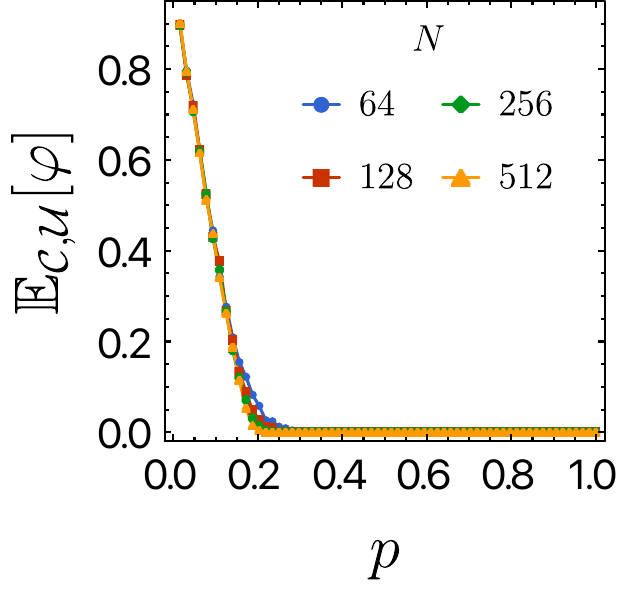}
    }
    \caption{\label{fig:RCC_free_entropy}
    The expectation \(\mathbb E_{\mathcal C,\mathcal U}[\varphi]\) as a function of the gate-activation probability \(p\) for RCCs with \(N=64,128,256,512\). Panels (a)--(c) show the \(q=1\) and long-range \(q\)-unitary models, while panels (d)--(e) show the \(q\)-local model. Each data point is averaged over both random code realizations and random coherent-error realizations.}
\end{figure}

As shown in Fig.~\ref{fig:RCC_free_entropy}, the large-\(N\) behavior is
\begin{equation}
    \mathbb E_{\mathcal C,\mathcal U}[\varphi]
    \to
    \begin{cases}
        F_{\mathrm{RCC}}(p), & p < p_c, \\[3pt]
        0, & p > p_c,
    \end{cases}
\end{equation}
where \(F_{\mathrm{RCC}}(p)>0\). Thus, once \(p\) exceeds \(p_c\), the data
are consistent with a subextensive number of independent syndrome constraints
and a syndrome entropy density approaching one.  This does not rule out an
\(O(1)\) or otherwise subextensive set of residual correlations and therefore
does not, by itself, show that the syndrome carries no useful decoding
information.

Taken together, these findings indicate that for \(p<p_c\), the syndrome
branches remain concentrated near the original logical basis sector, whereas
for \(p>p_c\) the syndrome loses an extensive density of constraints and the logical
stabilizer structure changes extensively in ensemble mean.  For \(q>1\), the
change is typical and the state-level Pauli-frame return probability is
exponentially small; for \(q=1\), concentration has not been established.
The channel-level coherent information nevertheless remains close to its
maximal value in the thermodynamic limit. Thus, for the RCC as well, the
coherent-error-induced
phase transition marks the onset of a logically scrambled phase; constructing
a state-independent recovery algorithm for this regime is beyond the scope of
the present work.

\section{Conclusion and discussion}
\label{sec:conclusion}

In this work, we studied how coherent unitary errors affect encoded
information in quantum stabilizer codes. Our formulation is based on two
closely related objects: the syndrome-resolved post-measurement state and the
corresponding syndrome distribution. Within this framework, the
coherent-error-induced transition can be diagnosed from several complementary
perspectives: the change in the logical stabilizer structure of the syndrome
state, the state-level MAP Pauli-frame return probability for a chosen
logical-basis input, the coherent information of the quantum channel, and the local and global structure of the syndrome distribution.

We illustrated these ideas in two representative classes of quantum
error-correcting codes. For the toric code~\cite{dennis2002topological,
kitaev2003fault}, we found a critical error threshold \(p_c\) separating a
phase in which the post-measurement logical stabilizer group remains
concentrated on the original logical sector from a phase in which it spreads
over distinct logical sectors. This transition is accompanied by a drop in
the state-level MAP return probability, a reduction of the coherent
information of the uncorrected error-and-syndrome channel, and critical
behavior in both the quantum and classical conditional mutual informations.
The coherent information establishes that the toric-code transition includes genuine channel-level loss of recoverability above threshold and therefore differs from the setting of Ref.~\cite{cheng2024emergent}.

For finite-rate random stabilizer-code ensembles, including the HGP
code~\cite{tillich2013quantum, leverrier2015quantum} and the random Clifford
code~\cite{brown2013short, nelson2023fault, PhysRevResearch.6.023055,
PhysRevX.11.031066}, the same post-threshold scrambling mechanism has a
different scaling signature.  Above threshold, the logical-group diagnostic
indicates an \(O(K)\) rearrangement in ensemble mean.  At the same time, the
per-logical-qubit coherent
information remains asymptotically close to its maximal value, with the data
consistent with an \(O(1)\) ensemble-mean total deficit. This
scrambling is accompanied by the collapse of the reduced free-entropy
density, indicating that the syndrome loses an extensive density of
constraints. 

The \(p>p_c\) data for both the toric code and the finite-rate codes agree
with the global-random-instrument benchmark, in which the coherent-error
unitary is taken to be a global random unitary.  This agreement suggests that
the post-threshold phase behaves as a random-state-sampling problem: syndrome
measurement produces conditional branch states with the statistics of the
corresponding random-state ensemble.

More broadly, our results highlight that coherent unitary errors pose
challenges fundamentally different from those of incoherent Pauli noise.
Whereas Pauli errors primarily modify stabilizer signs, coherent unitaries can
alter the logical stabilizer structure itself.  In the post-threshold phase
\(p>p_c\), the observed agreement with the global-random-instrument benchmark
suggests a connection to random-circuit sampling: syndrome measurement samples
conditional states from an effectively random ensemble.  For generic
non-Clifford coherent errors, this sampling viewpoint may entail a transition
in classical simulability, as studies of shallow random circuits relate
measurement-induced entanglement to the boundary between efficiently
simulable and quantum-advantage regimes~\cite{Napp2022,Watts2025}.  This gives
a computational-complexity perspective complementary to the transfer-matrix
limitations associated with volume-law
entanglement~\cite{liu2023quantum,bao2024phases,McGinley2025}.
Because the present
Clifford model remains stabilizer-simulable, however, our data do not by
themselves establish classical hardness.  

Several important directions remain open. One is the development of decoders specifically tailored to coherent errors and logical scrambling, beyond conventional syndrome-based strategies. Another is to understand how general the post-threshold random-state sampling picture is beyond the Clifford setting: for sufficiently scrambling non-Clifford coherent errors, we expect the stabilizer statistics used here to be replaced by Haar/Page-type statistics~\cite{ZyczkowskiSommers2001,Page1993}, but establishing this beyond the stabilizer-simulable regime remains an open problem. It would also be interesting to explore whether controlled coherent operations followed by syndrome measurement can be used as a robust route to implementing logical gates inside protected code spaces. Finally, clarifying the relation between coherent-error-induced transitions and other monitored-system transitions may help establish a broader framework connecting quantum error correction, statistical mechanics, and information dynamics.

\begin{acknowledgments}
We gratefully acknowledge the computing resources provided by Research Services at Boston College and the invaluable assistance of Wei Qiu. We are grateful to Yaodong Li, Shengqi Sang, Tianci Zhou and Timothy Hsieh for their insightful discussions. This research was supported in part by the National Science Foundation under Grant No.~DMR-2219735.
\end{acknowledgments}

\appendix

\section{Pauli sampling on quantum stabilizer states}
\label{app:stab_measurement}

In this appendix, we derive the probability distribution of syndrome outcomes
produced by Pauli measurements on a stabilizer state and characterize the
corresponding post-measurement stabilizer structure. This appendix provides
the general measurement framework used throughout the main text. In
App.~\ref{app:map_delta}, we build on this structure to relate the logical
stabilizer group of a syndrome branch to the state-level MAP Pauli-frame
return probability for a fixed logical-basis input.

\subsection*{Setup}

Consider a commuting set of \(M\) Hermitian Pauli observables
\begin{equation}
    \mathcal{O}=\{O_1,\ldots,O_M\},
    \qquad
    O_i^\dagger=O_i,
    \quad
    O_i^2=I_N,
    \label{eq:pauli_observables}
\end{equation}
which are measured projectively.  The list may contain algebraically
dependent observables. The measurement outcomes are labeled by
\begin{equation}
    s=(s_1,\ldots,s_M)\in\{\pm1\}^M .
\end{equation}

Let the pre-measurement state \(\rho_G\) be an \(N\)-qubit pure stabilizer
state with signed \(+1\)-stabilizer group
\begin{equation}
    G=\langle g_1,\ldots,g_N\rangle .
\end{equation}
Thus \(g_j\rho_G=\rho_G\) for every generator, and since \(\rho_G\) is pure,
\(G\) has rank \(N\).

After measuring \(\mathcal{O}\), the post-measurement state \(\rho_s\)
conditioned on outcome \(s\) is again a pure stabilizer state.  Its full
signed physical stabilizer group \(\mathcal T_s\) is generated by two types
of operators:
\begin{itemize}
    \item the measured observables, with eigenvalues fixed by the outcomes \(s_i\),
    \item stabilizers inherited from the pre-measurement group that commute with all measured observables.
\end{itemize}
Below we select independent generators of each type.  Because the signs are
already part of the signed Pauli generators, every inherited element
\(g'\in G\) has eigenvalue \(+1\) on \(\rho_G\); no additional stabilizer
charge is attached to it.

\subsection*{Measurement outcome distribution}

Not all measurement outcomes are independent. They must satisfy constraints inherited from stabilizers in the original group \(G\) that can be written as products of the measured observables.

The allowed binary constraint vectors \(\Gamma_\alpha\) are determined by the
commutator matrix \(T\in\mathbb F_2^{M\times N}\), defined by
\begin{equation}
    T_{i,j}=
    \begin{cases}
        0, & [O_i,g_j]=0,\\
        1, & [O_i,g_j]\neq 0.
    \end{cases}
    \label{eq:commutator}
\end{equation}
A product \(\prod_i O_i^{\Gamma_{i,\alpha}}\) commutes with every generator \(g_j\) if and only if
\begin{equation}
    \sum_{i=1}^{M} \Gamma_{i,\alpha}\,T_{i,j}\equiv 0 \pmod 2
    \quad \forall j.
    \label{eq:comm_sol}
\end{equation}
Thus the compatible constraints are precisely the vectors
\(\Gamma_\alpha\in\ker_{\mathbb F_2}(T^\top)\), the left nullspace of \(T\).
For each basis vector of this nullspace, the measured Pauli product commutes
with the full pure-state stabilizer and hence equals a stabilizer up to a
sign:
\begin{equation}
    g_\alpha'
    =
    \epsilon_\alpha
    \prod_{i=1}^M O_i^{\,\Gamma_{i,\alpha}},
    \qquad
    g_\alpha'\in G,
    \quad
    \epsilon_\alpha\in\{\pm1\}.
    \label{eq:signed_measurement_relation}
\end{equation}
Because \(g_\alpha'\rho_G=\rho_G\), the corresponding outcomes satisfy
\begin{equation}
    \prod_{i=1}^{M} (s_i)^{\Gamma_{i,\alpha}}
    =
    \epsilon_\alpha .
    \label{eq:syndrome_constraint}
\end{equation}

In stabilizer systems, measurements of observables that are not fixed by the stabilizer structure yield equiprobable outcomes. By contrast, compatible observables reveal stabilizer charges and therefore impose constraints on the outcome pattern. As a result, the syndrome distribution takes the form
\begin{equation}
    \mathbb{P}[s]=Z^{-1}\prod_{\alpha=1}^{r_\Gamma}\delta_\alpha[s],
\end{equation}
with
\begin{equation}
    \delta_\alpha[s]
    =
    \frac{1}{2}
    \left(
        1+
        \epsilon_\alpha\prod_{i=1}^{M}(s_i)^{\Gamma_{i,\alpha}}
    \right),
\end{equation}
where
\begin{equation}
    r_\Gamma
    =
    \dim\ker_{\mathbb F_2}(T^\top)
    =
    M-\rank_{\mathbb F_2}T
\end{equation}
is the number of linearly independent constraints.

The normalization factor is the partition function
\begin{equation}
    Z=\sum_s \prod_{\alpha=1}^{r_\Gamma}\delta_\alpha[s]
    =2^{M-r_\Gamma} = 2^{\rank T}.
    \label{eq:partition_function}
\end{equation}

\subsection*{Post-measurement stabilizer structure}

We now characterize the stabilizers that survive the measurement.

Any stabilizer inherited from the pre-measurement group can be written as
\begin{equation}
    g_l'=\prod_{j=1}^N g_j^{X_{j,l}},
\end{equation}
where the binary coefficient vector
\(X_l=(X_{1,l},\ldots,X_{N,l})^\top\) must satisfy
\begin{equation}
    TX_l=0 .
\end{equation}
Thus the inherited post-measurement stabilizers are determined by the right
nullspace \(\ker_{\mathbb F_2}T\).

Since there are \(N\) original stabilizer generators and \(\operatorname{rank}(T)\) independent commutation constraints, the number of independent inherited stabilizers is
\begin{equation}
    n_g = N-\operatorname{rank}(T).
\end{equation}

In addition, there are
\begin{equation}
    n_o=\operatorname{rank}(T)
\end{equation}
independent measured observables whose outcomes are not fixed by the pre-measurement stabilizer structure. We denote these free measured observables by
\begin{equation}
    O_\mu'=\prod_{i=1}^M O_i^{\Theta_{i,\mu}},
    \quad
    s_\mu'=\prod_{i=1}^M s_i^{\Theta_{i,\mu}},
\end{equation}
where the \(\Theta_\mu\) span a complement of the constraint subspace in \(\mathbb{F}_2^M\), and \(s_\mu'\) is the corresponding measured eigenvalue.

The full post-measurement stabilizer group is therefore
\begin{equation}
    \mathcal T_s=\langle s_\mu' O_\mu',\; g_l' \rangle,
    \label{eq:post_mea_stab}
\end{equation}
with total rank
\begin{equation}
    \operatorname{rank}(\mathcal T_s)=n_g+n_o=N.
\end{equation}
Hence Eq.~\eqref{eq:post_mea_stab} indeed gives a complete generating set for the post-measurement pure stabilizer state.

For the stabilizer-check measurement considered in the main text,
\(\mathcal T_s\) contains the signed syndrome-sector stabilizer
\(\mathcal S_s=\langle s_1h_1,\ldots,s_{N-k}h_{N-k}\rangle\).
After discarding Pauli phases,
\(\underline{\mathcal S}_s=\underline{\mathcal S}\) is a subspace of
\(\underline{\mathcal T}_s\).  Row-reducing
\(\underline{\mathcal T}_s\) against \(\underline{\mathcal S}\) therefore
gives the phase-free post-measurement logical stabilizer subspace directly:
\begin{equation}
    \underline{\mathcal G}'_{\mathcal L}(s;c,\mathcal U)
    =
    \underline{\mathcal T}_s/
    \underline{\mathcal S}.
\end{equation}
No recovery operation is involved in this step.

For the signed logical group used in the branch-specific Pauli-frame
correction, conjugate \(\mathcal T_s\) by the chosen \(R_s^\dagger\) and then
quotient by \(\mathcal S\):
\begin{equation}
    \widetilde{\mathcal T}_s
    =
    R_s^\dagger\mathcal T_sR_s,
    \quad
    \mathcal G'_{\mathcal L}(s;c,\mathcal U)
    =
    \mathfrak q_{\mathcal L}(\widetilde{\mathcal T}_s)
    =
    \widetilde{\mathcal T}_s/\mathcal S,
\end{equation}
which is Eq.~\eqref{eq:postmeasurement_logical_group}.  This completes the
constructive passage from the commutator system to the signed logical group
used in the state-level correction.

Finally, if we average over all free measurement outcomes, we obtain the
nonselective state
\begin{equation}
    \rho_{\mathrm{ns}}
    =
    \sum_s \Pi_s \rho_G \Pi_s .
\end{equation}
Its stabilizer subgroup consists only of the inherited commuting stabilizers,
\begin{equation}
    \mathcal T_{\mathrm{ns}}
    =
    \left\langle
    g_l'=\prod_{j=1}^N g_j^{X_{j,l}}
    \,\middle|\,
    l=1,\dots,N-\operatorname{rank}(T)
    \right\rangle .
\end{equation}
In other words, averaging over the unfixed measurement outcomes removes the stabilizers associated with the free measured observables and retains only the stabilizer subgroup common to all syndrome branches.

\section{Random-instrument and random-state benchmarks}
\label{app:random_instrument_benchmark}

This appendix derives the random-instrument results quoted in
Sec.~\ref{sec:random_instrument}.  We first express coherent information in
terms of the syndrome-resolved logical POVM, then perform the global-Clifford
symplectic count, and finally give the global-Haar/Page benchmark.

\subsection*{Syndrome-resolved coherent information}

Let \(D=2^k\), let \(A_s\) be the decoded logical branch map in
Eq.~\eqref{eq:logical_branch_map}, and let
\begin{equation}
    E_s=A_s^\dagger A_s,
    \quad
    p_s^{(\pi)}=\frac{\Tr E_s}{D},
    \quad
    \widehat E_s=\frac{E_s}{\Tr E_s}
    \label{eq:app_normalized_logical_povm}
\end{equation}
for every nonzero branch.  Because the physical output sectors are mutually
orthogonal, the effective channel is isometrically equivalent to the flagged
logical channel
\begin{equation}
    \mathcal N_{\mathrm{flag}}(\rho)
    =
    \sum_s
    \ket{s}\!\bra{s}
    \otimes
    A_s\rho A_s^\dagger .
    \label{eq:app_flagged_channel}
\end{equation}
For the maximally mixed logical input, the normalized conditional Choi state
is pure, and either of its marginals has the same nonzero spectrum as
\(\widehat E_s\).  The classical syndrome entropy cancels between the two
terms in the coherent information, giving
\begin{equation}
    I_c
    =
    \sum_{s:p_s^{(\pi)}>0} p_s^{(\pi)} S(\widehat E_s).
    \label{eq:app_general_coherent_information}
\end{equation}
Equivalently, for the maximally mixed logical input, the reference--syndrome
state is
\begin{equation}
    \rho_{\mathsf R\mathsf S}
    =
    \sum_{s:p_s^{(\pi)}>0}
    p_s^{(\pi)}\,\widehat E_s^{\,T}
    \otimes\ket{s}\!\bra{s}_{\mathsf S},
\end{equation}
and
\begin{equation}
    k-I_c=I(\mathsf R:\mathsf S)_{\rho}.
    \label{eq:app_deficit_readout_identity}
\end{equation}
The coherent-information deficit is therefore exactly the information about
the reference acquired by the syndrome readout.  Equation
\eqref{eq:app_general_coherent_information} also proves that \(I_c=k\) if
and only if every nonzero \(E_s=p_s^{(\pi)}I_{\mathcal L}\), in agreement with
Proposition~\ref{prop:branch_unitarity}.

For a Clifford realization, Eq.~\eqref{eq:logical_instrument_normal_form}
implies
\begin{equation}
    \widehat E_s
    =
    \frac{
        C_{\mathrm{meas}}^\dagger
        P_{\lambda(s)}C_{\mathrm{meas}}
    }{
        2^{k-r}
    },
    \quad
    S(\widehat E_s)=k-r.
\end{equation}
Since the logical-measurement rank \(r\) is syndrome independent at fixed
Clifford realization,
\begin{equation}
    I_c=k-r.
    \label{eq:app_clifford_coherent_information}
\end{equation}

\subsection*{Uniform-global-Clifford count at fixed \(k\)}

We now compute the distribution of \(r\) when the physical Clifford is
uniformly random.  Ignore Pauli phases and let
\(H\in\mathbb F_2^{m\times 2N}\), with \(m=N-k\), be a full-rank binary
check matrix.  For a phase-free Pauli label
\(v(P)\in\mathbb F_2^{2N}\), define the row-vector convention for the
Clifford symplectic matrix by
\begin{equation}
    v(\mathcal U P\mathcal U^\dagger)
    =
    v(P)F_{\mathcal U},
    \qquad
    F_{\mathcal U}\in\operatorname{Sp}(2N,\mathbb F_2).
\end{equation}
The measured checks pulled backward through the error are therefore
\begin{equation}
    B=HF_{\mathcal U}^{-1}.
\end{equation}
Writing \(\mathcal S_{\mathrm{bin}}=\operatorname{row}(H)\) and using the
binary symplectic complement \(\mathcal S_{\mathrm{bin}}^\perp\), the
commuting logical Paulis learned by the syndrome form
\begin{equation}
    \mathcal L_{\mathrm{meas}}
    =
    \frac{
        \operatorname{row}(B)\cap\mathcal S_{\mathrm{bin}}^\perp
    }{
        \operatorname{row}(B)\cap\mathcal S_{\mathrm{bin}}
    },
    \quad
    r=\dim_{\mathbb F_2}\mathcal L_{\mathrm{meas}}.
    \label{eq:app_logical_measurement_subgroup}
\end{equation}
A uniform global Clifford makes \(\operatorname{row}(B)\) a uniformly
random \(m\)-dimensional isotropic subspace.

For later use, define the number of \(b\)-dimensional isotropic subspaces of
a \(2a\)-dimensional binary symplectic space,
\begin{equation}
    J_{a,b}
    =
    \prod_{i=0}^{b-1}
    \frac{2^{2(a-i)}-1}{2^{b-i}-1},
\end{equation}
and the Gaussian binomial coefficient
\begin{equation}
    \qbinom{a}{b}
    =
    \prod_{i=0}^{b-1}
    \frac{2^{a-i}-1}{2^{b-i}-1}.
\end{equation}
Choosing the intersection with \(\mathcal S_{\mathrm{bin}}\), passing to
the corresponding symplectic quotient, and summing the remaining transverse
isotropic subspaces with the binary \(q\)-binomial theorem gives the exact
number of global-Clifford images with logical-measurement rank \(r\):
\begin{equation}
    N_{m,k,r}
    =
    J_{k,r}
    \qbinom{m}{r}
    2^{r^2}
    \prod_{j=2k+1}^{2k+m-r}(1+2^j).
    \label{eq:app_exact_symplectic_count}
\end{equation}
These strata exhaust all allowed isotropic subspaces,
\begin{equation}
    \sum_{r=0}^{\min\{k,m\}}N_{m,k,r}=J_{m+k,m}.
\end{equation}

For fixed \(k,r\) and \(m\to\infty\), division by the \(r=0\) stratum
yields
\begin{equation}
    \lim_{m\to\infty}
    \frac{N_{m,k,r}}{N_{m,k,0}}
    =
    \frac{
        J_{k,r}\,
        2^{-2kr+r(r-1)/2}
    }{
        \prod_{\ell=1}^{r}(1-2^{-\ell})
    }.
    \label{eq:app_fixed_k_weights}
\end{equation}
For \(k=2\),
\begin{equation}
    J_{2,0}=1,
    \quad
    J_{2,1}=15,
    \quad
    J_{2,2}=15,
\end{equation}
and Eq.~\eqref{eq:app_fixed_k_weights} gives relative weights
\begin{equation}
    1:\frac{15}{8}:\frac{5}{16}
    =
    16:30:5.
\end{equation}
Normalizing and using Eq.~\eqref{eq:app_clifford_coherent_information},
\begin{equation}
    \Pr(r=0,1,2)
    =
    \frac{1}{51}(16,30,5),
    \quad
    \mathbb E_{\mathrm{Cl}}[I_c]
    =
    \frac{62}{51}.
    \label{eq:app_k2_clifford_result}
\end{equation}

\subsection*{Finite-rate global-Clifford limit}

Let \(K=R_{\mathrm{code}}N\) and \(m=N-K\), with both
\(K,m\to\infty\).  The exact count
in Eq.~\eqref{eq:app_exact_symplectic_count}, now with \(k=K\), obeys
\begin{equation}
    \begin{aligned}
        \frac{N_{m,K,r+1}}{N_{m,K,r}}
        &=
        \frac{
            (2^{2(K-r)}-1)(2^{m-r}-1)2^{2r+1}
        }{
            (2^{r+1}-1)^2(1+2^{2K+m-r})
        },\\
        &\rightarrow
        \frac{2}{(2^{r+1}-1)^2}.
    \end{aligned}
    \label{eq:app_finite_rate_ratio}
\end{equation}
Therefore \(r\) remains \(O(1)\) and has the limiting distribution
\begin{equation}
    \Pr_{\mathrm{Cl}}(r)
    =
    \frac{w_r}{Z_{\mathrm{Cl}}},
    \quad
    w_0=1,
    \quad
    w_r
    =
    \frac{2^r}{\prod_{j=1}^{r}(2^j-1)^2}.
    \label{eq:app_finite_rate_distribution}
\end{equation}
Numerically,
\begin{equation}
    Z_{\mathrm{Cl}}\approx3.462746619,
    \quad
    \mu_{\mathrm{Cl}}
    \equiv
    \mathbb E_{\mathrm{Cl}}[r]
    \approx0.850179831.
\end{equation}
Since \(I_c=K-r\),
\begin{align}
    \mathbb E_{\mathrm{Cl}}[i_c]
    &=
    1-\frac{\mu_{\mathrm{Cl}}}{R_{\mathrm{code}}N}+o(N^{-1})
    \notag\\
    &=
    \exp\!\left[
        -\frac{\mu_{\mathrm{Cl}}/R_{\mathrm{code}}}{N}
        +o(N^{-1})
    \right].
    \label{eq:app_finite_rate_clifford_scaling}
\end{align}
This is the discrete random-stabilizer counterpart of the Page result below.

\subsection*{Uniform-global-Haar and Page limit}

For the Haar benchmark, take \(D=2^K\) and
\(M_{\mathrm{syn}}=2^m\).  In the syndrome-decoded output basis defined by
\(\mathcal J\) in Eq.~\eqref{eq:syndrome_decoding_isometry}, decompose the
Haar-random encoding isometry into syndrome blocks,
\begin{equation}
    \mathcal J\mathcal U V
    =
    \begin{pmatrix}
        A_1\\ \vdots\\ A_{M_{\mathrm{syn}}}
    \end{pmatrix},
    \quad
    \sum_{s=1}^{M_{\mathrm{syn}}}A_s^\dagger A_s=I_D.
    \label{eq:app_haar_blocks}
\end{equation}
A Haar isometry can be obtained by normalizing an
\(M_{\mathrm{syn}}D\times D\) complex Ginibre matrix.  At large
\(M_{\mathrm{syn}}\), each normalized
block POVM element \(\widehat E_s\) therefore approaches the fixed-trace
Wishart ensemble obtained from a \(D\times D\) Ginibre
block~\cite{ZyczkowskiSommers2001}.  Equivalently, the conditional Choi
branch approaches a Haar-random pure state on
\(\mathbb C^D\otimes\mathbb C^D\).

Consequently, in the large-\(M_{\mathrm{syn}}\) limit,
Eq.~\eqref{eq:app_general_coherent_information} approaches Page's mean
entanglement entropy~\cite{Page1993},
\begin{align}
    \lim_{M_{\mathrm{syn}}\to\infty}
    \mathbb E_{\mathrm{Haar}}[I_c]
    &=
    \frac{H_{D^2}-H_D-(D-1)/(2D)}{\ln 2},
    \notag\\
    D&=2^K,
    \label{eq:app_page_formula}
\end{align}
where \(H_n=\sum_{j=1}^{n}j^{-1}\).  When \(K,m\to\infty\) at fixed
\(0<R_{\mathrm{code}}<1\),
\begin{align}
    \mathbb E_{\mathrm{Haar}}[I_c]
    &=
    K-\frac{1}{2\ln 2}+o(1),
    \notag\\
    \mathbb E_{\mathrm{Haar}}[i_c]
    &=
    \exp\!\left[
        -\frac{1/(2R_{\mathrm{code}}\ln 2)}{N}
        +o(N^{-1})
    \right].
    \label{eq:app_finite_rate_haar_scaling}
\end{align}

The global-Clifford and global-Haar ensembles thus explain why a fully
scrambled finite-rate model can have an \(O(1)\) total
coherent-information deficit but only an \(O(N^{-1})\) deficit per logical
qubit.  They do not determine the threshold, critical exponents, or the
function \(h(p)\) for the local \(q\)-unitary ensembles; those remain
model-dependent properties to be tested numerically.

\section{Relation between logical-group change and the state-level MAP return probability}
\label{app:map_delta}

In this appendix, we relate the logical-group diagnostic introduced in the
main text to the state-level MAP return probability.  The initial
logical-basis state \(c\) and coherent-error realization \(\mathcal U\) are
fixed throughout and are suppressed in some expressions for readability.
The probability below is the logical-basis Born distribution of a syndrome
branch, not a posterior over unknown physical error classes or a
state-independent channel-recovery probability.

Let the initial logical code state be stabilized by
\begin{equation}
    \mathcal{G}_{\mathcal L}=\langle \bar g_1,\dots,\bar g_k\rangle.
\end{equation}
Choose conjugate logical classes
\(\{\overline{\tilde g}_i\}_{i=1}^k\).  Their induced Hermitian logical
operators are \(\widehat g_i\) and \(\widehat{\tilde g}_i\), as in
Eq.~\eqref{eq:induced_logical_operator}.  To keep the formulas readable, only
in this appendix we abbreviate these induced operators as \(g_i\) and
\(\tilde g_i\); likewise, \(h'_j\) below denotes an induced logical operator,
whereas an underline denotes the phase-free image of its logical class.  The
induced operators satisfy
\begin{equation}
    [g_i,g_j]=[\tilde g_i,\tilde g_j]=0,
    \quad
    g_i\tilde g_j = (-1)^{\delta_{ij}} \tilde g_j g_i.
\end{equation}
Thus the corresponding classes form a logical symplectic basis modulo the
physical stabilizer group \(\mathcal S\).

For a fixed syndrome branch \(s\), let
\(\mathcal G'_{\mathcal L}(s;c,\mathcal U)
=\langle\bar h'_1,\dots,\bar h'_k\rangle\)
denote the post-measurement logical stabilizer group.  After phases are
discarded, each generator admits the unique binary expansion
\begin{equation}
    \underline h'_j
    =
    \prod_{i=1}^k \underline g_i^{\,b_{ij}}
    \prod_{i=1}^k \underline{\tilde g}_i^{\,a_{ij}},
    \quad
    a_{ij},b_{ij}\in \mathbb F_2.
\end{equation}
We define the logical commutator matrix
\begin{equation}
    (T_{\mathcal L})_{ij}
    =
    \begin{cases}
        0, & [g_i,h'_j]=0,\\
        1, & [g_i,h'_j]\neq 0.
    \end{cases}
\end{equation}
Since \(g_i\) anticommutes only with \(\tilde g_i\), one immediately has
\begin{equation}
    (T_{\mathcal L})_{ij}=a_{ij}.
\end{equation}

Now consider the combined projective logical subspace
\begin{equation}
    G_{\mathrm{comb.}}^{\mathrm{sf}}(s;c,\mathcal U)
    =
    \operatorname{span}_{\mathbb F_2}
    \left(
        \underline{\mathcal G}_{\mathcal L}(c),
        \underline{\mathcal G}'_{\mathcal L}(s;c,\mathcal U)
    \right).
\end{equation}
Here the underlines denote the phase-free logical subspaces defined in
Sec.~\ref{sec:recoverability}, after the physical stabilizer redundancy has
been removed.  Relative to the initial projective logical stabilizer
subspace, only the
\(\tilde g_i\)-components of the generators of
\(\underline{\mathcal G}'_{\mathcal L}(s;c,\mathcal U)\) contribute new independent
logical directions. Hence
\begin{equation}
\begin{aligned}
    \Delta_{\mathrm{Logi.}}(s;c,\mathcal U)
    &\equiv
    \dim_{\mathbb F_2}G_{\mathrm{comb.}}^{\mathrm{sf}}(s;c,\mathcal U)
    -
    \dim_{\mathbb F_2}\underline{\mathcal G}_{\mathcal L}(c)\\
    &=
    \rank_{\mathbb F_2}(a_{ij})
    =
    \rank_{\mathbb F_2} T_{\mathcal L}.
\end{aligned}
    \label{eq:delta_rankT_compact}
\end{equation}

We now connect this rank to the state-level MAP rule. After applying a fixed
syndrome representative that returns the branch \(s\) to the code space, we
expand the resulting state in the logical basis of the fixed input. Let
\begin{equation}
    \ell=(\ell_1,\dots,\ell_k)\in\mathbb F_2^k
\end{equation}
denote the logical label in the basis specified by the commuting observables
\(\{g_i\}\), with the convention that the eigenvalue of \(g_i\) is
\((-1)^{\ell_i}\).
At fixed coherent-error realization \(\mathcal U\), the logical-basis Born
weights define the state-level distribution
\(\mathbb P(\ell\mid s,c,\mathcal U)\). The MAP return probability is, by
definition,
\begin{equation}
    P_{\mathrm{rec}}^{\mathrm{opt}}(s;c,\mathcal U)
    =
    \max_{\ell\in\mathbb F_2^k}\mathbb P(\ell\mid s,c,\mathcal U).
    \label{eq:map_def_appendix}
\end{equation}

We now derive this return probability directly from the logical-basis
expansion of the post-measurement branch. After applying a fixed syndrome
representative that returns the branch \(s\) to the code space, the resulting
state is a stabilizer state within the \(k\)-qubit logical subspace. Let
\(\{ \ket{\ell} \}_{\ell\in\mathbb F_2^k}\) denote the simultaneous
eigenbasis of the commuting logical observables \(\{g_i\}\), so that
\(g_i\ket{\ell}=(-1)^{\ell_i}\ket{\ell}\). Expanding the branch state
in this basis,
\begin{equation}
    \ket{\psi_s}_{\mathcal L}
    =
    \sum_{\ell\in\mathbb F_2^k} \alpha_\ell \ket{\ell},
\end{equation}
the Born distribution entering the state-level MAP rule is
\begin{equation}
    \mathbb P(\ell\mid s,c,\mathcal U)=|\alpha_\ell|^2,
\end{equation}
and therefore, by definition,
\begin{equation}
    P_{\mathrm{rec}}^{\mathrm{opt}}(s;c,\mathcal U)
    =
    \max_{\ell\in\mathbb F_2^k} |\alpha_\ell|^2.
    \label{eq:map_def_coeff}
\end{equation}
If \(\widehat\ell\) is a maximizing label, choose a physical logical-Pauli
representative \(P_{s;c,\mathcal U}^{\mathrm{MAP}}\) whose induced operator
\(\widehat P_{s;c,\mathcal U}^{\mathrm{MAP}}\) maps
\(\ket{\widehat\ell}\) to \(\ket c\).  The state-specific correction in
Eq.~\eqref{eq:map_recovery_unitary} then maps that component to the input,
so its conditional fidelity is exactly the maximum in
Eq.~\eqref{eq:map_def_coeff}.

By App.~\ref{app:stab_measurement}, the post-measurement logical stabilizer constraints fix \(k-\rank T_{\mathcal L}\) independent combinations of the logical charges, while leaving \(\rank T_{\mathcal L}\) unfixed. Equivalently, in the logical basis \(\{\ket{\ell}\}\), the state \(\ket{\psi_s}_{\mathcal L}\) has support on exactly
\begin{equation}
    2^{\rank T_{\mathcal L}}
\end{equation}
distinct logical charge sectors. Since \(\ket{\psi_s}_{\mathcal L}\) is itself a stabilizer state, all nonzero coefficients in this expansion have the same magnitude. Hence
\begin{equation}
    |\alpha_\ell|^2
    =
    \begin{cases}
        2^{-\rank T_{\mathcal L}}, & \alpha_\ell\neq 0,\\
        0, & \alpha_\ell = 0.
    \end{cases}
\end{equation}
Substituting this into Eq.~\eqref{eq:map_def_coeff}, we obtain
\begin{equation}
    P_{\mathrm{rec}}^{\mathrm{opt}}(s;c,\mathcal U)
    =
    2^{-\rank T_{\mathcal L}}.
\end{equation}
Using Eq.~\eqref{eq:delta_rankT_compact}, this becomes
\begin{equation}
    P_{\mathrm{rec}}^{\mathrm{opt}}(s;c,\mathcal U)
    =
    2^{-\Delta_{\mathrm{Logi.}}(s;c,\mathcal U)}.
    \label{eq:map_delta_final_branch}
\end{equation}

For the Clifford coherent-error settings studied in this work, the binary
stabilizer structure is independent of both the logical-basis label \(c\) and
the allowed syndrome branch \(s\) once the disorder realization
\(\mathcal U\) is fixed. Indeed, \(c\) fixes only the signs of the initial
logical stabilizer generators, while \(s\) fixes only the signs of the
measured generators. Conjugation by the Pauli syndrome representative also
affects only signs. None of these operations changes the binary symplectic
vectors that determine \(T_{\mathcal L}\). In the language of
App.~\ref{app:stab_measurement}, changing \(c\) changes the affine offsets of
the syndrome constraints but not their null space or rank; all syndromes
within the resulting affine solution set have equal probability. Therefore,
for every input label \(c\) and every syndrome \(s\) with
\(\mathbb P[s\mid c,\mathcal U]>0\),
\begin{equation}
    T_{\mathcal L}(s;c,\mathcal{U})\equiv T_{\mathcal L}(\mathcal{U}),
    \quad
    \Delta_{\mathrm{Logi.}}(s;c,\mathcal U)
    \equiv
    \Delta_{\mathrm{Logi.}}(\mathcal{U}),
\end{equation}
and Eq.~\eqref{eq:map_delta_final_branch} reduces to
\begin{equation}
    P_{\mathrm{rec}}^{\mathrm{opt}}(\mathcal{U})
    =
    2^{-\rank T_{\mathcal L}(\mathcal{U})}
    =
    2^{-\Delta_{\mathrm{Logi.}}(\mathcal{U})}.
    \label{eq:map_delta_final_signfree}
\end{equation}
Averaging over disorder realizations then gives
\begin{equation}
    \mathbb E_{\mathcal{U}}\!\left[P_{\mathrm{rec}}^{\mathrm{opt}}\right]
    =
    \mathbb E_{\mathcal{U}}\!\left[2^{-\Delta_{\mathrm{Logi.}}(\mathcal{U})}\right].
\end{equation}

For finite-rate random stabilizer codes with \(K\) logical qubits, writing
\begin{equation}
    \delta_{\mathrm{Logi.}}=\frac{1}{K}\Delta_{\mathrm{Logi.}},
\end{equation}
one equivalently has
\begin{equation}
    P_{\mathrm{rec}}^{\mathrm{opt}}(\mathcal C,\mathcal{U})
    =
    2^{-K(\mathcal C)\,
    \delta_{\mathrm{Logi.}}(\mathcal C,\mathcal{U})}.
\end{equation}
Thus, if \(\delta_{\mathrm{Logi.}}(\mathcal C,\mathcal U)\) converges in
probability to a strictly positive post-threshold value, the state-level MAP
return probability is exponentially small in the number of logical qubits
for a typical realization.

\bibliography{ref}

@article{preskill2018quantum,
  title={Quantum computing in the NISQ era and beyond},
  author={Preskill, John},
  journal={Quantum},
  volume={2},
  pages={79},
  year={2018},
  publisher={Verein zur F{\"o}rderung des Open Access Publizierens in den Quantenwissenschaften}
}

@article{kitaev2003fault,
  title={Fault-tolerant quantum computation by anyons},
  author={Kitaev, A Yu},
  journal={Annals of physics},
  volume={303},
  number={1},
  pages={2--30},
  year={2003},
  publisher={Elsevier}
}

@article{fowler2012surface,
  title={Surface codes: Towards practical large-scale quantum computation},
  author={Fowler, Austin G and Mariantoni, Matteo and Martinis, John M and Cleland, Andrew N},
  journal={Physical Review A—Atomic, Molecular, and Optical Physics},
  volume={86},
  number={3},
  pages={032324},
  year={2012},
  publisher={APS}
}

@article{tillich2013quantum,
  title={Quantum LDPC codes with positive rate and minimum distance proportional to the square root of the blocklength},
  author={Tillich, Jean-Pierre and Z{\'e}mor, Gilles},
  journal={IEEE Transactions on Information Theory},
  volume={60},
  number={2},
  pages={1193--1202},
  year={2013},
  publisher={IEEE}
}

@inproceedings{leverrier2015quantum,
  title={Quantum expander codes},
  author={Leverrier, Anthony and Tillich, Jean-Pierre and Z{\'e}mor, Gilles},
  booktitle={2015 IEEE 56th Annual Symposium on Foundations of Computer Science},
  pages={810--824},
  year={2015},
  organization={IEEE}
}

@article{dennis2002topological,
  title={Topological quantum memory},
  author={Dennis, Eric and Kitaev, Alexei and Landahl, Andrew and Preskill, John},
  journal={Journal of Mathematical Physics},
  volume={43},
  number={9},
  pages={4452--4505},
  year={2002},
  publisher={American Institute of Physics}
}

@inproceedings{brown2013short,
  title={Short random circuits define good quantum error correcting codes},
  author={Brown, Winton and Fawzi, Omar},
  booktitle={2013 IEEE International Symposium on Information Theory},
  pages={346--350},
  year={2013},
  organization={IEEE}
}

@article{nelson2023fault,
  title={Fault-Tolerant Quantum Memory using Low-Depth Random Circuit Codes},
  author={Nelson, Jon and Bentsen, Gregory and Flammia, Steven T and Gullans, Michael J},
  journal={arXiv preprint arXiv:2311.17985},
  year={2023}
}

@article{PhysRevResearch.6.023055,
  title = {Low-depth random Clifford circuits for quantum coding against Pauli noise using a tensor-network decoder},
  author = {Darmawan, Andrew S. and Nakata, Yoshifumi and Tamiya, Shiro and Yamasaki, Hayata},
  journal = {Phys. Rev. Res.},
  volume = {6},
  issue = {2},
  pages = {023055},
  numpages = {14},
  year = {2024},
  month = {Apr},
  publisher = {American Physical Society},
  doi = {10.1103/PhysRevResearch.6.023055},
  url = {https://link.aps.org/doi/10.1103/PhysRevResearch.6.023055}
}

@article{PhysRevX.11.031066,
  title = {Quantum Coding with Low-Depth Random Circuits},
  author = {Gullans, Michael J. and Krastanov, Stefan and Huse, David A. and Jiang, Liang and Flammia, Steven T.},
  journal = {Phys. Rev. X},
  volume = {11},
  issue = {3},
  pages = {031066},
  numpages = {23},
  year = {2021},
  month = {Sep},
  publisher = {American Physical Society},
  doi = {10.1103/PhysRevX.11.031066},
  url = {https://link.aps.org/doi/10.1103/PhysRevX.11.031066}
}

@article{chen2024nishimori,
  title={Nishimori transition across the error threshold for constant-depth quantum circuits},
  author={Chen, Edward H and Zhu, Guo-Yi and Verresen, Ruben and Seif, Alireza and B{\"a}umer, Elisa and Layden, David and Tantivasadakarn, Nathanan and Zhu, Guanyu and Sheldon, Sarah and Vishwanath, Ashvin and others},
  journal={Nature Physics},
  pages={1--7},
  year={2024},
  publisher={Nature Publishing Group}
}

@article{PhysRevLett.131.200201,
  title = {Nishimori's Cat: Stable Long-Range Entanglement from Finite-Depth Unitaries and Weak Measurements},
  author = {Zhu, Guo-Yi and Tantivasadakarn, Nathanan and Vishwanath, Ashvin and Trebst, Simon and Verresen, Ruben},
  journal = {Phys. Rev. Lett.},
  volume = {131},
  issue = {20},
  pages = {200201},
  numpages = {9},
  year = {2023},
  month = {Nov},
  publisher = {American Physical Society},
  doi = {10.1103/PhysRevLett.131.200201},
  url = {https://link.aps.org/doi/10.1103/PhysRevLett.131.200201}
}

@article{knill1998resilient,
  title={Resilient quantum computation: error models and thresholds},
  author={Knill, Emanuel and Laflamme, Raymond and Zurek, Wojciech H},
  journal={Proceedings of the Royal Society of London. Series A: Mathematical, Physical and Engineering Sciences},
  volume={454},
  number={1969},
  pages={365--384},
  year={1998},
  publisher={The Royal Society}
}

@article{RevModPhys.87.307,
  title = {Quantum error correction for quantum memories},
  author = {Terhal, Barbara M.},
  journal = {Rev. Mod. Phys.},
  volume = {87},
  issue = {2},
  pages = {307--346},
  numpages = {40},
  year = {2015},
  month = {Apr},
  publisher = {American Physical Society},
  doi = {10.1103/RevModPhys.87.307},
  url = {https://link.aps.org/doi/10.1103/RevModPhys.87.307}
}

@article{PhysRevX.2.021004,
  title = {Strong Resilience of Topological Codes to Depolarization},
  author = {Bombin, H. and Andrist, Ruben S. and Ohzeki, Masayuki and Katzgraber, Helmut G. and Martin-Delgado, M. A.},
  journal = {Phys. Rev. X},
  volume = {2},
  issue = {2},
  pages = {021004},
  numpages = {10},
  year = {2012},
  month = {Apr},
  publisher = {American Physical Society},
  doi = {10.1103/PhysRevX.2.021004},
  url = {https://link.aps.org/doi/10.1103/PhysRevX.2.021004}
}

@article{bravyi2018correcting,
  title={Correcting coherent errors with surface codes},
  author={Bravyi, Sergey and Englbrecht, Matthias and K{\"o}nig, Robert and Peard, Nolan},
  journal={npj Quantum Information},
  volume={4},
  number={1},
  pages={55},
  year={2018},
  publisher={Nature Publishing Group UK London}
}

@article{zhao2021analytic,
  title={An analytic study of the independent coherent errors in the surface code},
  author={Zhao, Yuanchen and Liu, Dong E},
  journal={arXiv preprint arXiv:2112.00473},
  year={2021}
}

@article{marton2023coherent,
  title={Coherent errors and readout errors in the surface code},
  author={M{\'a}rton, {\'A}ron and Asb{\'o}th, J{\'a}nos K},
  journal={Quantum},
  volume={7},
  pages={1116},
  year={2023},
  publisher={Verein zur F{\"o}rderung des Open Access Publizierens in den Quantenwissenschaften}
}

@article{PhysRevLett.131.060603,
  title = {Coherent-Error Threshold for Surface Codes from Majorana Delocalization},
  author = {Venn, Florian and Behrends, Jan and B\'eri, Benjamin},
  journal = {Phys. Rev. Lett.},
  volume = {131},
  issue = {6},
  pages = {060603},
  numpages = {6},
  year = {2023},
  month = {Aug},
  publisher = {American Physical Society},
  doi = {10.1103/PhysRevLett.131.060603},
  url = {https://link.aps.org/doi/10.1103/PhysRevLett.131.060603}
}

@article{darmawan2017tensor,
  title   = {Tensor-Network Simulations of the Surface Code under Realistic Noise},
  author  = {Darmawan, Andrew S. and Poulin, David},
  journal = {Phys. Rev. Lett.},
  volume  = {119},
  pages   = {040502},
  year    = {2017},
  doi     = {10.1103/PhysRevLett.119.040502}
}

@article{behrends2024surface,
  title = {The Surface Code beyond Pauli Channels: Logical Noise Coherence, Information-Theoretic Measures, and Errorfield-Double Phenomenology},
  author = {Behrends, Jan and B\'eri, Benjamin},
  journal = {PRX Quantum},
  volume = {6},
  issue = {4},
  pages = {040350},
  year = {2025},
  month = {Dec},
  publisher = {American Physical Society},
  doi = {10.1103/psf5-b6j2},
  url = {https://link.aps.org/doi/10.1103/psf5-b6j2}
}

@article{schultz2021schwarma,
  title = {{SchWARMA}: A Model-Based Approach for Time-Correlated Noise in Quantum Circuits},
  author = {Schultz, Kevin and Quiroz, Gregory and Titum, Paraj and Clader, B. D.},
  journal = {Phys. Rev. Research},
  volume = {3},
  pages = {033229},
  year = {2021},
  publisher = {American Physical Society},
  doi = {10.1103/PhysRevResearch.3.033229},
  url = {https://link.aps.org/doi/10.1103/PhysRevResearch.3.033229}
}

@article{marshall2025leakage,
  title = {Incoherent Approximation of Leakage in Quantum Error Correction},
  author = {Marshall, Jeffrey and Kafri, Dvir},
  journal = {Phys. Rev. Applied},
  volume = {23},
  pages = {054025},
  year = {2025},
  publisher = {American Physical Society},
  doi = {10.1103/PhysRevApplied.23.054025},
  url = {https://link.aps.org/doi/10.1103/PhysRevApplied.23.054025}
}

@article{bao2024phases,
  title={Phases of decodability in the surface code with unitary errors},
  author={Bao, Yimu and Anand, Sajant},
  journal={arXiv preprint arXiv:2411.05785},
  year={2024}
}

@article{cheng2024emergent,
  title = {Emergent Unitary Designs for Encoded Qubits from Coherent Errors and Syndrome Measurements},
  author = {Cheng, Zihan and Huang, Eric and Khemani, Vedika and Gullans, Michael J. and Ippoliti, Matteo},
  journal = {PRX Quantum},
  volume = {6},
  issue = {3},
  pages = {030333},
  year = {2025},
  publisher = {American Physical Society},
  doi = {10.1103/bnld-2chd},
  url = {https://link.aps.org/doi/10.1103/bnld-2chd}
}

@article{liu2023quantum,
  title={Quantum Entanglement Phase Transitions and Computational Complexity: Insights from {Ising} Models},
  author={Liu, Hanchen and Ravindranath, Vikram and Chen, Xiao},
  journal={Physical Review B},
  volume={111},
  number={2},
  pages={024312},
  year={2025},
  publisher={American Physical Society},
  doi={10.1103/PhysRevB.111.024312}
}

@article{McGinley2025,
  title={Measurement-Induced Entanglement and Complexity in Random Constant-Depth 2D Quantum Circuits},
  author={McGinley, Max and Ho, Wen Wei and Malz, Daniel},
  journal={Physical Review X},
  volume={15},
  number={2},
  pages={021059},
  year={2025},
  publisher={American Physical Society},
  doi={10.1103/PhysRevX.15.021059}
}

@article{PhysRevA.77.042322,
  title = {Statistical mechanical models and topological color codes},
  author = {Bombin, H. and Martin-Delgado, M. A.},
  journal = {Phys. Rev. A},
  volume = {77},
  issue = {4},
  pages = {042322},
  numpages = {10},
  year = {2008},
  month = {Apr},
  publisher = {American Physical Society},
  doi = {10.1103/PhysRevA.77.042322},
  url = {https://link.aps.org/doi/10.1103/PhysRevA.77.042322}
}

@article{PhysRevA.76.022304,
  title = {Measurement-based quantum computation with the toric code states},
  author = {Bravyi, Sergey and Raussendorf, Robert},
  journal = {Phys. Rev. A},
  volume = {76},
  issue = {2},
  pages = {022304},
  numpages = {10},
  year = {2007},
  month = {Aug},
  publisher = {American Physical Society},
  doi = {10.1103/PhysRevA.76.022304},
  url = {https://link.aps.org/doi/10.1103/PhysRevA.76.022304}
}

@article{Fisher2022qey,
    author = "Fisher, Matthew P. A. and Khemani, Vedika and Nahum, Adam and Vijay, Sagar",
    title = "{Random Quantum Circuits}",
    eprint = "2207.14280",
    archivePrefix = "arXiv",
    primaryClass = "quant-ph",
    doi = "10.1146/annurev-conmatphys-031720-030658",
    journal = "Ann. Rev. Condensed Matter Phys.",
    volume = "14",
    pages = "335--379",
    year = "2023"
}

@article{chubb2021statistical,
  title={Statistical mechanical models for quantum codes with correlated noise},
  author={Chubb, Christopher T and Flammia, Steven T},
  journal={Annales de l’Institut Henri Poincar{\'e} D},
  volume={8},
  number={2},
  pages={269--321},
  year={2021}
}

@inproceedings{aharonov1997fault,
  title={Fault-tolerant quantum computation with constant error},
  author={Aharonov, Dorit and Ben-Or, Michael},
  booktitle={Proceedings of the twenty-ninth annual ACM symposium on Theory of computing},
  pages={176--188},
  year={1997}
}

@article{koenig2014efficiently,
  title={How to efficiently select an arbitrary Clifford group element},
  author={Koenig, Robert and Smolin, John A},
  journal={Journal of Mathematical Physics},
  volume={55},
  number={12},
  year={2014},
  publisher={AIP Publishing}
}

@book{nielsen2010quantum,
  title={Quantum computation and quantum information},
  author={Nielsen, Michael A and Chuang, Isaac L},
  year={2010},
  publisher={Cambridge university press}
}

@article{PhysRevA.54.2629,
  title = {Quantum data processing and error correction},
  author = {Schumacher, Benjamin and Nielsen, M. A.},
  journal = {Phys. Rev. A},
  volume = {54},
  issue = {4},
  pages = {2629--2635},
  numpages = {0},
  year = {1996},
  month = {Oct},
  publisher = {American Physical Society},
  doi = {10.1103/PhysRevA.54.2629},
  url = {https://link.aps.org/doi/10.1103/PhysRevA.54.2629}
}

@article{PhysRevA.55.1613,
  title = {Capacity of the noisy quantum channel},
  author = {Lloyd, Seth},
  journal = {Phys. Rev. A},
  volume = {55},
  issue = {3},
  pages = {1613--1622},
  numpages = {0},
  year = {1997},
  month = {Mar},
  publisher = {American Physical Society},
  doi = {10.1103/PhysRevA.55.1613},
  url = {https://link.aps.org/doi/10.1103/PhysRevA.55.1613}
}

@book{mezard2009information,
  title={Information, physics, and computation},
  author={Mezard, Marc and Montanari, Andrea},
  year={2009},
  publisher={Oxford University Press}
}

@article{mackay1997near,
  title={Near Shannon limit performance of low density parity check codes},
  author={MacKay, David JC and Neal, Radford M},
  journal={Electronics letters},
  volume={33},
  number={6},
  pages={457--458},
  year={1997}
}

@article{colmenarez2024accurate,
  title={Accurate optimal quantum error correction thresholds from coherent information},
  author={Colmenarez, Luis and Huang, Ze-Min and Diehl, Sebastian and M{\"u}ller, Markus},
  journal={Physical Review Research},
  volume={6},
  number={4},
  pages={L042014},
  year={2024},
  publisher={APS}
}

@article{skinner2019measurement,
  title={Measurement-induced phase transitions in the dynamics of entanglement},
  author={Skinner, Brian and Ruhman, Jonathan and Nahum, Adam},
  journal={Physical Review X},
  volume={9},
  number={3},
  pages={031009},
  year={2019},
  publisher={APS}
}

@article{gullans2020dynamical,
  title={Dynamical purification phase transition induced by quantum measurements},
  author={Gullans, Michael J and Huse, David A},
  journal={Physical Review X},
  volume={10},
  number={4},
  pages={041020},
  year={2020},
  publisher={APS}
}

@article{li2019measurement,
  title={Measurement-driven entanglement transition in hybrid quantum circuits},
  author={Li, Yaodong and Chen, Xiao and Fisher, Matthew PA},
  journal={Physical Review B},
  volume={100},
  number={13},
  pages={134306},
  year={2019},
  publisher={APS}
}

@article{li2018quantum,
  title={Quantum Zeno effect and the many-body entanglement transition},
  author={Li, Yaodong and Chen, Xiao and Fisher, Matthew PA},
  journal={Physical Review B},
  volume={98},
  number={20},
  pages={205136},
  year={2018},
  publisher={APS}
}

@article{PhysRevResearch.6.013137,
  title = {Surface codes, quantum circuits, and entanglement phases},
  author = {Behrends, Jan and Venn, Florian and B\'eri, Benjamin},
  journal = {Phys. Rev. Res.},
  volume = {6},
  issue = {1},
  pages = {013137},
  numpages = {15},
  year = {2024},
  month = {Feb},
  publisher = {American Physical Society},
  doi = {10.1103/PhysRevResearch.6.013137},
  url = {https://link.aps.org/doi/10.1103/PhysRevResearch.6.013137}
}

@article{turkeshi2024error,
  title={Error-resilience phase transitions in encoding-decoding quantum circuits},
  author={Turkeshi, Xhek and Sierant, Piotr},
  journal={Physical Review Letters},
  volume={132},
  number={14},
  pages={140401},
  year={2024},
  publisher={APS}
}

@article{eckstein2024robust,
  title={Robust teleportation of a surface code and cascade of topological quantum phase transitions},
  author={Eckstein, Finn and Han, Bo and Trebst, Simon and Zhu, Guo-Yi},
  journal={PRX Quantum},
  volume={5},
  number={4},
  pages={040313},
  year={2024},
  publisher={APS}
}

@article{wang2025decoherence,
  title={Decoherence-induced self-dual criticality in topological states of matter},
  author={Wang, Qingyuan and Vasseur, Romain and Trebst, Simon and Ludwig, Andreas WW and Zhu, Guo-Yi},
  journal={arXiv preprint arXiv:2502.14034},
  year={2025}
}

@article{putz2025learning,
  title={Learning transitions in classical Ising models and deformed toric codes},
  author={P{\"u}tz, Malte and Garratt, Samuel J and Nishimori, Hidetoshi and Trebst, Simon and Zhu, Guo-Yi},
  journal={arXiv preprint arXiv:2504.12385},
  year={2025}
}

@article{iyer2015hardness,
  title={Hardness of decoding quantum stabilizer codes},
  author={Iyer, Pavithran and Poulin, David},
  journal={IEEE Transactions on Information Theory},
  volume={61},
  number={9},
  pages={5209--5223},
  year={2015},
  publisher={IEEE},
  doi={10.1109/TIT.2015.2422294}
}

@article{chamberland2017hard,
  title={Hard decoding algorithm for optimizing thresholds under general Markovian noise},
  author={Chamberland, Christopher and Wallman, Joel J. and Beale, Stefanie and Laflamme, Raymond},
  journal={Phys. Rev. A},
  volume={95},
  issue={4},
  pages={042332},
  year={2017},
  publisher={American Physical Society},
  doi={10.1103/PhysRevA.95.042332},
  url={https://link.aps.org/doi/10.1103/PhysRevA.95.042332}
}

@article{beale2018quantum,
  title={Quantum Error Correction Decoheres Noise},
  author={Beale, Stefanie J. and Wallman, Joel J. and Guti{\'e}rrez, Mauricio and Brown, Kenneth R. and Laflamme, Raymond},
  journal={Phys. Rev. Lett.},
  volume={121},
  issue={19},
  pages={190501},
  year={2018},
  publisher={American Physical Society},
  doi={10.1103/PhysRevLett.121.190501},
  url={https://link.aps.org/doi/10.1103/PhysRevLett.121.190501}
}

@article{PhysRevLett.127.235701,
  title = {Topological Order and Criticality in $(2+1)\mathrm{D}$ Monitored Random Quantum Circuits},
  author = {Lavasani, Ali and Alavirad, Yahya and Barkeshli, Maissam},
  journal = {Phys. Rev. Lett.},
  volume = {127},
  issue = {23},
  pages = {235701},
  numpages = {6},
  year = {2021},
  month = {Dec},
  publisher = {American Physical Society},
  doi = {10.1103/PhysRevLett.127.235701},
  url = {https://link.aps.org/doi/10.1103/PhysRevLett.127.235701}
}

@article{hchr-rqq9,
  title = {Mixed-State Topological Order under Coherent Noise},
  author = {Lee, Seunghun and Moon, Eun-Gook},
  journal = {PRX Quantum},
  volume = {6},
  issue = {3},
  pages = {030355},
  numpages = {26},
  year = {2025},
  month = {Sep},
  publisher = {American Physical Society},
  doi = {10.1103/hchr-rqq9},
  url = {https://link.aps.org/doi/10.1103/hchr-rqq9}
}

@article{gskb-t5ql,
  title = {Statistical Mechanical Mapping and Maximum-Likelihood Thresholds for the Surface Code under Generic Single-Qubit Coherent Errors},
  author = {Behrends, Jan and B\'eri, Benjamin},
  journal = {PRX Quantum},
  volume = {6},
  issue = {4},
  pages = {040305},
  numpages = {17},
  year = {2025},
  month = {Oct},
  publisher = {American Physical Society},
  doi = {10.1103/gskb-t5ql},
  url = {https://link.aps.org/doi/10.1103/gskb-t5ql}
}

@article{iverson2020coherence,
  title={Coherence in logical quantum channels},
  author={Iverson, Joseph K and Preskill, John},
  journal={New Journal of Physics},
  volume={22},
  number={7},
  pages={073066},
  year={2020},
  publisher={IOP Publishing}
}

@article{Page1993,
  title={Average entropy of a subsystem},
  author={Page, Don N.},
  journal={Physical Review Letters},
  volume={71},
  number={9},
  pages={1291--1294},
  year={1993},
  publisher={American Physical Society},
  doi={10.1103/PhysRevLett.71.1291}
}

@article{DahlstenPlenio2006,
  title={Exact entanglement probability distribution of bi-partite randomised stabilizer states},
  author={Dahlsten, Oscar C. O. and Plenio, Martin B.},
  journal={Quantum Information and Computation},
  volume={6},
  number={6},
  pages={527--538},
  year={2006},
  eprint={quant-ph/0511119},
  archivePrefix={arXiv}
}

@article{SmithLeung2006,
  title={Typical entanglement of stabilizer states},
  author={Smith, Graeme and Leung, Debbie},
  journal={Physical Review A},
  volume={74},
  number={6},
  pages={062314},
  year={2006},
  publisher={American Physical Society},
  doi={10.1103/PhysRevA.74.062314}
}

@article{ZyczkowskiSommers2001,
  title={Induced measures in the space of mixed quantum states},
  author={{\.Z}yczkowski, Karol and Sommers, Hans-J{\"u}rgen},
  journal={Journal of Physics A: Mathematical and General},
  volume={34},
  number={35},
  pages={7111--7125},
  year={2001},
  publisher={IOP Publishing},
  doi={10.1088/0305-4470/34/35/335}
}

@article{bombin2023logicalblocks,
  title={Logical Blocks for Fault-Tolerant Topological Quantum Computation},
  author={Bombin, Hector and Dawson, Chris and Mishmash, Ryan V. and Nickerson, Naomi and Pastawski, Fernando and Roberts, Sam},
  journal={PRX Quantum},
  volume={4},
  number={2},
  pages={020303},
  year={2023},
  publisher={American Physical Society},
  doi={10.1103/PRXQuantum.4.020303},
  eprint={2112.12160},
  archivePrefix={arXiv},
  primaryClass={quant-ph}
}

@article{Napp2022,
  title={Efficient Classical Simulation of Random Shallow 2D Quantum Circuits},
  author={Napp, John C. and La Placa, Rolando L. and Dalzell, Alexander M. and Brand{\~a}o, Fernando G. S. L. and Harrow, Aram W.},
  journal={Physical Review X},
  volume={12},
  number={2},
  pages={021021},
  year={2022},
  publisher={American Physical Society},
  doi={10.1103/PhysRevX.12.021021}
}

@article{Watts2025,
  title={Quantum Advantage from Measurement-Induced Entanglement in Random Shallow Circuits},
  author={Watts, Adam Bene and Gosset, David and Liu, Yinchen and Soleimanifar, Mehdi},
  journal={PRX Quantum},
  volume={6},
  number={1},
  pages={010356},
  year={2025},
  publisher={American Physical Society},
  doi={10.1103/PRXQuantum.6.010356}
}

\end{document}